%% file: main.tex
\documentclass[conference]{IEEEtran}
\IEEEoverridecommandlockouts

\usepackage{cite}
\usepackage{amsmath,amssymb,amsfonts}
\usepackage{algorithm}
\usepackage{algpseudocode}
\usepackage{graphicx}
\usepackage[table]{xcolor}
\usepackage{tikz}
\usepackage{subfig} 
\usepackage{fmtcount}
\usepackage{xspace}
\usepackage{url}
\usepackage{acronym}
\usepackage{adjustbox}
\usepackage{float}
\usepackage{latexsym}
\usepackage{amsmath}
\usepackage{courier}
\usepackage{diagbox}
\usepackage{comment}
\usepackage{longtable}
\usepackage{tabularx}
\usepackage{cancel}
\usepackage{array}
\usepackage{multirow}
\usepackage{booktabs}
\newcommand{\sysname}{\textsf{DeepLeak}\xspace}
\renewcommand\fbox{\fcolorbox{black}{lightgray}}

\usepackage{tcolorbox}
\tcbset{
  colback=gray!5!white,
  colframe=gray!40!black,
  boxrule=0.5pt,
  arc=2pt,
  left=4pt, right=4pt, top=4pt, bottom=4pt
}
\newenvironment{summarybox}{\begin{tcolorbox}\itshape}{\end{tcolorbox}}
\usepackage{algorithm}
\usepackage{algpseudocode}

\usepackage{amsmath,amssymb}

\algrenewcommand\algorithmicrequire{\textbf{Require:}}
\algrenewcommand\algorithmicensure{\textbf{Ensure:}}

\begin{document}

% \title{\sysname{}: All Explanations Are Leaky, But Some Are More Leaky Than Others}

\title{\sysname{}: Privacy Enhancing Hardening of Model Explanations Against Membership Leakage}

\author{
\IEEEauthorblockN{
Firas Ben Hmida\IEEEauthorrefmark{1}, 
Zain Sbeih\IEEEauthorrefmark{2}, 
Philemon Hailemariam\IEEEauthorrefmark{3}, 
Birhanu Eshete\IEEEauthorrefmark{4}
}
\IEEEauthorblockA{
\IEEEauthorrefmark{1}fbhmida@umich.edu, \;
\IEEEauthorrefmark{2}zsbeih@umich.edu, \;
\IEEEauthorrefmark{3}philemon@umich.edu, \;
\IEEEauthorrefmark{4}birhanu@umich.edu
}
\IEEEauthorblockA{
Department of Computer and Information Science \\
University of Michigan-Dearborn, Michigan, USA
}
}

\maketitle  

\begin{abstract}
Machine learning (ML) explainability is central to algorithmic transparency in high-stakes settings such as predictive diagnostics and loan approval. However, these same domains require rigorous privacy guaranties, creating tension between interpretability and privacy. Although prior work has shown that explanation methods can leak membership information, practitioners still lack systematic guidance on selecting or deploying explanation techniques that balance transparency with privacy.\\
We present \sysname{}, a system to audit and mitigate privacy risks in post-hoc explanation methods. \sysname{} advances the state-of-the-art in three ways: (1) comprehensive leakage profiling: we develop a stronger explanation-aware membership inference attack (MIA) to quantify how much representative explanation methods leak membership information under default configurations; (2) lightweight hardening strategies: we introduce practical, model-agnostic mitigations, including sensitivity-calibrated noise, attribution clipping, and masking, that substantially reduce membership leakage while preserving explanation utility; and (3) root-cause analysis: through controlled experiments, we pinpoint algorithmic properties (e.g., attribution sparsity and sensitivity) that drive leakage.
Evaluating 15 explanation techniques across four families on image benchmarks, \sysname{} shows that default settings can leak up to 74.9\% more membership information than previously reported. Our mitigations cut leakage by up to 95\% (minimum 46.5\%) with only $\leq$3.3\% utility loss on average. \sysname{} offers a systematic, reproducible path to safer explainability in privacy-sensitive ML.
\end{abstract}

\input{intro.tex}
\input{background.tex}
\input{related}
\input{approach.tex}

\input{evaluation.tex}
\input{conclusions.tex}

\bibliographystyle{plain}
\bibliography{bib}

\input{appendix.tex}

\end{document}

%% file: intro.tex
\section{Introduction}\label{chap:intro}

Machine learning (ML) models are increasingly being deployed in high‑stakes settings, ranging from medical diagnostics~\cite{tjoa2020survey} to credit scoring~\cite{arrieta2020explainable} and cybersecurity \cite{guidotti2018survey}, where decisions must be accurate and explainable to a human. Post hoc explanation methods, which generate human interpretable attributions of model inference, have thus become crucial in debugging a model, algorithmic recourse, detecting bias, satisfying regulatory requirements, and enabling expert oversight~\cite{ribeiro2016why,lundberg2017unified}. At the same time, these domains often handle sensitive personal data, introducing stringent privacy mandates (e.g., HIPAA, GDPR). However, recent work has shown that explanations can leak private training examples via membership inference attacks, compromising the privacy guarantees that practitioners seek to uphold~\cite{shokri2021privacy,PleaseTellmeMore,zhao2021inversion, duddu2022inferring, luo2022feature}.

Despite compelling evidence of membership information leakage from explanations, there is no unified methodology to (i) establish membership leakage benchmarks across diverse explanation methods, (ii) systematically reduce membership leakage without compromising explainability, or (iii) analyze \emph{why} certain methods leak more than others. 
As a result, ML practitioners and domain experts lack clear guidance on which explanation methods to use and under what safeguards to balance explainability and privacy. In the absence of such guidance, privacy leaks may go undetected or explainability may be unnecessarily sacrificed, outcomes that are undesirable in mission-critical domains such as predictive diagnostics, financial forecasting, and autonomous driving.

In a recent work, Liu et al. \cite{PleaseTellmeMore} demonstrated a novel explanation-guided membership inference attack based on perturbation of important features, which resulted in the highest membership leakage from model explanations. While their attack demonstrates that explanations can exacerbate membership leakage, even in black‑box settings, it has several key limitations. First, it focuses exclusively on attribution‑based explanation methods and evaluates only seven techniques (e.g., SmoothGrad, SHAP, LIME). Second, beyond a perturbation‑guided attack, the study does not propose mitigations or guidance on how to reduce membership leakage without sacrificing explainability. Third, although the paper correlates overfitting and explanation quality with membership leakage, it lacks analysis of the underlying causes of leakage across explanation families. Fourth, all experiments are conducted using default explanation configurations, without exploring the parameter settings practitioners often adjust (e.g., number of perturbation samples in LIME~\cite{ribeiro2016why} or kernel width in SHAP~\cite{lundberg2017unified}), missing an opportunity to understand how tuning impacts the privacy–explainability trade‑off.

Motivated by these gaps, in this paper we present \sysname{}, a practical system for profiling, diagnosing, and hardening post‑hoc explanation methods against membership leakage. \sysname{} advances the state-of-the-art~\cite{PleaseTellmeMore} on three fronts: $(i)$ \textbf{systematic membership leakage profiling}: We develop a more effective membership inference attack against a broad spectrum of explanation methods to establish a membership leakage benchmark under default configurations, which advances the state-of-the-art~\cite{PleaseTellmeMore} by a large margin; $(ii)$ \textbf{privacy enhancing hardening}: We design lightweight hardening, such as calibrated noise injection and gradient clipping, which are easily integrated into existing explanation pipelines; and $(iii)$ \textbf{root-cause analysis}: Through controlled ablation studies, we correlate key algorithmic parameters with observed membership leakage rates, revealing the mechanisms by which explanations expose membership information. 

We validate \sysname{} on 15 explanation methods that span four families on models trained on image classification datasets:  CIFAR‑10~\cite{krizhevsky2012cifar}, CIFAR-100~\cite{krizhevsky2012cifar}, and GTSRB~\cite{stallkamp2011gtsrb}. 
Our evaluation provides a comprehensive picture of privacy risks in ML explainability. First, we show that widely used explanation methods, under default configurations, can {\bf leak up to 74.9\% more membership information} than previously reported~\cite{PleaseTellmeMore}. Second, our improved explanation-only membership inference attack achieves \textbf{state-of-the-art accuracy} without relying on predicted labels or confidence scores, underscoring the severity of explanation-driven leakage. Third, we demonstrate that lightweight, model-agnostic defenses such as attribution clipping, sensitivity-calibrated noise injection, and masking achieve \textbf{46.5--95\% membership leakage reduction} with a negligible \textbf{$\leq$3.3\%} drop in explanation utility, and combining these defenses yields the strongest privacy–utility trade-offs. Finally, our root-cause analysis attributes leakage to factors such as attribution sparsity, gradient outliers, and sensitivity to input perturbations, offering actionable guidance for privacy enhancing deployment of explanation methods.

This paper makes the following key contributions:
\begin{itemize}
    \item \textbf{\sysname{} approach}: A unified toolkit for  membership leakage auditing and hardening of post‑hoc explanations.
    
    \item \textbf{Comprehensive benchmark}: An extensive membership leakage profile of representative explanation methods across four families, establishing a new baseline for membership leakage of ML explanation methods.
    
    \item \textbf{Practical mitigations}: Simple, effective mitigation techniques that reduce privacy risks by orders of magnitude while retaining explanation utility.
    
    \item \textbf{Mechanistic insights}: A root‑cause analysis that pinpoints which algorithmic features drive membership leakage, informing principled defense design.
    
    \item \textbf{Reproducible open-source code}: To foster future research at the intersection of ML privacy and explainability, we release reproducible \sysname{} code at:\\ \url{https://https://github.com/um-dsp/DeepLeak}.
\end{itemize}

%% file: background.tex
\section{Background}
\label{sec:background}
We review ML explanation methods, whose outputs serve as the attack surface, and membership inference attacks, which we use as proxy for auditing membership leakage.

\subsection{ML Explanation Methods}
Broadly, explanation methods fall into two complementary categories: inherent (ante-hoc) and post-hoc approaches. 
An inherently interpretable model satisfies
transparency of structure with a fairly simple decision function (e.g., linear, rule-based) and feature-level interpretability whereby the influence of each feature on the output is explicitly accessible. Although inherently interpretable models are favored in regulated or safety-critical settings, they may sacrifice accuracy or expressiveness on complex tasks where high-capacity models such as deep neural networks (DNNs) excel. Post-hoc explanation methods generate explanations by quantifying the contributions of individual input features to the output of a model\cite{sundararajan2017axiomatic, selvaraju2017grad}. 

In this work, we consider post‑hoc explanation methods $E(f,x)$ that produce attributions for $f(x)$. Let \( f: \mathbb{R}^d \rightarrow \mathbb{R}^C \) be a model that maps input \( x \in \mathbb{R}^d \) to a vector of class scores $C$, \( f_c(x) \) denote the scalar output (e.g., logit or probability) for class \( c \), \( \frac{\partial f_c(x)}{\partial x} \) denote the gradient of the output with respect to the input, \( x' \) be a baseline input (e.g., zero vector), and \( \odot \) denote the element-wise  product. In this work, we categorize them into four families and provide formal definitions.
     
\textbf{Gradient-based explanation methods} compute gradients of the model’s output with respect to input features to determine feature importance. Saliency Map score ~\cite{simonyan2014deep} is computed as \(\text{Saliency}(x) = \left| \frac{\partial f_c(x)}{\partial x} \right|\), where $\frac{\partial f_c(x)}{\partial x}$ measures sensitivity of the model output to changes in each input feature and $|.|$ captures element-wise absolute value to measure feature importance. 
    Deconvolution score~\cite{deconv} computed as \(\text{Deconv}(x) = \left. \frac{\partial f_c(x)}{\partial x} \right|_{\text{deconv rules}}\) is similar to Saliency map but modifies the backward pass through ReLU layers and passes gradients only for activations that were positive in the forward pass. 
    Input×Gradient~\cite{InputXGradient} computed as \(
    \text{InputGrad}(x) = x \odot \frac{\partial f_c(x)}{\partial x}\) highlights input dimensions that are large and whose gradients strongly influence the output.
    SmoothGrad~\cite{smilkov2017smoothgrad} computed as \(
    \text{SmoothGrad}(x) = \frac{1}{n} \sum_{i=1}^{n} \left| \frac{\partial f_c(x + \mathcal{N}_i)}{\partial x} \right|\) adds Gaussian noise $\mathcal{N}_i \sim \mathcal{N}(0, \sigma^2)$ to the input and averages the resulting gradients. It reduces visual noise and sharpens the saliency map. 
    VarGrad \cite{vargrad}  is a variance-based extension of SmoothGrad that measures the variability of explanation scores under noise perturbations. It is defined as $
E_{\text{vg}}(x) = \mathcal{V}\big(E(x + g_i)\big)
$. The noise vectors $\mathbf{g}_i$ are sampled independently from a normal distribution $\mathcal{N}(0, \sigma^2)$ and the variance $\mathcal{V}$ quantifies how sensitive the explanations are to these perturbations.
    Integrated Gradients~\cite{sundararajan2017axiomatic} computed as \(
    \text{IG}(x) = (x - x') \odot \int_{\alpha=0}^1 \frac{\partial f_c(x' + \alpha (x - x'))}{\partial x} \, d\alpha
    \) attribute prediction differences between \( x \) and a baseline \( x' \) along a straight path and it does so by aggregating gradients along the path from $x'$ to $x$.
    DeepLIFT~\cite{shrikumar2019learning} calculated as \(\text{DeepLIFT}(x) = \Delta x \odot \frac{\Delta f_c}{\Delta x}\) uses differences from a reference input \( x' \) instead of gradients, \( \Delta x = x - x' \), and \( \Delta f_c \) is the difference in output. DeepLIFT avoids gradient saturation by assigning contribution scores. Guided Backpropagation~\cite{springenberg2015striving} computed as \(\text{GuidedBP}(x) = \left. \frac{\partial f_c(x)}{\partial x} \right|_{\text{guided rules}}\) combines saliency and deconvolution. Backpropagation through ReLU passes gradient only if both the forward activation and the backward gradient are positive.

    \textbf{Perturbation-based explanation methods} modify input features and observe the corresponding changes in model predictions to assess feature relevance.
    Let \( x_{i} \) be the input where the $i^{th}$ feature is occluded (e.g., set to zero or blurred) and \( S \subseteq \{1, \dots, d\} \) be a subset of input features.  Occlusion Sensitivity~\cite{Uchiyama_2023_WACV} calculated as \(\text{Occlusion}_i(x) = f_c(x) - f_c(x_{i}) \) measures the change in prediction when feature \( i \) is occluded where \( x_{i} \) is input with feature \( i \) removed or replaced (e.g., with baseline value). A large drop in \( f_c \) implies feature \( i \) is important.
    SHAP (SHapley Additive exPlanations)~\cite{lundberg2017unified} uses game-theoretic perturbations to compute feature importance as \(
    \text{SHAP}_i(x) = \sum_{S \subseteq \{1, \dots, d\} \setminus \{i\}} \frac{|S|!(d - |S| - 1)!}{d!} \left[ f_c(x_{S \cup \{i\}}) - f_c(x_S) \right]\), where \( x_S \) is input where only features in \( S \) are present (others set to a baseline). It fairly distributes the total prediction among input features using Shapley values from cooperative game theory and it also accounts for feature interactions by averaging marginal contributions over all possible subsets.
    Anchors~\cite{anchors} is based on the idea of an {\em Anchor}, a set of input conditions (feature values) that, when fixed, are sufficient to keep the model’s prediction stable. It is computed as \(\text{Anchor}(x) = \text{Minimal feature set } A \subseteq \{1, \dots, d\} \text{ s.t. } P(f(x') = f(x) \mid x'_A = x_A) \geq \tau\), where \( x'_A = x_A \) is perturbed inputs that preserve the anchor conditions, \( \tau \) is a precision threshold. It provides human-interpretable if-then rules explaining why the model predicted class \( c \).
    
    \textbf{Representation-guided explanation methods} use intermediate feature representations in DNNs to provide saliency maps. 
    Gradient-weighted Class Activation Mapping (Grad-CAM)~\cite{selvaraju2017grad} produces coarse localization maps that highlight important regions in a convolutional layer for a given class. Let $A^k \in \mathbb{R}^{H \times W}$ denote the $k^{th}$ feature map (activation) in a convolutional layer for input $x$ and $s_y(x) = \log f_y(x)$ be the score (e.g., logit) for class $y$, Grad-CAM computes the importance weight $\alpha_k^y$ for each channel $k$ as the spatially averaged gradient of the class score with respect to the feature map:
\(\alpha_k^y = \frac{1}{H \times W} \sum_{i=1}^{H} \sum_{j=1}^{W} \frac{\partial s_y(x)}{\partial A_{i,j}^k}.\) The final class-discriminative localization map is computed as: \(
    \phi^{\text{Grad-CAM}}(x) = \mathrm{ReLU} \left( \sum_k \alpha_k^y A^k \right),\)
where ReLU ensures that only the features positively influencing the class are visualized.
Grad-CAM++~\cite{chattopadhay2018grad++} extends Grad-CAM by providing better localization and handling multiple object instances. It improves the weighting scheme by incorporating higher-order partial derivatives to compute more accurate weights for each feature in the feature maps.

\textbf{Approximation-based explanation methods} fit interpretable surrogate models (e.g., decision trees, linear regression models) to approximate local decision boundaries. 
    LIME~\cite{ribeiro2016why} explanation is obtained by solving the optimization problem: \(
    \phi^{\text{LIME}}(x) = \arg\min_{g \in \mathcal{G}} \mathcal{L}(f, g, \pi_x) + \Omega(g),\)
where: $\mathcal{L}(f, g, \pi_x)$ is a loss function that measures the fidelity of $g$ in approximating $f$ in the local neighborhood of $x$, weighted by a locality kernel $\pi_x$, $\pi_x(z)$ assigns higher weights to perturbed samples $z$ that are closer to $x$, and $\Omega(g)$ is a regularization term that penalizes the complexity of the interpretable model $g$.
In practice, LIME perturbs the input instance $x$ to generate a dataset $\{z_i\}$, obtains predictions $f(z_i)$, and fits $g$ to minimize: \(
    \mathcal{L}(f, g, \pi_x) = \sum_{i} \pi_x(z_i) (f(z_i) - g(z_i))^2\). The explanation $\phi^{\text{LIME}}(x)$ is then the feature attribution vector from the fitted surrogate $g$. SHAP~\cite{lundberg2017unified} also approximates the Shapley values by averaging marginal contributions over all possible subsets of features.
  ProtoDash~\cite{protodash} is a prototype-based method that selects a small set of representative examples (prototypes) from a dataset $\mathcal{Z} = \{z_1, z_2, \ldots, z_n\}$ to best summarize a target distribution (e.g., instances with the same prediction as $x$). The explanation is a sparse set of prototypes with associated nonnegative weights indicating their relevance to the input.
Let $\mu_p$ be the mean embedding of a target sample distribution $p$ (e.g., all instances predicted the same as $x$) in a reproducing kernel Hilbert space (RKHS), and let $\mu_q(S, \mathbf{w}) = \sum_{i \in S} w_i \phi(z_i)$ be the weighted empirical kernel mean embedding over selected prototypes $z_i$ from $\mathcal{Z}$, ProtoDash selects a subset $S \subseteq \{1, \ldots, n\}$ and weights $\mathbf{w} \in \mathbb{R}^{|S|}_{\geq 0}$ that maximize the similarity (inner product) between $\mu_p$ and $\mu_q$:
\(
    \phi^{\text{ProtoDash}}(x) = \arg\max_{S, \mathbf{w} \geq 0} \langle \mu_p, \mu_q(S, \mathbf{w}) \rangle = \arg\max_{S, \mathbf{w} \geq 0} \sum_{i \in S} w_i \langle \mu_p, \phi(z_i) \rangle
\), subject to constraints:\\
\( |S| \leq m, \quad w_i \geq 0 \; \forall i,\)
where $m$ is the maximum number of prototypes. A greedy algorithm is used to select the prototypes that best align with the target distribution in kernel space. The final explanation consists of the selected prototypes $\{z_i\}_{i \in S}$ and their weights $\{w_i\}$, which highlight representative examples that influence $f(x)$.

\subsection{Membership Inference Attacks}
Membership inference attacks (MIAs) pose a significant privacy risk in ML models trained on sensitive data, allowing the leakage of private information~\cite{shokri2017membership,liu2022membership,salem2018ml,song2021systematic,carlini2022membership,nasr2019comprehensive,zhang2023byzantine,liu2021encodermi,chen2020gan,chen2021unlearning,yuan2022membership,li2022auditing}. Given a model $f$ trained on a training set $D$, a sample $x$, and auxiliary information $\mathcal{I}$ available to the adversary, an attack model $\mathcal{A}$ aims to infer membership as:
 \begin{equation}
    \mathcal{A} : (x, f, \mathcal{I}) \rightarrow \{0,1\},
 \end{equation}
where $\mathcal{A}$ outputs 1 if $x\in D$ (member) and 0 otherwise (non-member) \cite{shokri2017membership}. 

In practice, most MIAs operate in a black-box setting, where the adversary has access only to the model's output probabilities but lacks direct knowledge of its parameters \cite{carlini2022membership, song2019auditing}. A common strategy for such attacks is to train \textit{shadow models} that mimic the behavior of $f$ on auxiliary training data drawn from the same distribution as $D$. The shadow models are then used to generate attack data to train $\mathcal{A}$ \cite{shokri2017membership, salem2018ml}. Other approaches rely on statistical properties such as loss entropy \cite{yeom2018privacy} or loss trajectories during model training \cite{liu2021membership}. Recent work has introduced a more advanced MIA called the Likelihood Ratio Attack (LiRA), which assesses per-sample vulnerability by training multiple (hundreds) shadow models per instance and analyzing confidence distributions \cite{carlini2022membership}. 

When explanations are exposed, the adversary’s auxiliary information $\mathcal{I}$ expands to include explanation outputs and $\mathcal{A}$ is denoted as:
 \begin{equation}
\mathcal{A} : (x, f, \mathcal{I}, \mathcal{E}) \rightarrow \{0,1\},
 \end{equation}
where $\mathcal{E}$ denotes a post‑hoc explanation function. For example, $\mathcal{E}$ may output an attribution map $\phi\in\mathbb{R}^d$ such that $\phi_i$ scores feature $i$. 
Recent work~\cite{PleaseTellmeMore} shows that supplementing $f(x)$ with $\phi$ can increase the effectiveness of MIA.

%% file: related.tex
\section{Related Work}\label{sec:related}
% Recent work on explanation methods and privacy leakage shows that explanations can serve as vectors for leaking sensitive information.
Shokri et al. \cite{shokri2021privacy} were the first to demonstrate membership leakage of model explanations. They use a threshold-based attack that leverages explanation variance to differentiate members vs. nonmembers. Their intuition is that when a model is confident about a prediction, small perturbations will not change the model's output; therefore, the feature attributions are low, leading to lower explanation variance. They showed that gradient-based explanations, due to their high variance, are the most susceptible to MIAs. They explored the trade-off between privacy and explainability by analyzing perturbation-based explanations which are more resistant to such attacks but come at the cost of lower-quality explanations.  
% They consider two scenarios: optimal threshold and shadow models. For the shadow model approach, they train shadow models on surrogate data to determine the selection threshold. 

Liu et al. \cite{PleaseTellmeMore} built the most effective MIAs on attribution-based explanations, through a model-based attack that uses explanations to perturb inputs. The intuition is that members should have lower prediction probability scores under perturbations compared to non-members. They attribute differences in explanations to the generalization gap between training and testing data. They also noted that there is a difference in the degree of membership leakage between different explanation methods, inferring that explanation methods with greater accuracy potentially pose a higher risk of membership leakage. 

Counterfactual explanations, which are hypothetical data samples that provide insights into decision boundaries, also pose privacy risks. Pawelczyk et al.\cite{pawelczyk2023privacy} show that in algorithmic recourse, counterfactual explanations intended to help users reverse a bad decision of the model can be abused to launch MIAs. Furthermore, Naretto et al. \cite{naretto2022evaluating} show that membership leakage risks extend to global explanation methods.

\sysname{} performs a comprehensive analysis to examine the root causes of membership leakage, but also includes lightweight hardening strategies to reduce membership leakage while maintaining explanation utility. This additional contribution makes \sysname{} the first framework for diagnosing and minimizing privacy risks across explanation methods. 

Beyond MIAs, additional privacy risks of model explanations have also been explored. Zhao et al. \cite{zhao2021inversion} extended the privacy risks of model explanations to model-inversion attacks: reconstructing sensitive information (e.g., faces) from model explanations. They found that activation-based (saliency map) explanations leak more privacy than sensitivity-based (gradient) explanations. Duddu et al. \cite{duddu2022inferring} focus on the privacy risks of attribute inference attacks on model explanations, where an adversary can infer sensitive attributes (e.g., race and gender) of individual data records from their corresponding model explanations. 
Luo et al.\cite{luo2022feature} analyze the privacy risks of Shapley‐values-based model explanations by introducing two feature‐inference attacks, reconstructing private model inputs from their shapley explanation values, and validating their effectiveness across leading ML-as-a-service platforms.  Wang et al. \cite{wang2022dualcf} address the issue of security and privacy of model extraction attacks with counterfactual explanations. 

Differential privacy (DP) has been utilized to mitigate the privacy risks of model explanations. Patel et al.~\cite{patel2022dp} developed DP for computing model explanations, via an adaptive DP-SGD algorithm that uses a minimal privacy budget but provides accurate explanations. However, they also applied DP-SGD on the model, which resulted in utility loss. Yang et al.~\cite{yang2022dp} proposed a DP algorithm to derive counterfactual explanations with DP guarantee.

%% file: approach.tex
\section{\sysname{} Approach}\label{sec:approach}
\begin{figure*}[t!]
    \centering   
    \includegraphics[scale =0.25]{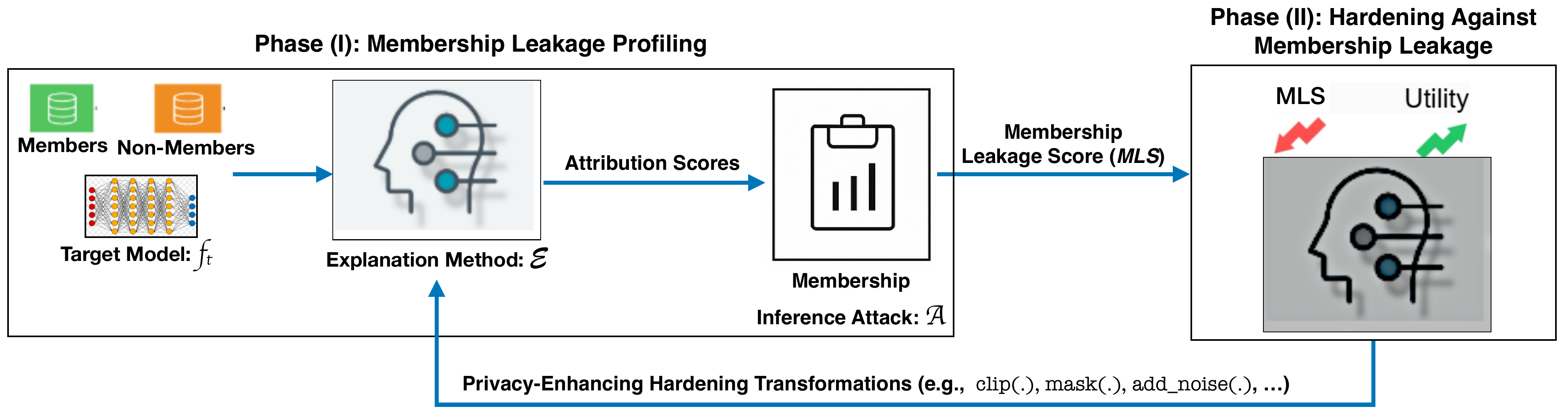}
     
    \caption{\textbf{\sysname{} system overview.}}
    
    \label{fig:leakyfarm}
\end{figure*}

\textbf{Overview.} Figure \ref{fig:leakyfarm} shows an overview of the \sysname{} approach. It comprises two main steps: (I) {\em Membership Leakage Profiling} and (II) {\em Privacy-Enhancing Hardening Against Membership Leakage}. In (I), we develop a more effective explanation-guided MIA than prior work~\cite{PleaseTellmeMore}. This stage establishes a membership leakage benchmark for a broad spectrum of explanation methods deployed under default configurations. In \S \ref{subsec:leakage-results}, we show that our attack is more powerful than state-of-the-art explanation-guided MIA~\cite{PleaseTellmeMore}, which proves the existence of more powerful attacks against explanation APIs, and hence the need for effective mitigation strategies.
Using membership leakage measurements, we correlate key algorithmic parameters of explanation methods with observed leakage rates to reveal mechanisms by which explanations expose membership signals. Then in (II) we develop lightweight hardening methods tailored to the nature of each explanation family. We design our hardening methods in a manner that significantly reduces membership leakage with no/minimal reduction on explanation utility. In \S \ref{subsec:hardening-results}, we show the effectiveness of our hardening strategies across families of explanation methods. In  \S \ref{subsec:root-cause-analysis}, we draw insights as to the root causes of membership leakage in model explanations across families of explanation methods. Our analysis suggests new insights compared to previous work \cite{PleaseTellmeMore}.

\subsection{Membership Leakage Profiling}\label{subsec:leakage-profile}
The default configurations of post-hoc explanation methods are primarily aimed at offering high utility explanations. Establishing privacy risk implications of such default configurations/parameters of explanation methods offers a benchmark of leakage under default configurations and informs our pursuit for privacy-enhancing hardening of explanation methods against leakage without compromising explanation utility.

Let $f_t$ be a model trained on dataset $D =(x_1, y_1) ... (x_n, y_n)$ where each $x_i$ is of dimension $d$. Let $\mathcal{E}$ be an explanation method (function) that, given a model $f_t$, a test input $x$, and a prediction $f_t(x)$, generates an attribution score vector $\phi = [\phi_1, ..., \phi_d]$ where each $\phi_i$ scores feature $i$ of $x$. To realize membership leakage profiling, we need to launch a MIA against $f_t$ by exploiting attribution scores of inputs generated by $\mathcal{E}$ as potential sources of membership signal. This process is repeated for as many $\mathcal{E}$'s and MIAs as needed.

\textbf{Threat Model.}
The \textit{adversary’s goal} is to use feature attribution scores from $\mathcal{E}(f_t(x))$ to determine whether or not $x$ was used to train $f_t$.
With regards to \textit{adversary’s knowledge and capabilities}, we assume that the adversary has black-box API access to $f_t$ with access only to $f_t(x)$, and the corresponding attribution map $\phi$ with no visibility into the details of the explanation method and its configurations. We also assume that the adversary has access to an auxiliary dataset sampled from the same distribution as $D$. This will give the attacker the possibility to collect another auxiliary data that is out-of-distribution compared to $D$. The adversary also knows the target model architecture. These assumptions are realistic and are in line with previous threat models for MIAs~\cite{liu2022membership,  carlini2022membership,li2022auditing,salem2018ml, shokri2017membership}. We will relax these assumptions in the ablation study by using disjoint datasets and different model architectures between the target model and the shadow model.

\textbf{Attack Details.} 
\label{sec:attackdetails}
The rationale of our MIA is that defenders can audit leakage using publicly known attacks by simulating a worst-case attack. With this rationale, our MIA is inspired by \cite{shokri2021privacy}, which demonstrated explanation-guided MIA. Based on our threat model, we partition each dataset into 4 subsets: $D^{\mathit{train}}_{\mathit{target}}$, $D^{\mathit{train}}_{\mathit{shadow}}$, $D^{\mathit{test}}_{\mathit{target}}$, and $D^{\mathit{test}}_{\mathit{shadow}}$. $D^{\mathit{train}}_{\mathit{target}}$ is used to train the target model $f_t$ and all samples in it are \textbf{members}. $D^{\mathit{test}}_{\mathit{target}}$ consists of samples that are treated as \textbf{non-members} with respect to the target model.
$D^{\mathit{train}}_{\mathit{shadow}} \subset D^{\mathit{train}} _{\mathit{target}}$ is used to train a shadow model $f_s$ that mimics $f_t$. 
$D^{\mathit{test}}_{\mathit{shadow}}$ is designated as non-members with the same size as  $D^{\mathit{train}}_{\mathit{shadow}}$, and is disjoint with $D^{\mathit{test}}_{\mathit{target}}$ so as to test the generalization of our MIA against $f_t$ on members and non-members.

 We craft a shadow-model MIA based on Shokri et al.~\cite{shokri2021privacy}, but with two key modifications motivated by our baseline results. In the canonical shadow attack the shadow models are trained and evaluated on data drawn from the same distribution as $f_t$.  When we followed that recipe the attack achieved poor performance (Table \ref{tab:basline}). We hypothesize that the attack’s weakness stems from a mismatch between (a) the distributional relationship between members and non-members that the adversary assumes during shadow training and (b) the actual relationship produced by the target model where the training and test data could come from different distributions. To better match realistic conditions and give the attack model a stronger leverage on the membership signal, we train and test each shadow model using shadow training set \(D^{\mathit{train}}_{\mathit{shadow}}\) and shadow test set \(D^{\mathit{test}}_{\mathit{shadow}}\), respectively,  such that members are drawn from the same distribution as training set of $f_t$ and non-members are drawn from a held-out out-of-distribution data. More precisely, for each shadow model we (i) sample \(D^{\mathit{train}}_{\mathit{shadow}}\) and \(D^{\mathit{test}}_{\mathit{shadow}}\), (ii) train a shadow model \(f_s\) on \(D^{\mathit{train}}_{\mathit{shadow}}\), and (iii) compute attribution vectors \(\phi(x)=\mathcal{E}(f_s,x)\) for samples in \(D^{\mathit{train}}_{\mathit{shadow}}\) (labeled ($y=1$ = member) and \(D^{\mathit{test}}_{\mathit{shadow}}\) (labeled $y=0$ = non-member). The labeled attribution dataset \(\{(\phi(x),y)\}\) is used to train the attack model $\mathcal{A}$ that predicts membership from explanation outputs. 
 
 Because adversaries in practice explore multiple design choices (e.g., random seeds, shadow model architectures), we repeat the entire shadow-model-training $\rightarrow$ attribution $\rightarrow$ attack-model-training pipeline across multiple random seeds and configurations, and report the attack variant that performs best on a held-out validation set.

\begin{table}[t!]
\centering
\caption{\textbf{Effectiveness of baseline MIA}. Measured via TPR@0.1\%FPR for different explanation methods using same distribution for train and test of shadow model on CIFAR-10.}
\begin{tabular}{lc}
\toprule
\textbf{Explanation Method} & \textbf{TPR@0.1\%FPR} \\ 
\midrule
SmoothGrad   & 0.07\%  \\ 
VarGrad      & 0.1\%  \\ 
IG           & 0.52\%  \\ 
GradCAM      & 0.11\%  \\ 
GradCAM++    & 0.52\%  \\ 
SHAP         & 0.41\%  \\ 
LIME         & 0.37\%  \\ 
\bottomrule\\
\end{tabular}
\label{tab:basline}
\end{table}

\textbf{Membership Leakage Score.}
\label{sec:pls}
The effectiveness of MIAs is realistically measured by the true-positive rate (TPR) an adversary achieves at a low false-positive rate (FPR)~\cite{carlini2022membership,AerniZT24}. Accordingly, we define the \emph{Membership Leakage Score} (MLS) of an explanation method $\mathcal{E}$ on a target model $f_t$ as the TPR at a (low) target FPR level (e.g., $0.1\%$) when the attack has access to \emph{only} explanation vectors $\phi(x)=\mathcal{E}(f_t,x)$. This isolates leakage that originates from explanations, unlike previous works (e.g., ~\cite{PleaseTellmeMore}) which combine explanations with additional model outputs (e.g., predicted labels and probability vectors) as the attacker's information $\mathcal{I}$. Formally, the $MLS$ score is calculated against the target model $f_t$ as:
\begin{equation}\label{eq:mls}
MLS =   TPR(\mathcal{A}(\mathcal{E}(\mathbf{X_{f_t}})) \mid FPR = \epsilon)
\end{equation}

\noindent where, $\mathcal{E}(x)$ is the explanation vector, $\text{TPR}(\mathcal{A}(\mathcal{E}(x)))$ is the true positive rate of the attack, $\mathbf{X_{f_t}} \subset (D^{\mathit{train}} _{\mathit{target}} \cup D^{\mathit{test}}_{\mathit{target}})$, and $\epsilon$ is the FPR target. $\text{FPR} \leq \epsilon$ ensures low false positives.

\textbf{Explanation Utility.}
\label{sec:utli}
We consider the utility of an explanation method as its ability to retain informative explanations under privacy-preserving transformations. In this context, higher sensitivity indicates a greater change in explanations, which implies a reduction in utility. Conversely, lower sensitivity corresponds to higher utility, as the explanations remain closer to their original form.
We compared {\em (in)fidelity} and {\em sensitivity} from ~\cite{yeh2019infidelitysensitivityexplanations}, and we found that the (in)fidelity metric behaved inconsistently across different explanation methods (e.g., some produced values in the range [10,100] while others in the range [1,000,40,000]). In light of the inconsistency, we opted to use the more stable metric, explanation \emph{sensitivity}, as a proxy to assess the stability of model explanations. Sensitivity measures the degree of explanation changes to subtle input perturbations, which impacts the reliability of model explanations. High sensitivity implies that small variations in the input can lead to significant shifts in the explanation, making models more susceptible to adversarial manipulation and membership leakage  attacks~\cite{ghorbani2019interpretation, PleaseTellmeMore}. In \sysname{}, we measure the sensitivity pre- and post-hardening to assess the trade-off between hardening against leakage and the utility of the explanation. The lower sensitivity of an explanation to a membership leakage hardening countermeasure implies that the explanation method reduces membership leakage without reducing the utility of the explanation.

We define the sensitivity of an explanation method $\mathcal{E}$ denoted as $\mathbb{S}(\mathcal{E}, f_t, x, r)$ as the maximum change between $\mathcal{E}(f_t,x)$ and $\mathcal{E}(f_t,x+\delta)$ with respect to a small perturbation $\delta$ around input $x$, constrained within an $||L||_p$ (e.g., $L_\infty$) ball of radius $r$. It is computed as follows:

\begin{equation}\label{eq:sens}
\mathbb{S}(\mathcal{E}, f_t, x, r) = \max_{\|\delta - x\| \leq r} \|\mathcal{E}(f_t, x+\delta) - \mathcal{E}(f_t, x)\|
\end{equation}

 High sensitivity entails vulnerability to adversarial perturbations, where small, imperceptible changes in input significantly alter feature attributions. By minimizing $\mathbb{S}$, the stability of the explanation is improved, implying robust interpretability. 

\subsection{Hardening Against Membership Leakage}
\label{subsec:tailored-hardening}

Explanation-guided MIA is a statistical test that aims to distinguish two distributions: the explanation vectors (maps) for members versus non-members. An ideal hardening mitigation works by reducing the statistical separability (divergence) between these two distributions without compromising the utility of the explanation. To achieve this objective, previous work used differential privacy (DP)~\cite{Abadi_2016}. However, Patel and Shokri~\cite{patel2022dp} show that DP negatively affects the utility of the model by reducing the fidelity of the explanation. Given the well-documented cost of DP on model and explanation utility, in the hardening strategies we describe next, we aim to develop privacy enhancing hardening with no impact on model utility and acceptably minimal impact on explanation utility. 

Before developing our hardening methods, we conducted an in-depth analysis of 15 explanation methods. Based on our analysis, we categorize the explanation methods into two main classes: {\em parameterized} (those with tunable parameters) and {\em non-parameterized} (those without tunable parameters).   

\begin{algorithm}[!t]
\footnotesize
\caption{Hardening Parameterized Methods }\label{alg:privacy_explanation}
\begin{algorithmic}[1]

\Require $X$ (data), $f_t$ (target model), $\mathcal{E}$ (explanation method), $\mathcal{A}$ (attack model), $\mathcal{\phi}$ (attribution scores) , $\mathcal{T}$ (Number of Trials)
\Ensure $\theta^*$ (Optimized explanation parameters) 

% \State \textbf{Initialize} $\mathcal{O}$ (hardening Method)
\Function{Objective}{$\mathcal{E}, f_t,X$}
    \State \textbf{Sample} $\theta_p$
    \State $\phi_{\theta_{p}} \gets \mathcal{E}(f_t,X; \theta_{p})$
    \State $U \gets \mathbb{S}(\phi_{\theta_p})$ \Comment{Evaluate explanation utility}
    \State $MLS \gets \mathcal{A}(\phi_{\theta_p})$ \Comment{MIA attack} 
    \State \Return $(U_{\theta_p}, MLS_{\theta_p})$
\EndFunction

\Function{Optimizeexplanation}{$\mathcal{E}, f_t,X$}
    \State $\theta^* \gets \arg\min_{\theta} MLS, \arg\max_{\theta} U (\text{Objective($\mathcal{E}, f_t,X$)}, n=\mathcal{T})$
    \State \Return $\theta^*$ 
\EndFunction

\end{algorithmic}
\end{algorithm}

\begin{algorithm}[!t]
\footnotesize
\caption{Hardening Non-Parameterized Methods}
\label{alg:transfor}
\begin{algorithmic}[1]
\Require $X$ (data), $f_t$ (target model), $\phi$ (raw attributions), $\theta_p = (\sigma, c_{min}, c_{max}, \tau)$ (privacy params),  $\mathcal{T}$ (Number of Trials)
\Ensure $\theta^*$ (privacy-enhancing parameters)

\Function{PrivacyHardening}{$\mathcal{E}, f_t,X$}
    \State $\phi \gets \mathcal{E}(f_t,X)$
    \State \textbf{Sample} $\theta_p$
    \State $\phi \gets \text{clip}(\phi, c_{\min}, c_{\max})$  \Comment{Clamp Values}
    % \State $\phi \gets \text{round}(\phi, d)$  \Comment{Reduce Precision}
    \State $\phi[\phi < \tau] \gets 0$  \Comment{Mask Weak Attributions}
    \State $\phi \gets \phi + \mathcal{N}(0, \sigma^2)$  \Comment{Add Noise}
    \State $U \gets \mathbb{S}(\phi_{\theta_p})$ \Comment{Evaluate explanation utility}
    \State $MLS \gets \mathcal{A}(\phi_{\theta_p})$ \Comment{MIA attack} 
    \State \Return $(U_{\theta_p}, MLS_{\theta_p})$
\EndFunction

\Function{OptimizePrivacy}{$\mathcal{E}, f_t,X$}
    \State $\theta^* \gets \arg\min_{\theta} MLS, \arg\max_{\theta} U (\text{PrivacyHardening($\mathcal{E}, f_t,X$)}, n=\mathcal{T})$
    \State \Return $\theta^*$ 
\EndFunction
\end{algorithmic}
\end{algorithm}

\textbf{Hardening Parameterized Explanation Methods.}
In this category (\cite{smilkov2017smoothgrad,vargrad,sundararajan2017axiomatic,sundararajan2017axiomatic,selvaraju2017grad,chattopadhay2018grad++,lundberg2017unified,ribeiro2016why,Uchiyama_2023_WACV,protodash,anchors}), each explanation method relies on a set of parameters to generate explanations. These parameters were originally introduced by the researchers who developed the methods. In their respective works, they explore different configurations and report varying performance outcomes, ultimately recommending what we refer to as {\em default configuration} per an explanation method. Table~\ref{tab:param} in the Appendix shows, for each explanation method, the complete set of parameters along with their respective search space or options available for attribution generation. Using the default settings in the original works, we implemented all these explanation methods and observed that they resulted in significant membership leakage.

As detailed in Algorithm \ref{alg:privacy_explanation}, the hardening of a parameterized explanation method is an optimization problem with two objectives: (a) minimizing membership leakage score $MLS$ and (b) maximizing explanation utility $U$. Towards achieving these objectives, \sysname{} systematically explores different combinations of parameter values to reach an {\em ideal point}: a $MLS$ score of zero and the highest achievable score for $U$. This is achieved through an iterative search process in the search space of tunable parameters of the explanation method at hand. 

Initially (e.g., the first 5 iterations), Algorithm \ref{alg:privacy_explanation} identifies promising trajectories for tuning, i.e., whether a given parameter it to be increased or decreased based on observed performance. For example, if a parameter is numeric with defined upper and lower bounds, early iterations help infer the direction in which adjustments are likely to lead to achieving the optimization goals. Over time, this adaptive sampling leads to convergence to an optimal or near-optimal configuration for each explanation method, balancing privacy and utility.

Algorithm~\ref{alg:privacy_explanation} at its core is an objective function that evaluates a candidate set of parameters $\theta$ for a given explanation method $\mathcal{E}$ (line 2). It then computes the explanation vector $\phi_{\theta_p}$  by directly applying $\mathcal{E}$ with the sampled parameters (line 3). It then evaluates sensitivity (line 4). Next, membership leakage is measured by applying the attack model $\mathcal{A}$ on $\mathcal{E}$ and computing the $MLS$ (line 5). Our hardening method iteratively samples configurations and uses this dual-objective feedback to refine its search (lines 8–11). The hardening continues for a fixed number of trials (e.g., 20), and ultimately returns the parameter setting $\theta^*$ that minimizes $MLS$ while maximizing $U$.
This iterative approach allows \sysname{} to automatically discover the most privacy-preserving yet interpretable configurations of explanation methods. The initial iterations help map the performance landscape, guiding subsequent trials toward optimal regions in the parameter space.

\textbf{Hardening Non-Parameterized Explanation Methods:}
For non-parameterized explanation methods \cite{InputXGradient,simonyan2014deep,deconv,shrikumar2019learning,springenberg2015striving}, our approach to hardening against leakage is based on {\em transformations} applied on attribution scores. Algorithm \ref{alg:transfor} shows the details of our hardening method, with three transformation functions applied to attribution scores: clipping gradient values, masking low signal attributions, and calibrated noise injection.  Next, we describe these three transformations.

\textbf{Clipping Attribution Values:} Bounding attribution values (line 4) within a fixed range prevents extreme feature contributions from standing out. Since attributions values have negative and positive values, we allocate the lowest boundary for clipping negative values and highest boundary for the positive ones. Doing so ensures that training samples (members) do not have significantly different attribution (e.g., gradient) magnitudes compared to non-training samples (non-members), making it harder for the adversary to identify members.

 \textbf{Masking Low Signal Attributions:} Zeroing out low signal attributions below a set threshold removes weak, non-informative attribution values that could be used to infer hidden model behavior (line 5). This reduces the risk of attackers detecting tiny but consistent feature contributions unique to members/non-members.
 
\textbf{Calibrated Noise Injection:} Adding Gaussian noise (line 6) to attributions ensures that attackers cannot distinguish between members and non-members based on small variations in gradient sensitivity. It introduces randomness, making MIA less effective by preventing exact reconstruction of model responses as is done in privacy-preserving training schemes such as DP-SGD~\cite{Abadi_2016}. Unlike DP-SGD, where the sensitivity of gradients must be carefully computed to determine a fixed Gaussian noise that satisfies DP guarantees, \sysname{} directly parameterizes the Gaussian noise distribution and integrates this parameter into an optimization schema. This schema functions as a noise-level selector, dynamically balancing the privacy–utility trade-off: it enforces an upper bound on the noise parameter to satisfy privacy requirements while simultaneously ensuring that the utility of explanations is preserved. In doing so, sensitivity considerations are implicitly handled through the bounded parameter space, allowing us to achieve privacy protection without compromising the interpretability of the explanation method.

As shown in Algorithm \ref{alg:transfor}, we parameterize the attributions using $\theta_p = (\sigma, c_{\min}, c_{\max}, \tau)$ (line 3) and compute their utility $U$ (line 7) and privacy leakage $MLS$ (line 8). These metrics are returned jointly for evaluation (line 9). Finally, the optimization procedure (lines 11--14) selects the parameters $\theta^{*}$ that minimize leakage while maximizing utility.

The order and combination of these transformations dictate the membership leakage vs. explanation utility trade-off. 
% To evaluate each transformation's utility, we order them starting with applying attribution clipping and masking low signals. Specifically, we apply clipping using line 2: $\phi \leftarrow \text{clip}(\phi, c_{\min}, c_{\max})$, followed by masking using line 3: $\phi[\phi < \tau] \leftarrow 0$. Afterwards, we apply random Gaussian noise to fool the attacker each time they try to learn the patterns of members and non-members that the explanation method and model provide while generating explanations for a given input, as shown in line 4: $\phi \leftarrow \phi + \mathcal{N}(0, \sigma^2)$.
Our empirical analysis on how to order the three transformations is in \S \ref{sec:discussion}. We observe that applying only clipping and masking (lines 4–5) led to a reduction in membership leakage. However, the utility of explanation dropped. Therefore, we add calibrated Gaussian noise (line 6), as the clip$\rightarrow$mask$\rightarrow$add\_noise sequence proved effective in significantly reducing leakage with negligible reduction on explanation utility.

\textbf{Empirical hardening vs. formal privacy:} Although \sysname{} provides lightweight and model-agnostic mechanisms to reduce membership leakage from explanation outputs, these defenses are empirical in nature and do not offer formal guarantees of the likes of differential privacy (DP). Instead, our goal is to characterize and optimize the empirical privacy/utility tradeoffs that arise across a wide range of explanation methods. As we show in \S\ref{subsec:hardening-results}, carefully chosen transformations can substantially reduce leakage while preserving explanation utility. We position \sysname{} as a practical framework for auditing and mitigating explanation-induced leakage, complementary to formal DP-based approaches.

%% file: evaluation.tex
\section{Evaluation}\label{sec:eval}
Our validation of \sysname{} is guided by the following research questions:
\begin{itemize}
    \item \textbf{RQ1}: How leaky are explanation methods when deployed with default parameters/configurations?

    \item \textbf{RQ2}: How effective are explanation method hardening strategies in reducing membership leakage without compromising explanation utility? 
    
    \item \textbf{RQ3}: What are the root causes of leakage in explanation methods? Do the root causes vary among classes of explanation methods?
       
\end{itemize}

\subsection{Experimental Setup}

\textbf{Datasets}:
We use 3 benchmark datasets: CIFAR-10 \cite{krizhevsky2012cifar}, CIFAR-100 \cite{krizhevsky2012cifar},  and GTSRB \cite{stallkamp2011gtsrb} that are common in MIA studies. Following our attack setup in \S \ref{sec:attackdetails}, Table \ref{tab:data_splits} in the Appendix \S \ref{sec:appendix} shows how we split our datasets.

\textbf{Model Architectures}:
We deploy commonly used model architectures for image datasets. We use MobileVNET2 \cite{mobilevnet2} for CIFAR-10 and CIFAR-100 and ResNet-18 \cite{resnet18} for GTSRB, which are used by the state-of-the-art approach~\cite{PleaseTellmeMore}, with which we compare \sysname{}. In \S \ref{sec:discussion}, we evaluated the impact of using different model architectures (e.g. ResNet-18, DenseNET161\cite{huang2018denselyconnectedconvolutionalnetworks}) for the target and the shadow model against MobileVNET on CIFAR-10. For more details, Table~\ref{tab:model_acc} in Appendix \S \ref{sec:appendix} shows the performance of the target models trained on different datasets.

\textbf{Training Configurations.} We train each model for 100 epochs (to match our baseline~\cite{PleaseTellmeMore} as shown in Table \ref{tab:model_acc}) with a learning rate of 0.1. We also reduce the learning rate of the optimizer in a cosine annealing schedule to ensure better model convergence. Standard data augmentations and weight decays with a rate of 0.0001 are used to improve the generalization of the models.

\vspace{1em}

\textbf{Target Explanation Methods}:
We evaluated \sysname{} on 15 explanation methods. For head-to-head comparison, we used the seven explanation methods used  \cite{PleaseTellmeMore} as our baseline for MIA against the explanation methods. The methods are: SmoothGrad, VarGrad, Integrated Gradients, GradCAM, GradCAM++, SHAP, and LIME. In Figures \ref{fig:XAI_pareto}, \ref{fig:pareto1}, and \ref{fig:XAI_pareto_appendix} (Appendix), we present the hardening effectiveness results as pareto front plots for representative explanation methods. 

\textbf{Attack}: 
Our attack setup is straightforward. We run the attack 20 times using different random seeds. We pick the best seed that has the highest attack accuracy and TPR rate at low FPR. We train only one shadow model that serves as attribution score dataset generator to train the attack model.

\textbf{Evaluation Metrics}: Following recent studies~\cite{PleaseTellmeMore, shokri2021privacy}, we adopt TPR@0.1\%FPR, Balanced Accuracy, and Area Under the ROC Curve (AUC)~\cite{li2022auditing,salem2018ml,shokri2017membership}. Balanced Accuracy reflects the average of sensitivity and specificity on a dataset equally composed of members and non-members. Although this metric may not always be the most indicative, we include it for the sake of comparison with prior work~\cite{liu2022membership, carlini2022membership}. 

To measure hardening effectiveness, we use \text{MLS} (Equation \ref{eq:mls}) and the sensitivity change $\mathbb{S}$ (based on Equation \ref{eq:sens}) for a given explanation method. We measure the utility variation of the explanation method post-hardening by computing the percentage change in sensitivity as:
% \[
% \text{Sensitivity Change} = \frac{\text{Pre\_Hardening\_Sens}-\text{Post\_Hardening\_Sens}}{\text{Pre\_Hardening\_Sens}} \times 100
% \]
% A value close to 100\% indicates preserved Sensitivity, while a lower percentage indicates sensitivity decrease (utility Gain) and higher percentage (+100\%) means increase of sensitivity(utility loss) .
\scalebox{}{}
\[
\scalebox{1.1}{$
\Delta \mathbb{S} = \frac{\left| \mathbb{S}(Pre\_Hardening) - \mathbb{S}(Post\_Hardening) \right|}{\mathbb{S}(Pre\_Hardening)} \times 100
$}
\]

A higher $\Delta \mathbb{S}$ indicates a larger difference between the pre- and post-hardening MLS values. 
A downward arrow (\textcolor{blue}{$\downarrow$}) denotes a utility gain (i.e., decrease in sensitivity after hardening), 
while an upward arrow (\textcolor{red}{$\uparrow$}) indicates a utility loss (i.e., increase in sensitivity).

\subsection{Comparison with State-of-the-Art MIAs}
\label{subsec:leakage-results}

To address RQ1,  in Table~\ref{tab:mia_baselinecomp} we compare our MIA with our baseline~\cite{PleaseTellmeMore}. Our attack outperforms all recent works by 5\% on average in TPR@0.1\%FPR on CIFAR-10. We observe similar effectiveness on GTSRB by an average of 10\% across all explanation methods except for GradCAM. We believe the reason behind less performance of our attack on GradCAM is due to its usage of the last convolution layer to explain the prediction of the model, which is different from the rest of the explanation methods that explain the feature vector. 

We also outperform the state-of-the-art explanation-aware MIAs on CIFAR-100 on IG and GradCAM++ by 1\%. However, we fail to outperform their attack on other explanation methods like SmoothGrad and VarGrad, but, on balance, our results highlight a high degree of membership leakage in comparison to previously reported leakage across families of explanation methods.
{\small
\begin{summarybox}
\textbf{Summary:} Default explanation configurations leak \textbf{$\approx$75\% more membership information} than prior reports. 
Gradient-based methods are the most vulnerable due to high sensitivity. 
Leakage correlates strongly with model overfitting and attribution variance.
\end{summarybox}
}

\begin{table*}[h!]
\centering
\caption{\textbf{Comparison of our attack with state-of-the-art explanation-aware MIAs across datasets.}}
\label{tab:mia_baselinecomp}
\begin{adjustbox}{max width=0.95\textwidth}
\begin{tabular}{c|c|ccc|ccc|ccc}
    \toprule
    \multirow{2}{*}{\textbf{Attack Method}} & \multirow{2}{*}{\textbf{Explanation Method}} & 
    \multicolumn{3}{c}{\textbf{TPR@0.1\%FPR}} & 
    \multicolumn{3}{c}{\textbf{Balanced Accuracy}} & 
    \multicolumn{3}{c}{\textbf{Attack AUC}} \\
    & & CIFAR-10 & CIFAR-100 & GTSRB & CIFAR-10 & CIFAR-100 & GTSRB & CIFAR-10 & CIFAR-100 & GTSRB \\
    \midrule

    \multirow{7}{*}{Shokri et al. (expl.) \cite{shokri2021privacy}} 
    & SmoothGrad & 0.2\% & 0.4\% & 0.3\% & 0.607 & 0.741 & 0.500 & 0.642 & 0.799 & 0.679 \\
    & VarGrad    & 0.2\% & 0.3\% & 0.1\% & 0.616 & 0.754 & 0.658 & 0.638 & 0.814 & 0.692 \\
    & IG         & 0.3\% & 0.4\% & 0.1\% & 0.594 & 0.712 & 0.637 & 0.630 & 0.777 & 0.687 \\
    & GradCAM    & 0.3\% & 0.6\% & 0.3\% & 0.614 & 0.775 & 0.578 & 0.654 & 0.843 & 0.642 \\
    & GradCAM++  & 0.2\% & 0.6\% & 0.3\% & 0.621 & 0.765 & 0.623 & 0.659 & 0.842 & 0.613 \\
    & SHAP       & 0.2\% & 0.4\% & 0.2\% & 0.607 & 0.751 & 0.616 & 0.618 & 0.798 & 0.638 \\
    & LIME       & 0.1\% & 0.4\% & 0.2\% & 0.604 & 0.744 & 0.610 & 0.616 & 0.788 & 0.627 \\
    \midrule

    \multirow{7}{*}{Liu et al. \cite{PleaseTellmeMore} w/ loss traj.} 
    & SmoothGrad & 4.1\% & 16.3\% & 2.1\% & 0.652 & 0.876 & 0.619 & 0.750 & 0.961 & 0.708 \\
    & VarGrad    & 3.9\% & 16.4\% & 1.5\% & 0.656 & 0.885 & 0.617 & 0.745 & 0.957 & 0.704 \\
    & IG         & 3.8\% & 16.0\% & 1.7\% & 0.656 & 0.878 & 0.611 & 0.757 & 0.958 & 0.701 \\
    & GradCAM    & 3.9\% & 15.8\% & 1.9\% & 0.656 & 0.887 & 0.616 & 0.751 & 0.962 & 0.740 \\
    & GradCAM++  & 3.9\% & 16.5\% & 1.9\% & 0.632 & 0.897 & 0.616 & 0.750 & 0.962 & 0.707 \\
    & SHAP       & 3.9\% & 15.7\% & 1.3\% & 0.652 & 0.861 & 0.624 & 0.755 & 0.961 & 0.708 \\
    & LIME       & 4.0\% & 16.3\% & 1.3\% & 0.644 & 0.852 & 0.622 & 0.751 & 0.959 & 0.703 \\
    \midrule

    \multirow{7}{*}{\sysname{} (Ours)} 
    & SmoothGrad & \textbf{10.38\%} & 6.81\% & \textbf{6.3\%} &  \textbf{0.786} & 0.799 &  \textbf{0.631} &  \textbf{0.910} & 0.827 &  \textbf{0.751} \\
    & VarGrad    & \textbf{6.78\%} & 6.68\% & \textbf{3.5\%} &  \textbf{0.635} & 0.779 &  \textbf{0.619} &  \textbf{0.732} & 0.846 &  \textbf{0.726} \\
    & IG         & \textbf{4.88\%} &  \textbf{16.42\%} & \textbf{43.52\%} &  \textbf{0.793} &  \textbf{0.903} &  \textbf{0.968} &  \textbf{0.861} &  \textbf{0.986} &  \textbf{0.995} \\
    & GradCAM    & 0.95\% & 6.78\% & 0.74\% & 0.63 & 0.82 & 0.663 & 0.651 & 0.86 & 0.613 \\
    & GradCAM++  & \textbf{6.7\%} & \textbf{18.0\%} & \textbf{5.78\%} &  \textbf{0.703} &  \textbf{0.905} &  \textbf{0.729} &  \textbf{0.862} &  \textbf{0.983} &  \textbf{0.795} \\
    & SHAP       & \textbf{5.04\%} & 11.8\% & \textbf{28.83\%} &  \textbf{0.711} & 0.8393 &  \textbf{0.981} &  \textbf{0.857} & 0.884 &  \textbf{0.962} \\
    & LIME       & \textbf{10.26\%} & 10.5\% & \textbf{21.52\%} &  \textbf{0.751} & 0.8425 &  \textbf{0.935} &  \textbf{0.815} & 0.891 &  \textbf{0.947} \\
    \bottomrule
\end{tabular}
\end{adjustbox}
 \vspace{-1em}
\end{table*}

\subsection{Effectiveness of Hardening Strategies}
\label{subsec:hardening-results}
 
To address RQ2, we evaluated \sysname{} on 15 explanation methods for 3 different datasets on our 2 main metrics. Figures \ref{fig:pareto1}, \ref{fig:XAI_pareto}, and \ref{fig:XAI_pareto_appendix} illustrate the trade-offs between utility (sensitivity) and privacy risk (TPR@0.1\%FPR) across explanation methods on 3 datasets. Specifically, each plot has 60 points coming from 3 different seeds (20 points per seed). These 3 seeds were selected by having the attack model evaluated on 20 different seeds for each explanation method with default parameters, and the 3 highest TPR seeds were picked as baselines to harden. Furthermore, each individual point in these plots represents a different selection of parameters for a given explanation method with the color, size, and shape indicating the specific parameter values, as indicated by the legends. The red star marks the ideal point, located at (0,0), indicating no membership leakage (0\% TPR) and 0 sensitivity change. 

Focusing on the Pareto fronts, across the 15 different explanation methods and 3 different datasets, there is a great variance in the results and how these plots should be depicted. This variance stems from how these explanation methods calculate their attribution scores, which results from not only their natural explainability and privacy, but also the design of the method itself. For example, ProtoDash in Figure \ref{fig:XAI_pareto} (row 4, column 1) demonstrates an ideal Pareto front. As the sigma parameter increases, represented by the color gradient, the sensitivity value decreases while the TPR@0.1\%FPR increases, displaying a clear trade-off between the 2 metrics we are measuring. Some plots with similar perspectives include Deconvolution in CIFAR-10 in Figure \ref{fig:XAI_pareto_appendix} (row 2, column 1), with correlations between the parameters and both utility and privacy representing a trade-off of some sort at all points. On the other hand, explanation methods such as SHAP, Occlusion, and LIME show a different behavior across all 3 datasets. Specifically, their plots (Figure \ref{fig:XAI_pareto} (row 1, row 3) and Figure \ref{fig:pareto1} (row 3) seem to represent a mostly horizontal line which reveals that with a particular choice of parameter values, the TPR can be reduced while mostly preserving the utility, rather than an expected trade-off. Another varying example is with Integrated Gradients on the GTSRB dataset in Figure \ref{fig:pareto1} (row 2, column 3), where one can see the impact of the boolean parameter, \texttt{multiply\_by\_inputs}, and the direct trade-off between sensitivity and TPR between setting it to True (square) or False (diamond). However, looking at Integrated Gradients again but now on CIFAR-10 in Figure \ref{fig:pareto1} (row 2, column 1), tells a different story with setting multiply\_by\_inputs to True having both an increase in utility and privacy compared to setting it to False. Overall, what this says is that these plots all need to be interpreted individually due to the variance in not only explanation methods and their algorithms, but also the impact of parameters when it comes to different datasets. In conclusion, this analysis shows that all explanation methods benefit from careful hardening and tuning, underscoring the role of parameter selection in balancing utility and privacy.

 % Table \ref{tab:hardeningres} shows the impact of privacy-preserving transformations on various explanation methods. The Pre-Hardening MLS column shows membership leakage for default deployment of an explanation method. Overall, all methods experience a drop in TPR post-hardening. As can be seen from the $\Delta {S}$ column, \sysname{} not only maintains the utility but also reduces the sensitivity by 64\% on average across all explanation methods on the three datasets, with a range [0\%,98.7\%]. However, we lost utility for certain explanation methods such as IntegratedGradients: by 0.35\% on CIFAR-10 and 0.04\% on CIFAR-100. The average sensitivity increase (utility loss) is 3.3\%. Overall, \sysname{} successfully reduces membership leakage in explanation methods while preserving, and in some cases, improving the explanation methods' utility.  
 Table \ref{tab:hardeningres} shows the impact of privacy-preserving transformations on various explanation methods. The Pre-Hardening MLS column shows membership leakage for default deployment of an explanation method. Overall, all methods experience a drop in TPR post-hardening. As can be seen from the $\Delta {S}$ column, \sysname{} not only maintains the utility but also reduces the sensitivity by 64\% on average across all explanation methods on the three datasets, with a range [0\%,98.7\%]. However, we lost utility for certain explanation methods such as IntegratedGradients: by 0.35\% on CIFAR-10 and 0.04\% on CIFAR-100. The average sensitivity increase is 3.3\%. Compared to a DP-SGD \cite{steinke2023privacyauditing1training} baseline, \sysname{} achieves lower post-hardening MLS for several explanation methods such as SmoothGrad, GradCAM++, SHAP, and LIME at comparable or higher utility, while DP-SGD slightly outperforms \sysname{} for some settings (e.g., IntegratedGradients on CIFAR-100). For other methods such as VarGrad, both defenses are close in terms of leakage and utility, indicating that \sysname{} can match DP-SGD’s privacy benefits without retraining the target model and losing model utiliy which ranged from 2.5\% to 10\% across datasets. Overall, \sysname{} successfully reduces membership leakage in explanation methods while preserving, and in some cases, improving the utility of explanation methods.

\begin{table*}[ht]
\centering
\caption{\textbf{\small Comparison of pre- and post-hardening MLS and sensitivity change of \sysname{} vs. DP-SGD \cite{steinke2023privacyauditing1training}. CIFAR-10 (10 \% utility loss,  $\epsilon=8$,  $\delta=10^{-5}$). CIFAR-100 (8\% utility loss, $\epsilon=10$, $\delta=10^{-5}$). GTSRB (2.5\% utility loss,  $\epsilon=9$, $\delta=10^{-5}$).}}
\scalebox{0.7}{
\begin{tabular}{l ccc ccccccc ccccccc}
\toprule
 & \multicolumn{3}{c}{\textbf{Pre-Hardening MLS ((TPR@0.1\%FPR))}} 
 & \multicolumn{6}{c}{\textbf{Post-Hardening MLS (TPR@0.1\%FPR)}} 
 & \multicolumn{6}{c}{\textbf{Sensitivity Change: $\Delta \mathbb{S}$}} \\
\cmidrule(lr){2-4} \cmidrule(lr){5-10} \cmidrule(lr){11-16}
\textbf{Explanation Method} 
& CIFAR-10 & CIFAR-100 & GTSRB 
& \multicolumn{2}{c}{CIFAR-10} 
& \multicolumn{2}{c}{CIFAR-100} 
& \multicolumn{2}{c}{GTSRB} 
& \multicolumn{2}{c}{CIFAR-10} 
& \multicolumn{2}{c}{CIFAR-100} 
& \multicolumn{2}{c}{GTSRB} \\
\cmidrule(lr){5-6} \cmidrule(lr){7-8} \cmidrule(lr){9-10} \cmidrule(lr){11-12} \cmidrule(lr){13-14} \cmidrule(lr){15-16} 

 & & & 
 & Ours & DP-SGD 
 & Ours & DP-SGD 
 & Ours & DP-SGD 
 & Ours & DP-SGD 
 & Ours & DP-SGD 
 & Ours & DP-SGD \\
\midrule
SmoothGrad \cite{smilkov2017smoothgrad} 
& 10.38 & 10.38 & 6.31 
& 0.84 & \textbf{0.19} & 0.64 & \textbf{0.15} & 0.76 & \textbf{0.19}
& \textcolor{blue}{$\downarrow$  {22.44}\%} & \textcolor{blue}{$\downarrow$ 21.32\%}
& \textcolor{red}{$\uparrow$  {7.23}\%}    & \textcolor{red}{$\uparrow$ 7.59\%}
& \textcolor{blue}{$\downarrow$  {14.03}\%} & \textcolor{blue}{$\downarrow$ 13.33\%} \\

VarGrad \cite{vargrad} 
& 6.78 & 6.68 & 3.52 
& 0.27 & \textbf{0.14} & 1.62 & \textbf{0.96} & \textbf{0.00} & 0.39
& \textcolor{blue}{$\downarrow$  {6.76}\%} & \textcolor{blue}{$\downarrow$ 6.42\%}
& \textcolor{red}{$\uparrow$  {8.36}\%}    & \textcolor{red}{$\uparrow$ 8.78\%}
& \textcolor{blue}{$\downarrow$  {12.93}\%} & \textcolor{blue}{$\downarrow$ 12.28\%} \\

Integrated Gradients \cite{sundararajan2017axiomatic} 
& 4.88 & 16.42 & 43.52 
& \textbf{0.34} & 0.70 & 4.02 & \textbf{0.68} & 15.88 & \textbf{0.39}
& \textcolor{red}{$\uparrow$  {0.35}\%}    & \textcolor{red}{$\uparrow$ 10\%}
& \textcolor{red}{$\uparrow$  {0.0425}\%}  & \textcolor{red}{$\uparrow$ 24\%}
& \textcolor{blue}{$\downarrow$ 2.33\%} & \textcolor{blue}{$\downarrow$  {15}\%} \\

GradCAM \cite{selvaraju2017grad} 
& 0.95 & 6.78 & 0.39 
& \textbf{0.00} & 0.30 & 0.26 & \textbf{0.18} & \textbf{0.00} & 0.19
& \textcolor{blue}{$\downarrow$ 7.86\%}  & \textcolor{blue}{$\downarrow$  {23}\%}
& \textcolor{blue}{$\downarrow$ 24.77\%} & \textcolor{blue}{$\downarrow$  {29}\%}
& \textcolor{blue}{$\downarrow$ 17.31\%} & \textcolor{blue}{$\downarrow$  {35}\%} \\

GradCAM++ \cite{chattopadhay2018grad++} 
& 6.70 & 18.00 & 3.72
& 1.44 & \textbf{0.80} & 3.60 & \textbf{0.59} & 0.26 & \textbf{0.19}
& \textcolor{blue}{$\downarrow$ 0.0733\%} & \textcolor{blue}{$\downarrow$  {86}\%}
& \textcolor{blue}{$\downarrow$ 0.0016\%} & \textcolor{blue}{$\downarrow$  {32}\%}
& 0.00\%                                  & \textcolor{blue}{$\downarrow$  {99}\%} \\

SHAP \cite{lundberg2017unified} 
& 5.04 & 11.80 & 28.83 
& 0.00 & 0.00 & 0.06 & \textbf{0.06} & \textbf{0.29} & 8.23
& \textcolor{blue}{$\downarrow$  {0.97}\%}  & \textcolor{blue}{$\downarrow$ 0.92\%}
& \textcolor{blue}{$\downarrow$  {4.79}\%}  & \textcolor{blue}{$\downarrow$ 4.55\%}
& \textcolor{blue}{$\downarrow$  {7.92}\%}  & \textcolor{blue}{$\downarrow$ 7.52\%} \\

LIME \cite{ribeiro2016why} 
& 10.26 & 10.26 & 21.52 
& 0.12 & \textbf{0.00} & 1.07 & \textbf{0.12} & \textbf{0.13} & 18.47
& \textcolor{blue}{$\downarrow$  {3.72}\%}  & \textcolor{blue}{$\downarrow$ 3.53\%}
& \textcolor{blue}{$\downarrow$  {1.77}\%}  & \textcolor{blue}{$\downarrow$ 1.68\%}
& \textcolor{blue}{$\downarrow$  {1.79}\%}  & \textcolor{blue}{$\downarrow$ 1.70\%} \\
\hline
\\
Saliency Map \cite{simonyan2014deep} 
& 5.24 & 18.84 & 5.84 
& \textbf{0.06} & 0.32 & \textbf{0.00} & 0.42 & \textbf{0.00} & 0.35
& \textcolor{blue}{$\downarrow$  {51.06}\%} & \textcolor{blue}{$\downarrow$ 48\%}
& \textcolor{blue}{$\downarrow$  {98.7}\%}  & \textcolor{blue}{$\downarrow$ 63\%}
& \textcolor{blue}{$\downarrow$  {91.2}\%}  & \textcolor{blue}{$\downarrow$ 46\%} \\

Guided BackProp \cite{springenberg2015striving} 
& 10.94 & 45.32 & 6.64 
& 1.34 & \textbf{0.80} & \textbf{0.48} & 0.63 & \textbf{0.00} & 0.06
& \textcolor{blue}{$\downarrow$  {42.61}\%} & \textcolor{blue}{$\downarrow$ 35\%}
& \textcolor{blue}{$\downarrow$ 41.46\%} & \textcolor{blue}{$\downarrow$  {62}\%}
& \textcolor{blue}{$\downarrow$  {88.96}\%} & \textcolor{blue}{$\downarrow$ 35.71\%} \\
DeepLIFT \cite{shrikumar2019learning} 
& 1.00 & 11.82 & 37.61 
& \textbf{0.00} & 0.06 & \textbf{0.00} & 0.14 & \textbf{0.19} & 0.27
& \textcolor{red}{$\uparrow$  {0.07}\%}    & \textcolor{red}{$\uparrow$ 0.08\%}
& \textcolor{red}{$\uparrow$  {1.53}\%}     & \textcolor{red}{$\uparrow$ 1.61\%}
& \textcolor{blue}{$\downarrow$  {70.57}\%}  & \textcolor{blue}{$\downarrow$ 67.04\%} \\

InputXGrad \cite{InputXGradient} 
& 2.40 & 24.68 & 46.37 
& \textbf{0.18} & 0.80 & \textbf{0.00} & 0.93 & \textbf{0.00} & 0.20
& \textcolor{blue}{$\downarrow$  {29.79}\%}  & \textcolor{blue}{$\downarrow$ 28.30\%}
& \textcolor{blue}{$\downarrow$  {64.07}\%}  & \textcolor{blue}{$\downarrow$ 60.87\%}
& \textcolor{blue}{$\downarrow$  {34.42}\%}  & \textcolor{blue}{$\downarrow$ 32.70\%} \\

Deconvolution \cite{deconv} 
& 16.08 & 48.48 & 5.84 
& \textbf{0.18} & 0.52 & 0.83 & \textbf{0.28} & \textbf{0.00} & 0.06
& \textcolor{blue}{$\downarrow$  {30.93}\%}  & \textcolor{blue}{$\downarrow$ 29.38\%}
& \textcolor{blue}{$\downarrow$  {66.71}\%}  & \textcolor{blue}{$\downarrow$ 63.37\%}
& \textcolor{blue}{$\downarrow$  {88.67}\%}  & \textcolor{blue}{$\downarrow$ 35.71\%} \\

Occlusion \cite{Uchiyama_2023_WACV} 
& 1.10 & 0.64 & 0.53
& 0.14 & \textbf{0.06} & 0.22 & \textbf{0.13} & 0.16 & \textbf{0.13}
& \textcolor{red}{$\uparrow$  {2.22}\%}      & \textcolor{red}{$\uparrow$ 2.33\%}
& \textcolor{blue}{$\downarrow$  {4.54}\%}   & \textcolor{blue}{$\downarrow$ 4.31\%}
& \textcolor{red}{$\uparrow$  {6.83}\%}      & \textcolor{red}{$\uparrow$ 50\%} \\

ProtoDash \cite{protodash} 
& 14.90 & 14.90 & 52.96 
& \textbf{0.00} & 13.84 & \textbf{0.00} & 11.60 & \textbf{0.00} & 14.15
& \textcolor{blue}{$\downarrow$  {54.78}\%}  & \textcolor{blue}{$\downarrow$ 52.04\%}
& 0.00\%                                  & \textcolor{blue}{$\downarrow$ 0\%}
& \textcolor{blue}{$\downarrow$  {0.0137}\%} & \textcolor{blue}{$\downarrow$ 0.01\%} \\
Anchors \cite{anchors} 
& 45.44 & 43.32 & 80.80 
& 14.16 & 41.79 & 20.02 & 62.32 & 0.00 & 67.37
& \textcolor{blue}{$\downarrow$  {75.02}\%}  & \textcolor{blue}{$\downarrow$ 71.27\%}
& \textcolor{blue}{$\downarrow$  {3.355}\%}  & \textcolor{blue}{$\downarrow$ 3.19\%}
& \textcolor{blue}{$\downarrow$  {99.31}\%}  & \textcolor{blue}{$\downarrow$ 75\%} \\
\bottomrule
\end{tabular}
}
\label{tab:hardeningres}
\end{table*}

{\small
\begin{summarybox}
\textbf{Summary:} Attribute value clipping, sensitivity-calibrated noise, and masking reduce membership leakage by 
\textbf{46.5\% to 95\%} with only \textbf{$\leq$3.3\%} explanation utility loss. Layering defenses yields the strongest privacy–utility trade-offs.
\end{summarybox}
}
\subsection{Root-Cause Analysis}\label{subsec:root-cause-analysis}
\begin{figure*}[htbp]
    % \vspace{-0.5cm}
    \centering
     % Set caption format for subfloats in this figure
    \captionsetup[subfloat]{labelformat=empty}
    
    % Set subrefformat for subcaptions in this figure
\noindent
\rotatebox{90}{\makebox[3.5cm][c]{\textbf{\scriptsize Saliency Map}}}
     \subfloat[]{\includegraphics[width=0.27\textwidth]{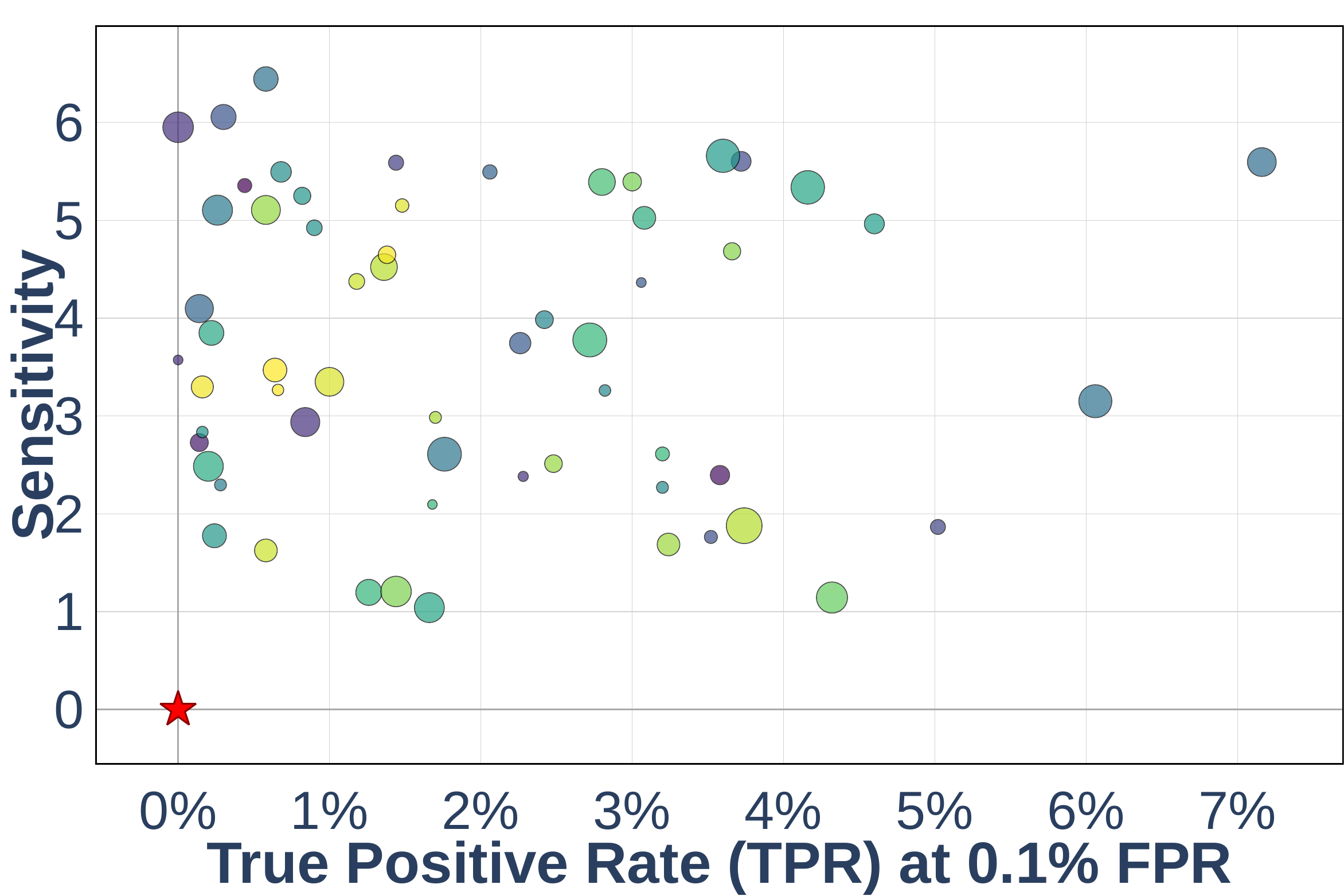}}
    \subfloat[]{\includegraphics[width=0.27\textwidth]{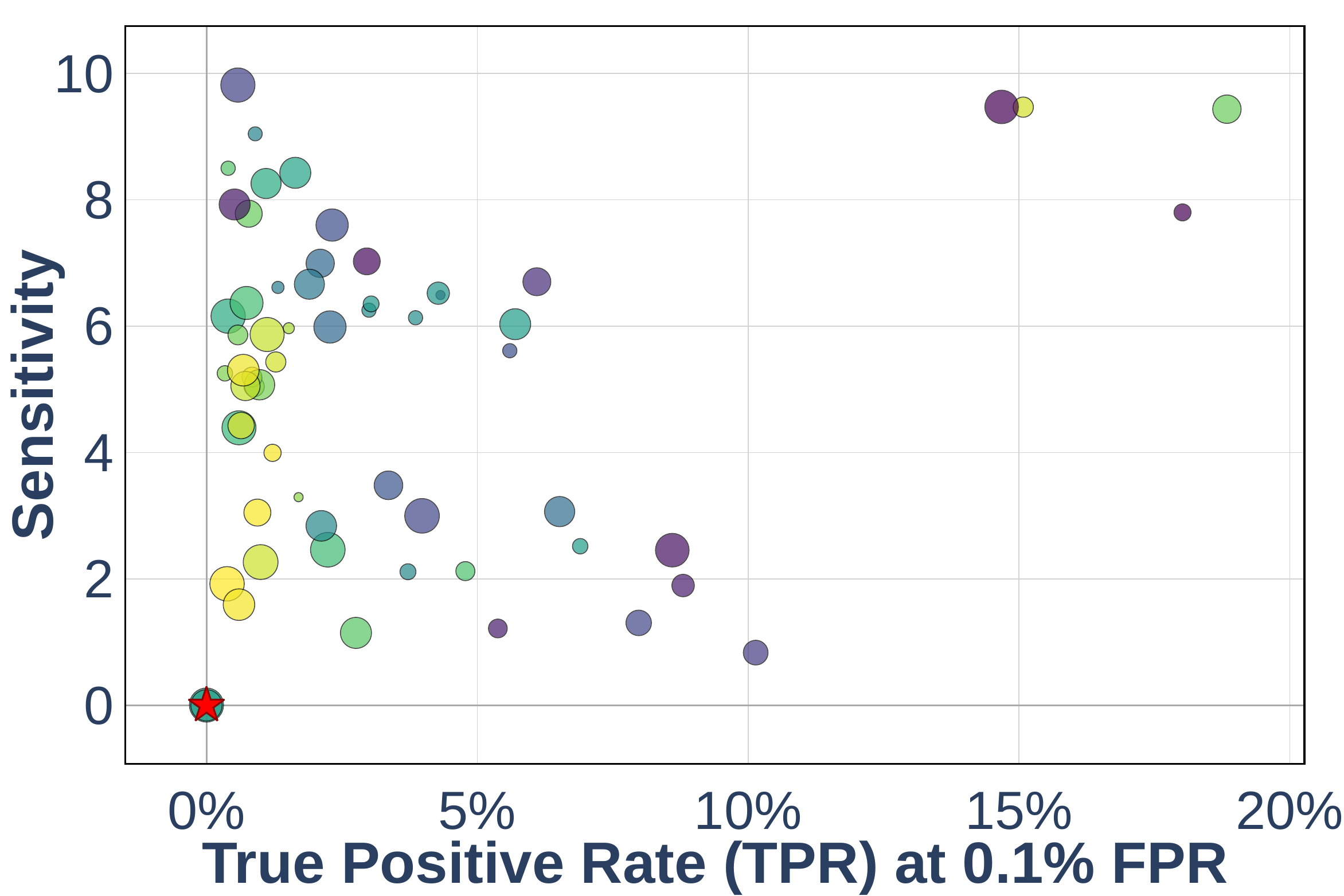}}
    \subfloat[]{\includegraphics[width=0.27\textwidth]{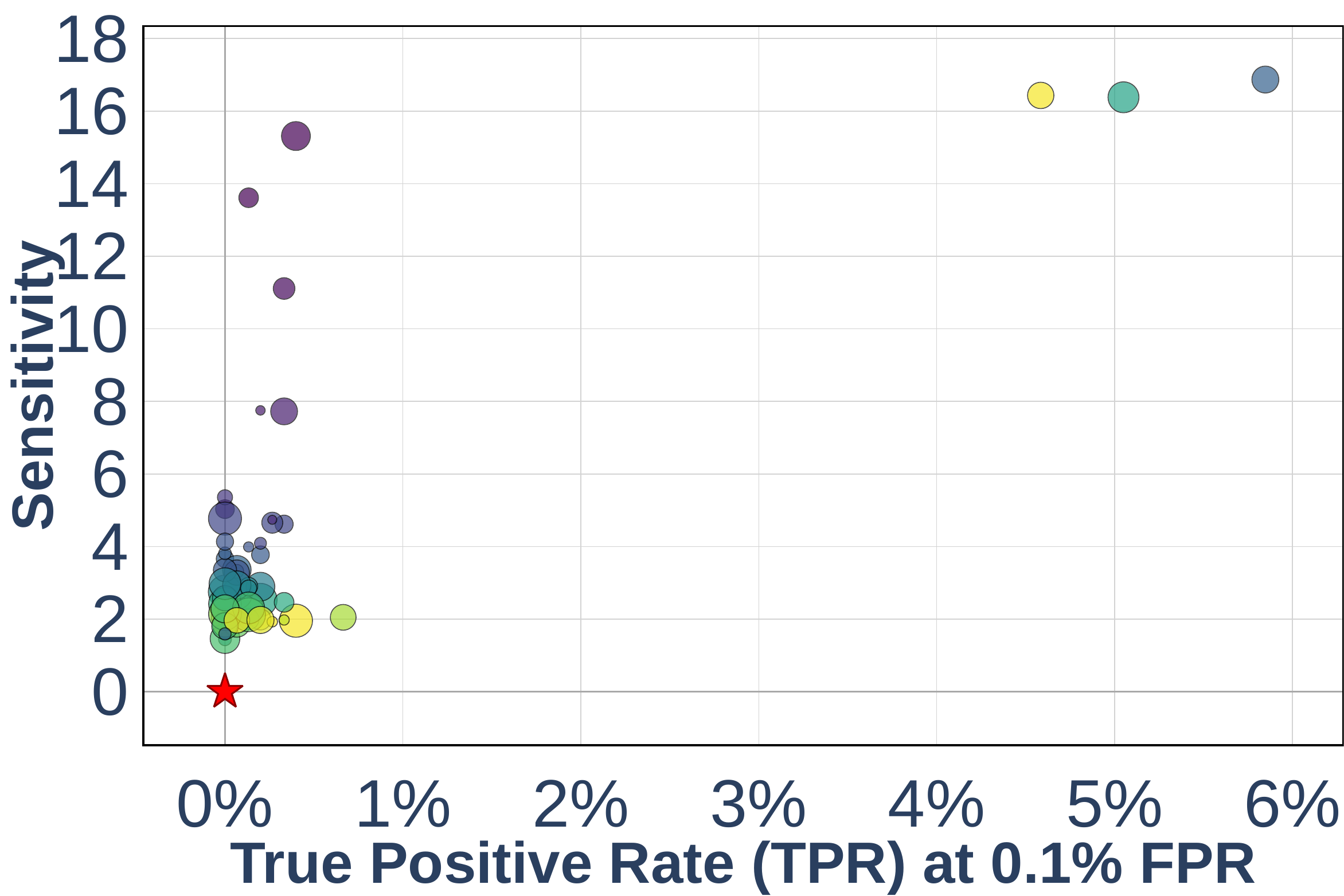}}
    \subfloat[]{\includegraphics[width=0.16\textwidth]{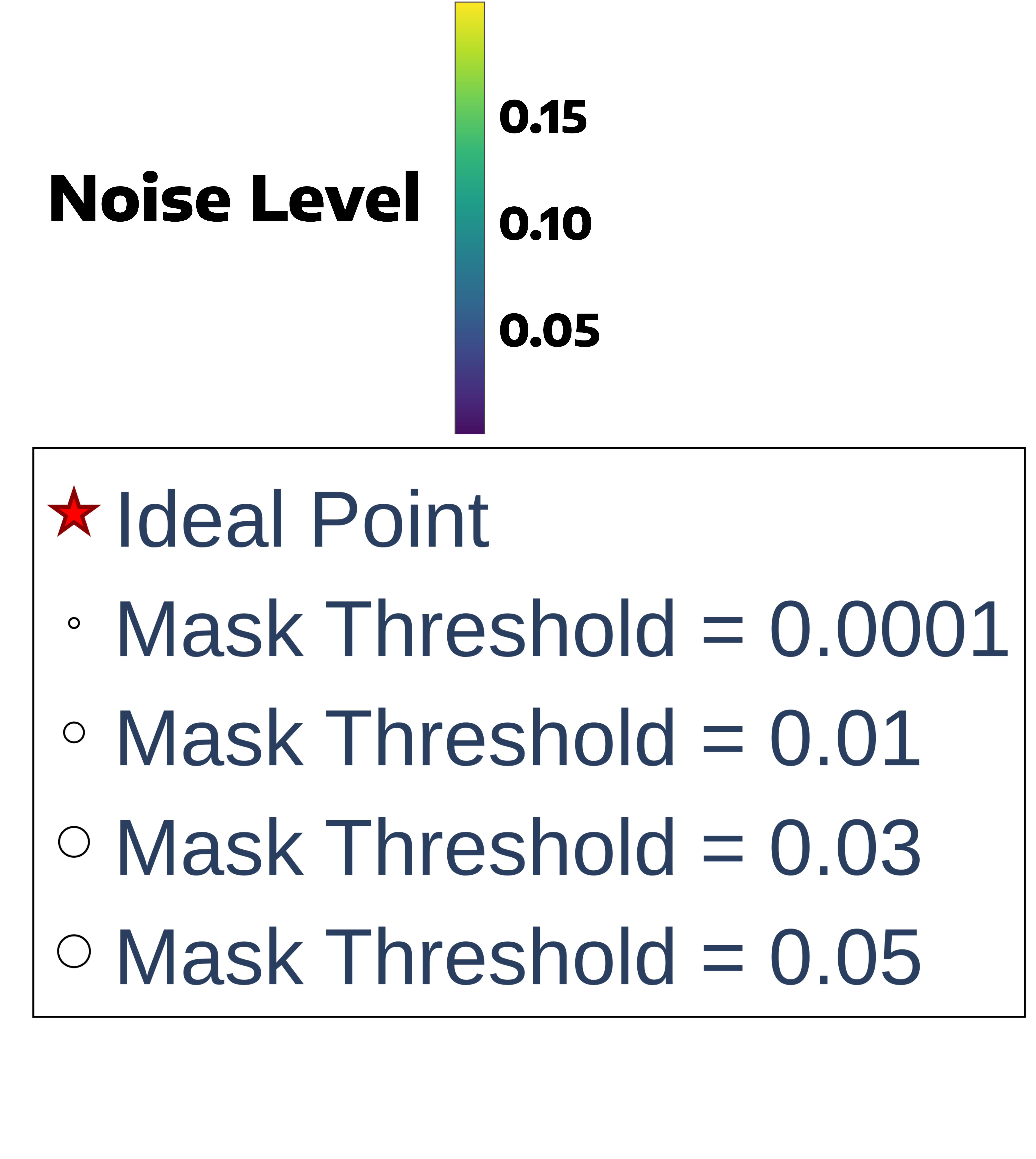}}
     \\
     \vspace{-0.8cm}

     \noindent
\rotatebox{90}{\makebox[3.5cm][c]{\textbf{\scriptsize Integrated Gradients}}}
          \subfloat[]{\includegraphics[width=0.27\textwidth]{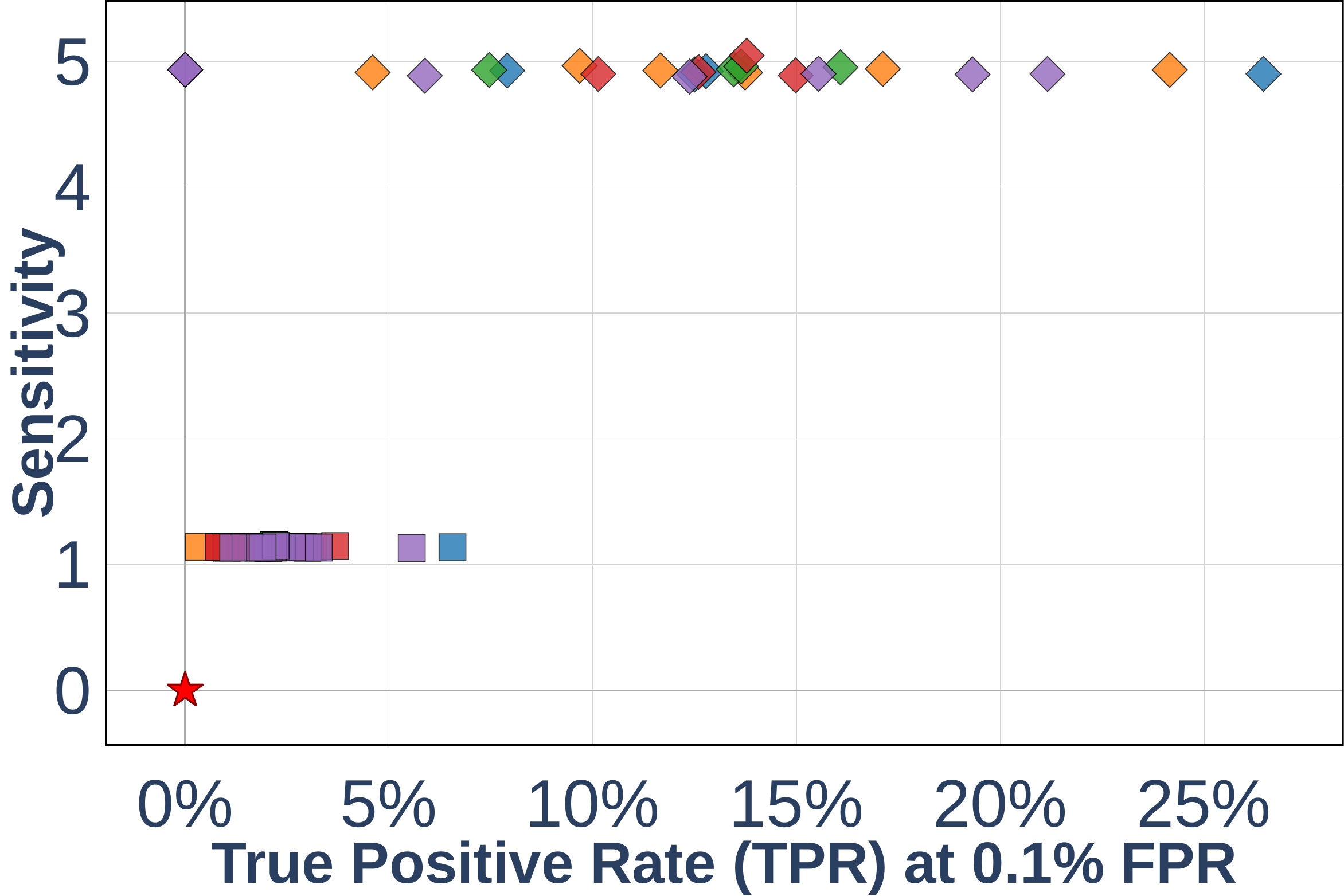}}
    \subfloat[]{\includegraphics[width=0.27\textwidth]{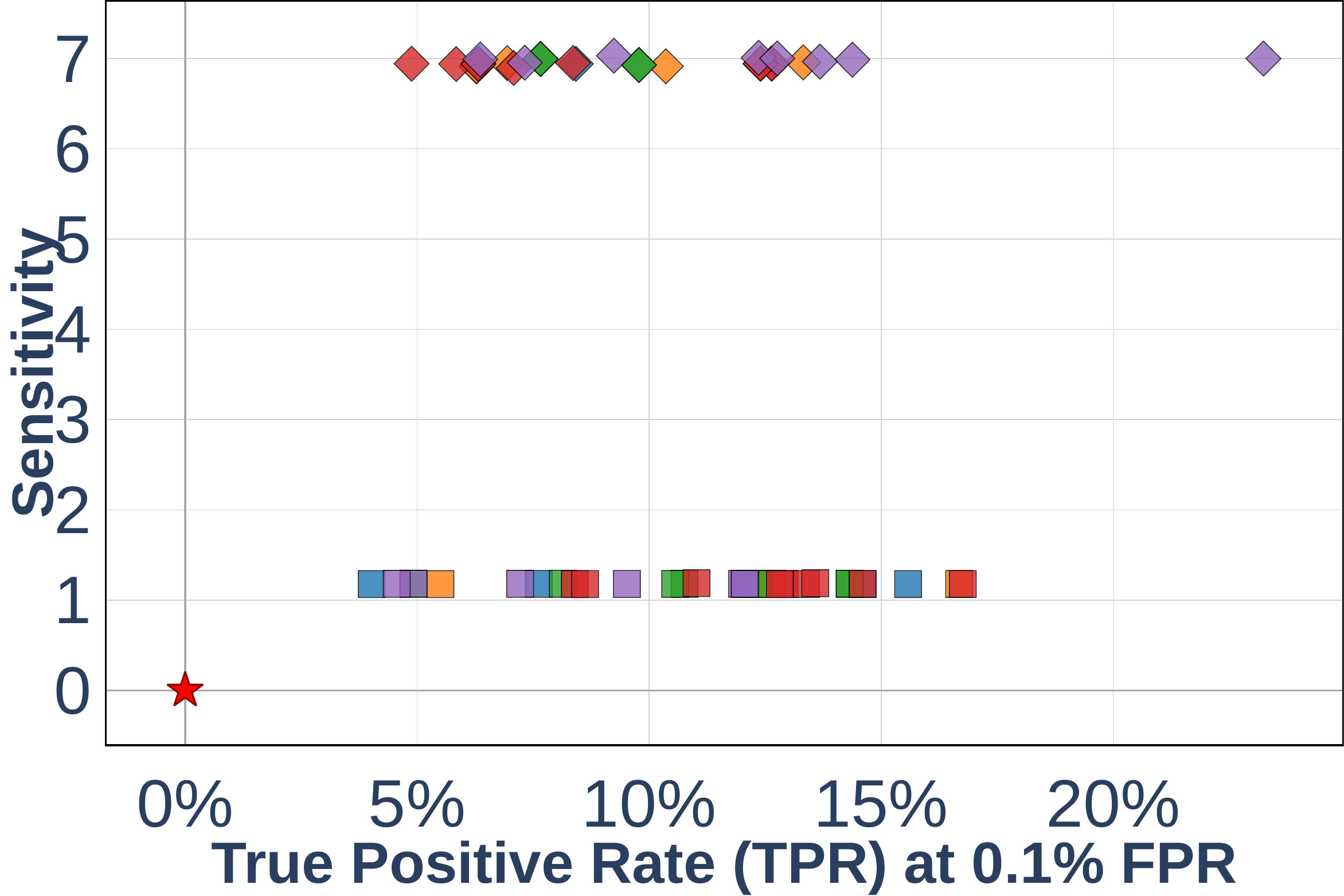}}
    \subfloat[]{\includegraphics[width=0.27\textwidth]{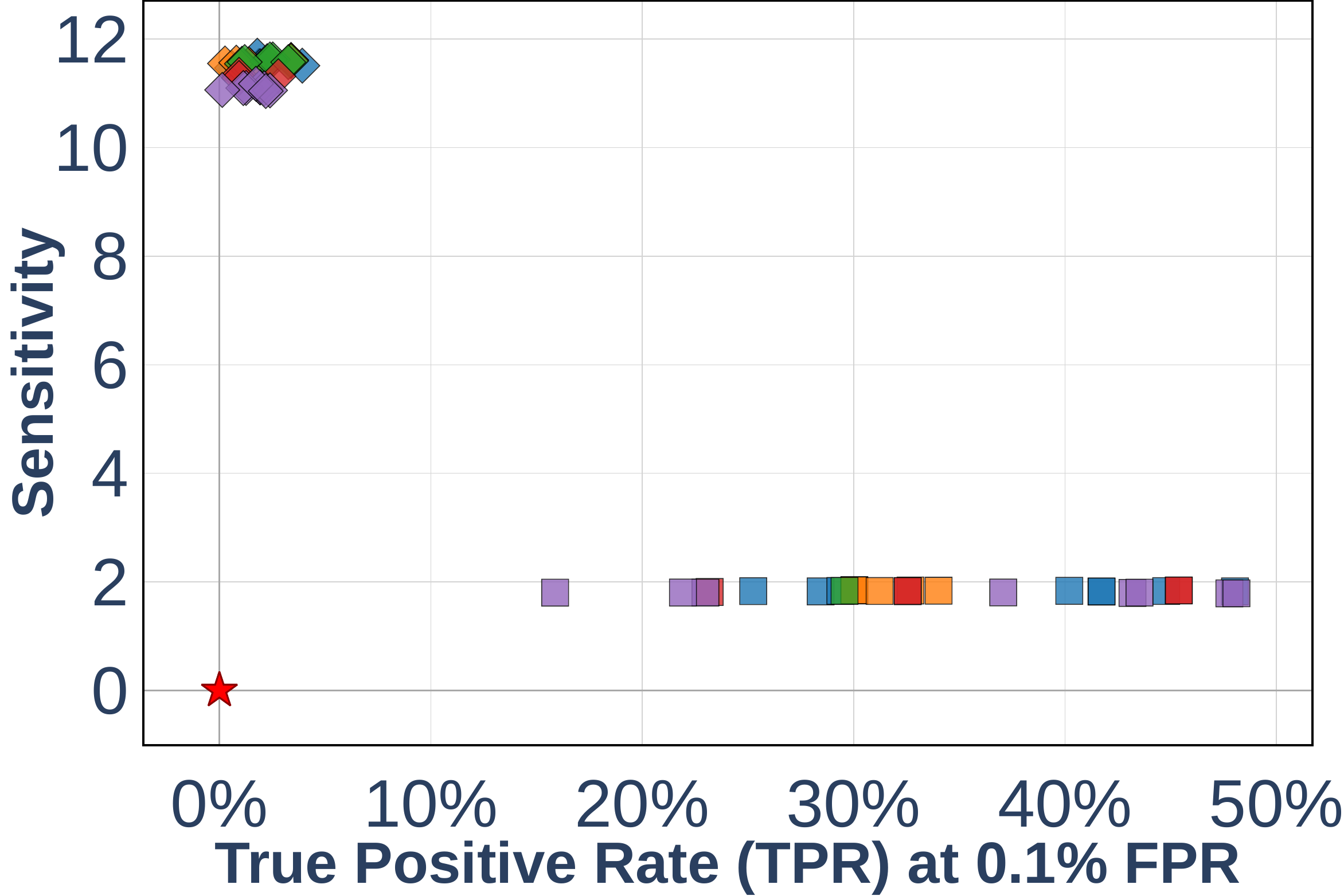}}
    \subfloat[]{\includegraphics[width=0.16\textwidth]{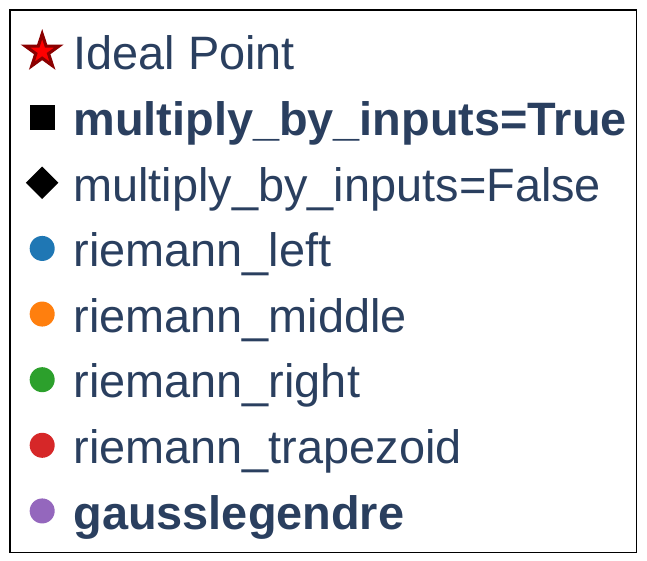}}
         \\
         \vspace{-0.8cm}
     \noindent
\rotatebox{90}{\makebox[3.5cm][c]{\textbf{\scriptsize SHAP}}}
          \subfloat[CIFAR-10]{\includegraphics[width=0.27\textwidth]{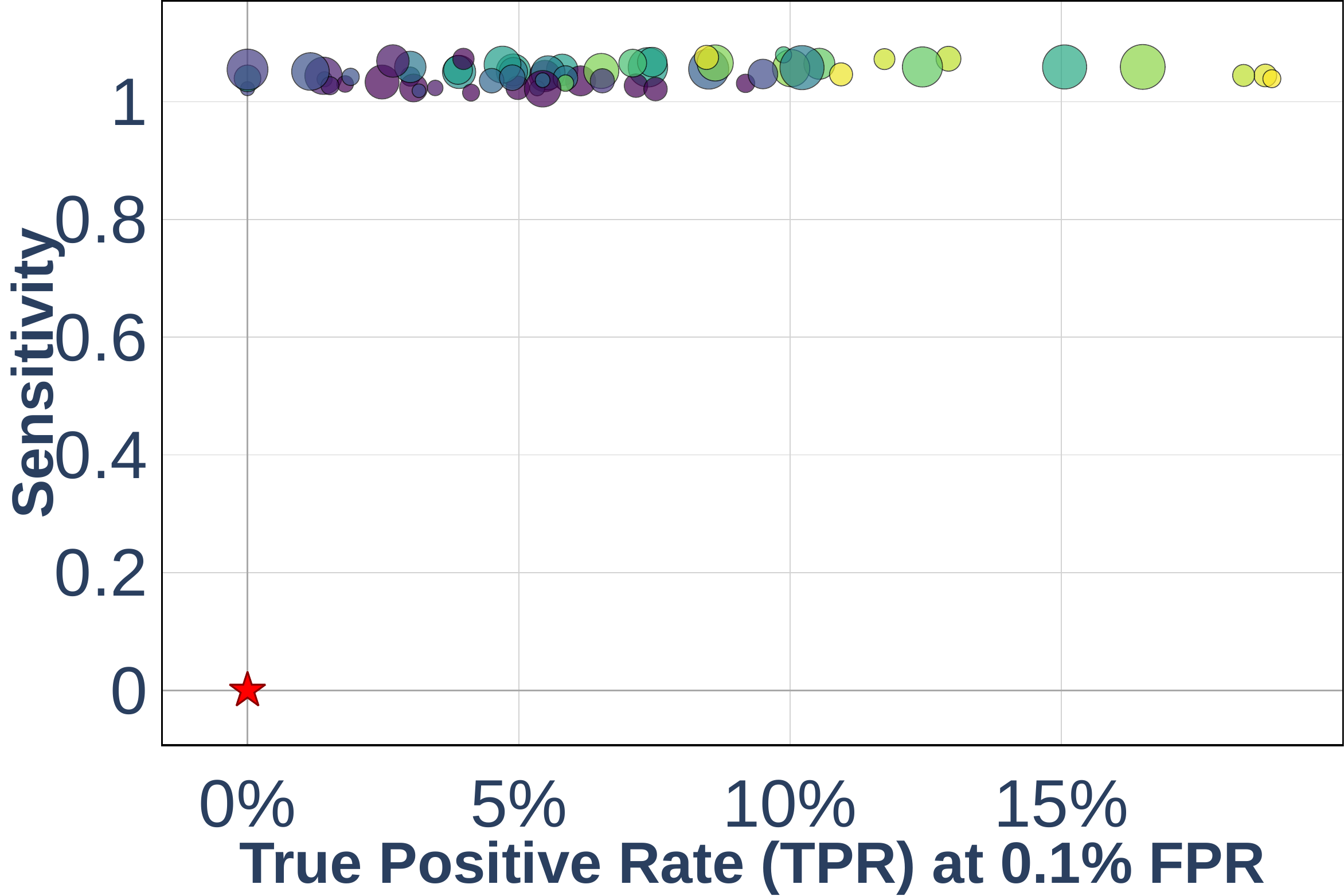}}
    \subfloat[CIFAR-100]{\includegraphics[width=0.27\textwidth]{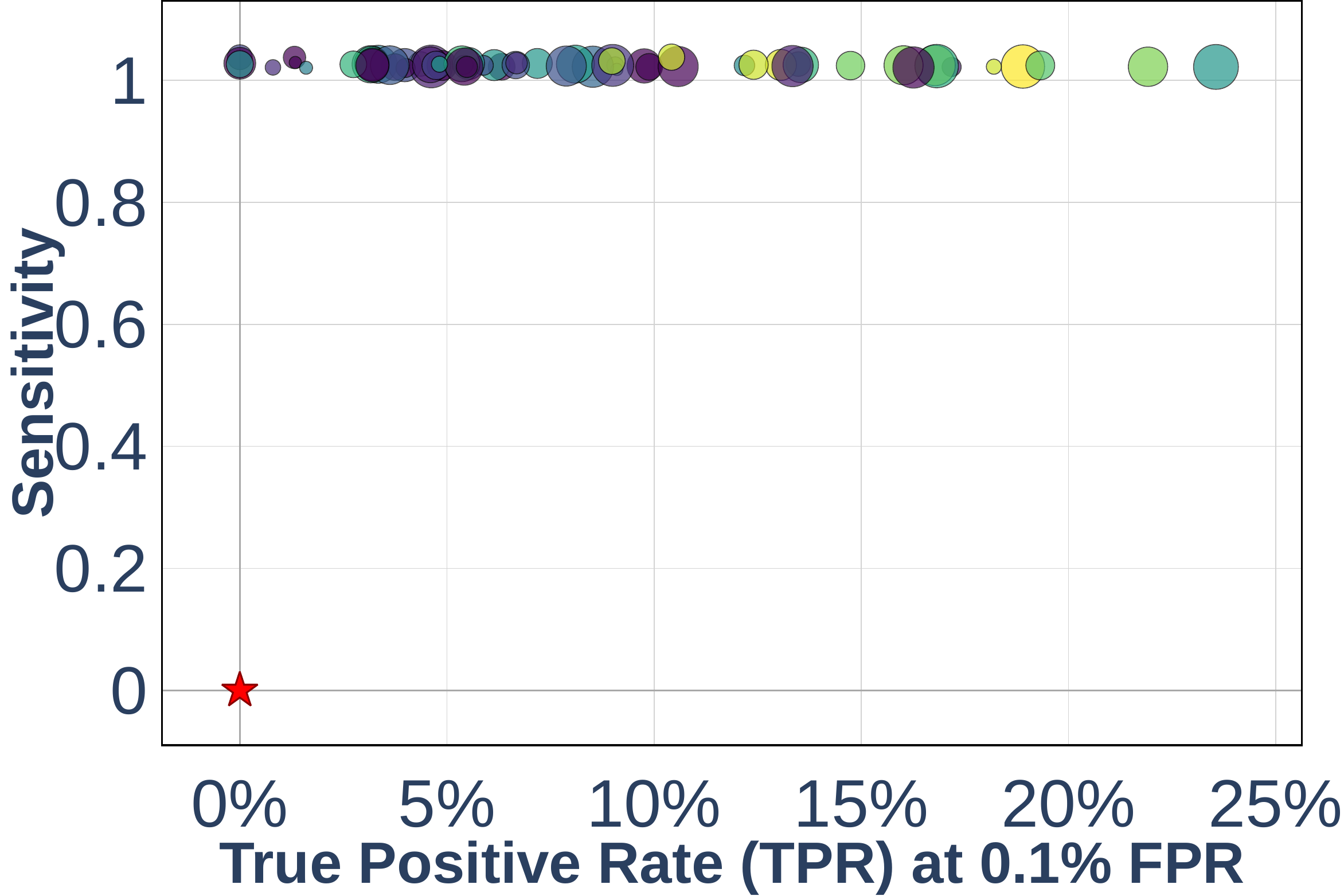}}
    \subfloat[GTSRB]{\includegraphics[width=0.27\textwidth]{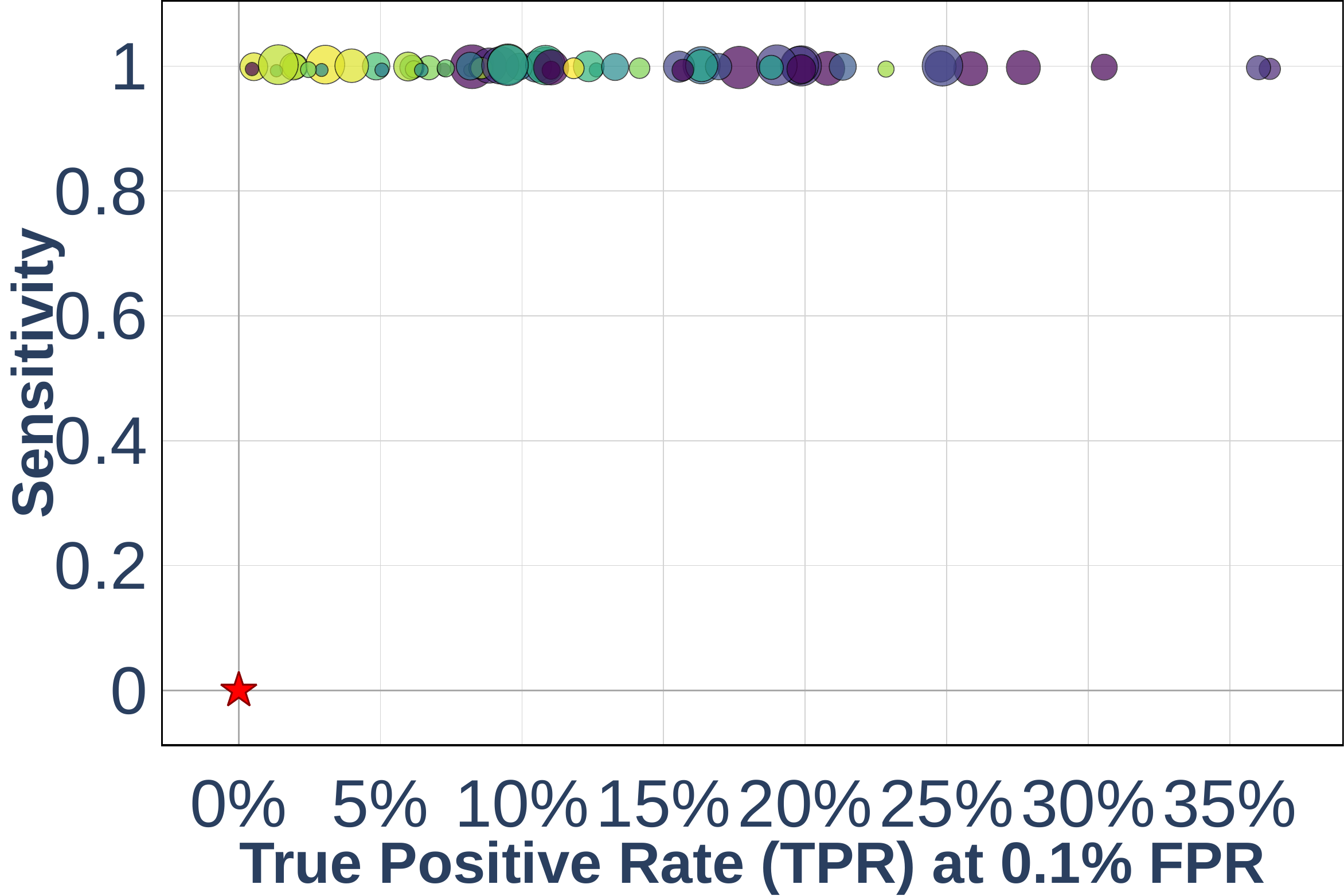}}
    \subfloat[Legend]{\includegraphics[width=0.17\textwidth]{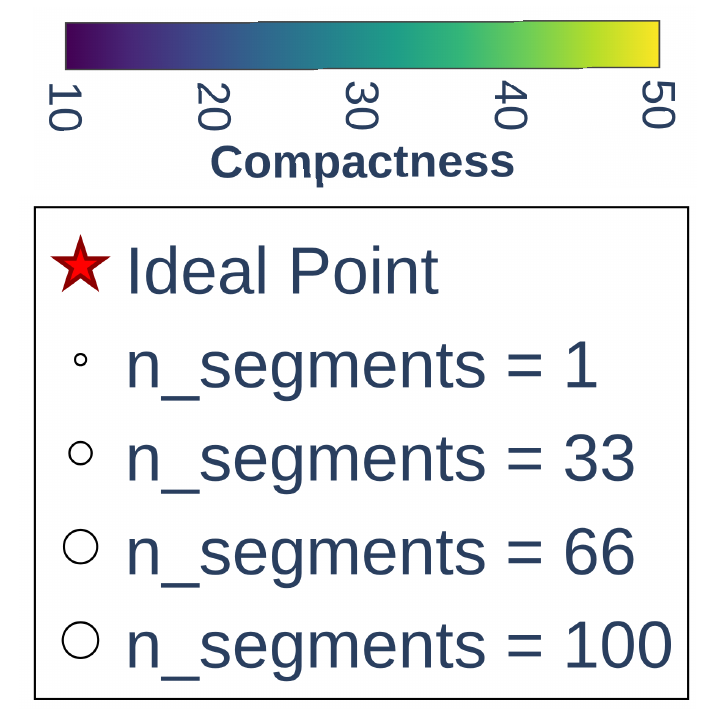}}
     % \vspace{-1em}
        \caption{\textbf{\sysname{} hardening optimization Pareto fronts for Saliency Map, Integrated Gradients, and SHAP.}}
      % \vspace{-1em}
      \label{fig:pareto1}
\end{figure*}
\begin{figure*}[t!]
   % \vspace{-0.5cm}
    \centering
     % Set caption format for subfloats in this figure
    \captionsetup[subfloat]{labelformat=empty}
    
    % Set subrefformat for subcaptions in this figure
\noindent
\rotatebox{90}{\makebox[3.5cm][c]{\textbf{\scriptsize Occlusion}}}
          \subfloat[]{\includegraphics[width=0.27\textwidth]{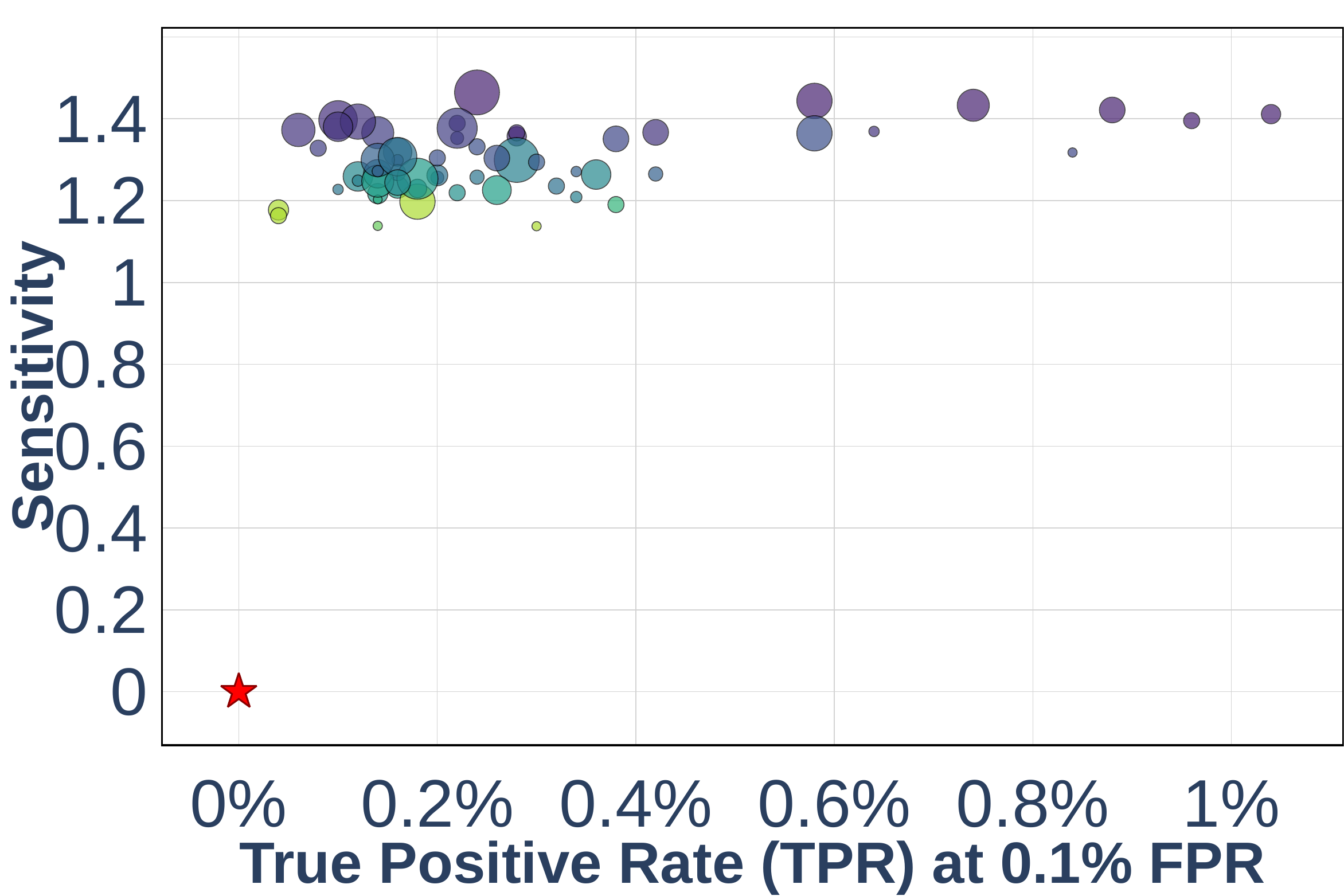}}
\subfloat[]{\includegraphics[width=0.27\textwidth]{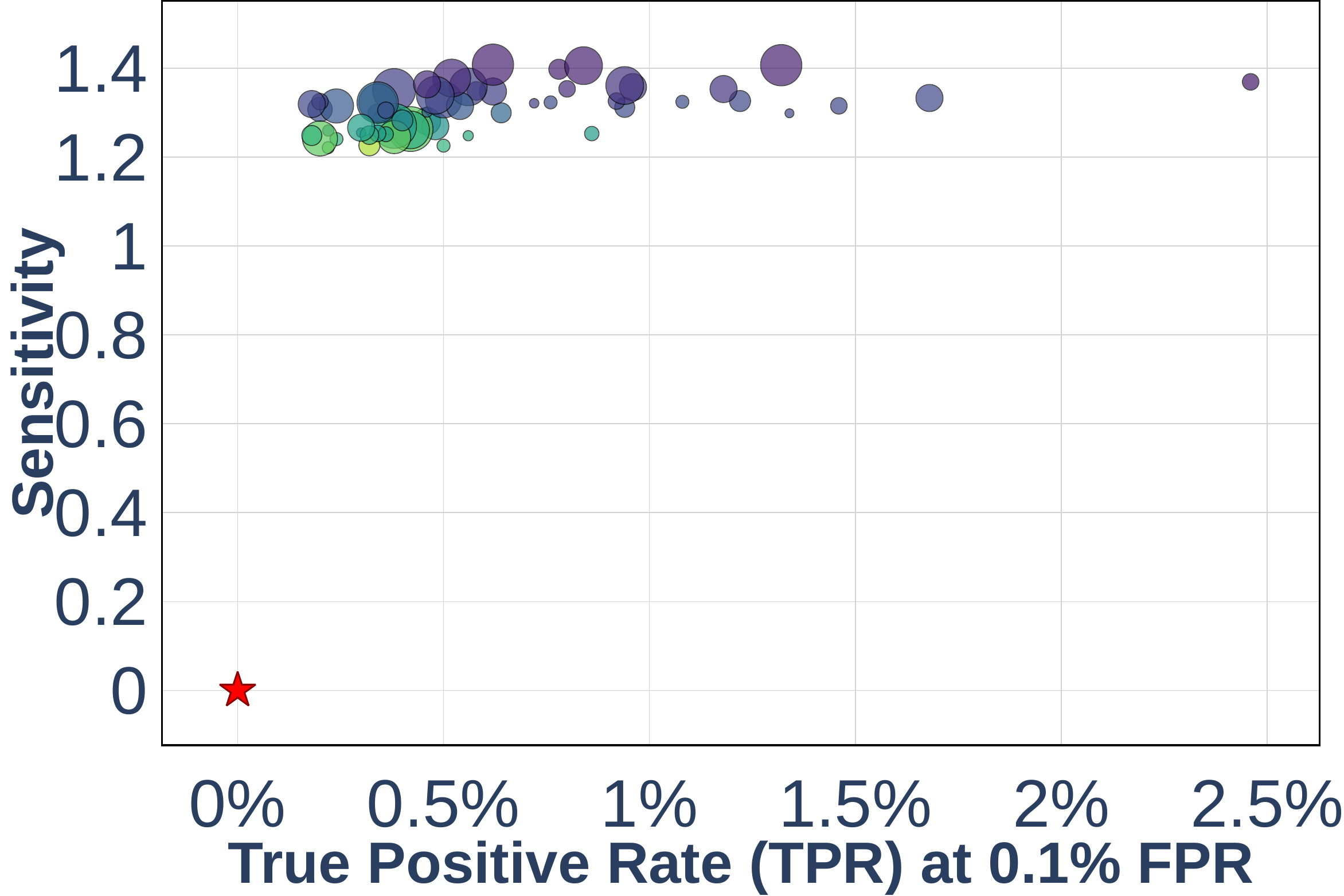}}
\subfloat[]{\includegraphics[width=0.27\textwidth]{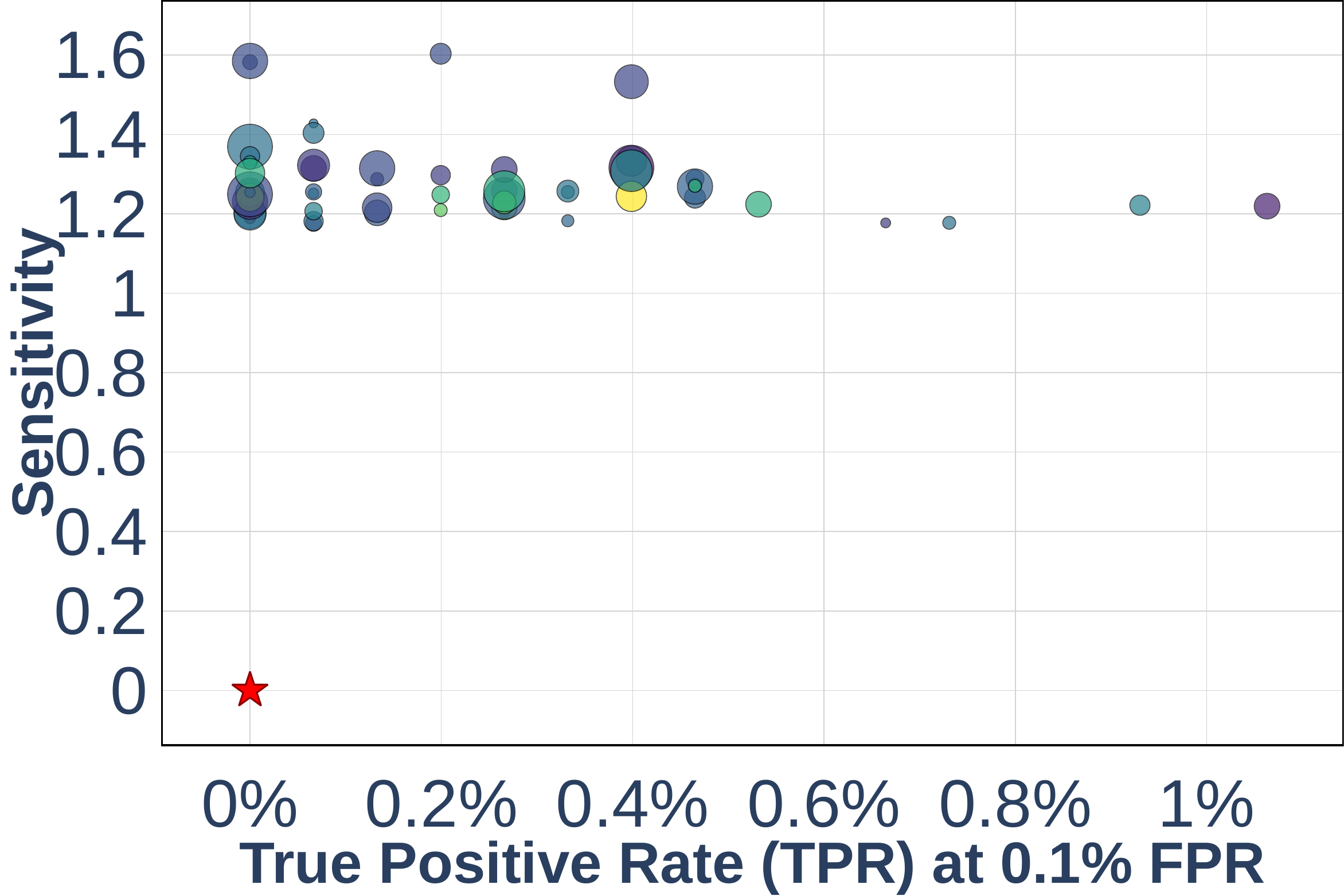}}
\subfloat[]{\includegraphics[width=0.17\textwidth]{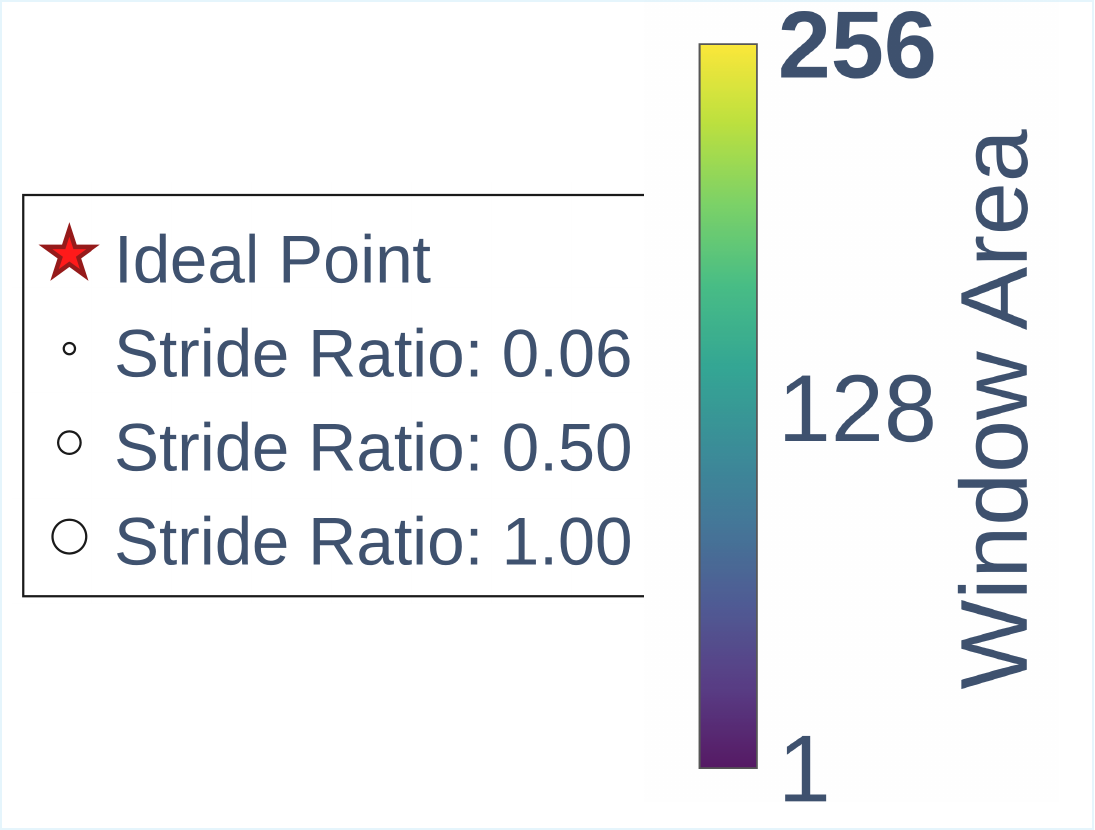}}
     \\
     \vspace{-0.8cm}
     \noindent
\rotatebox{90}{\makebox[3.5cm][c]{\textbf{\scriptsize GradCam++}}}
               \subfloat[]{\includegraphics[width=0.27\textwidth]{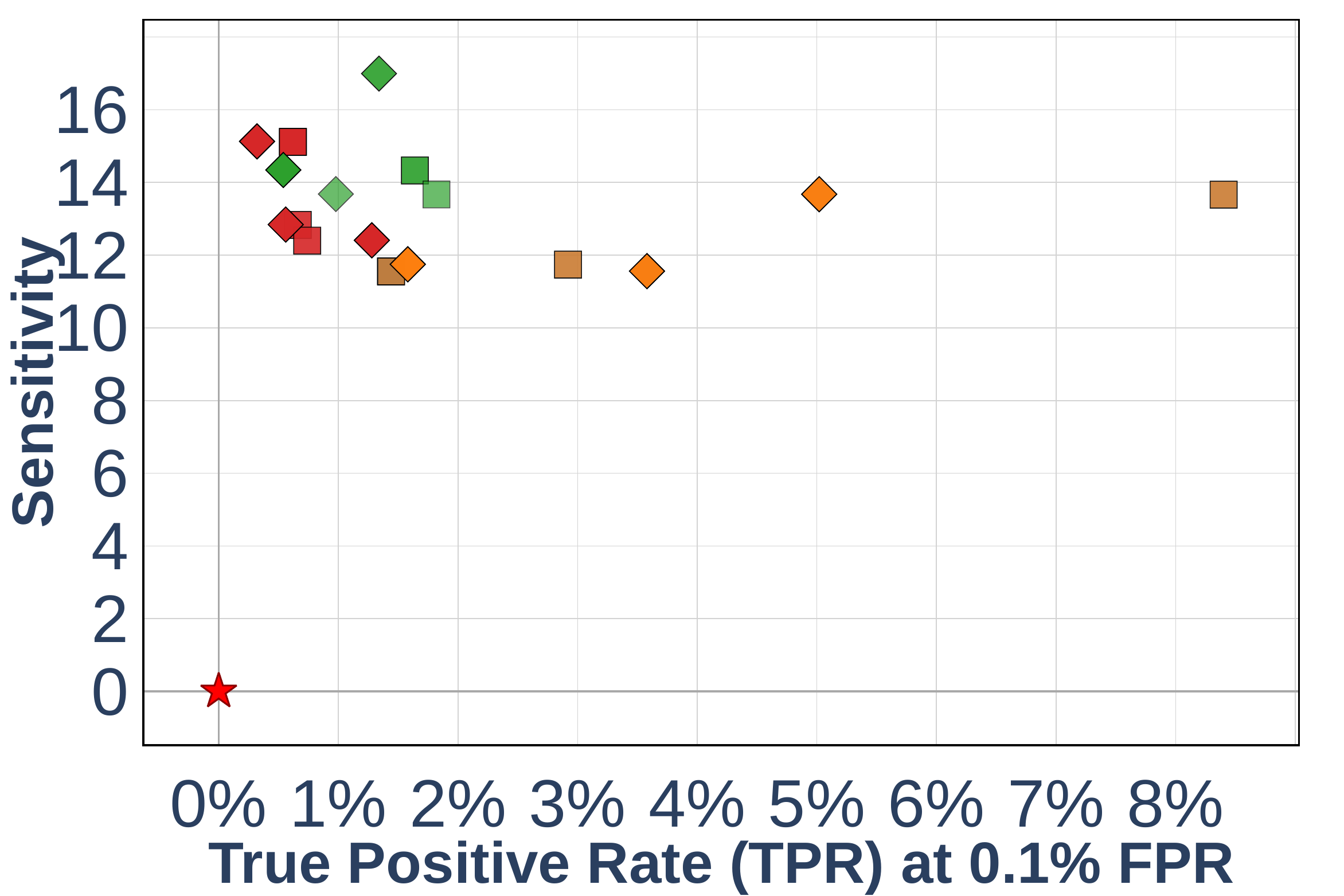}}
\subfloat[]{\includegraphics[width=0.27\textwidth]{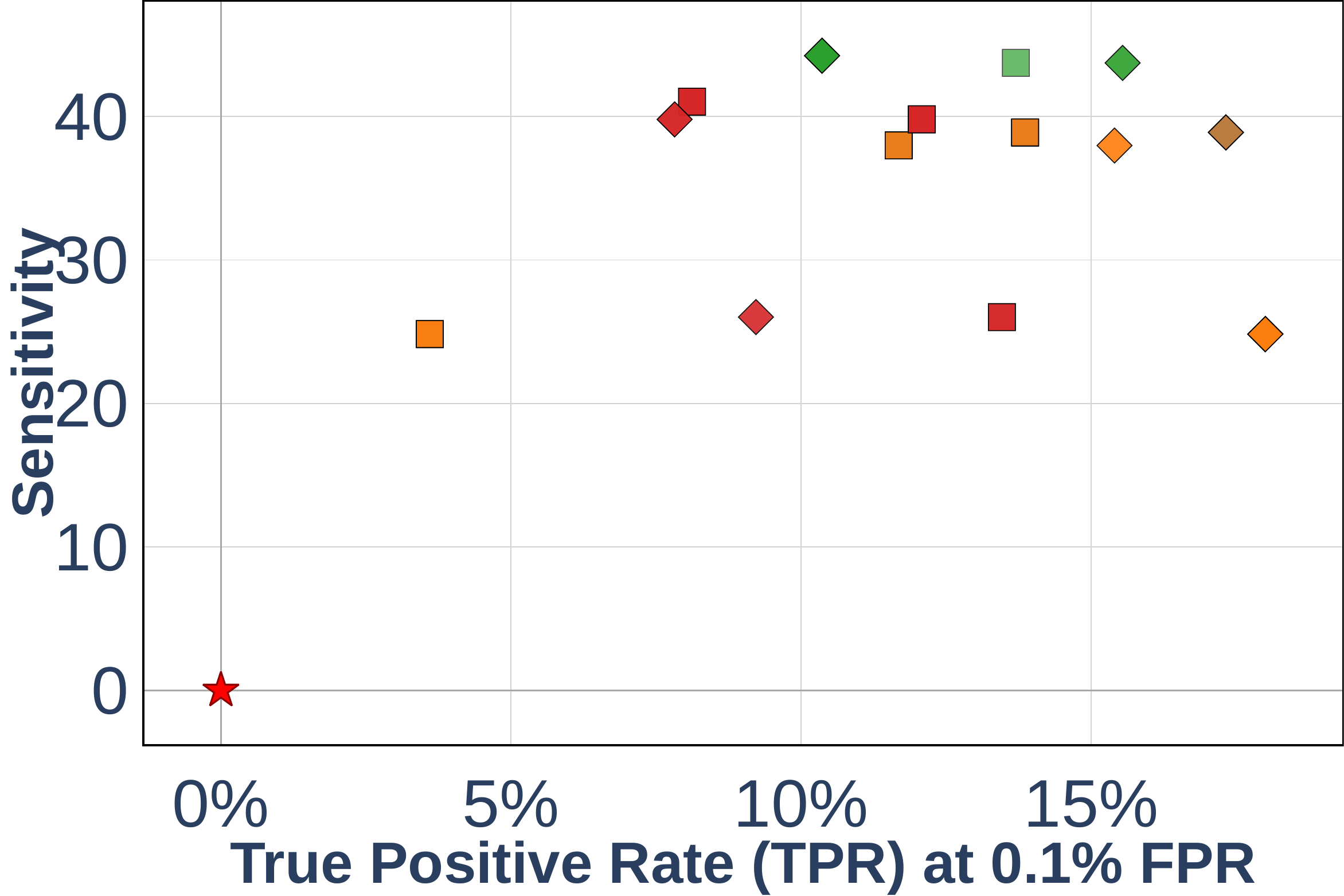}}
\subfloat[]{\includegraphics[width=0.27\textwidth]{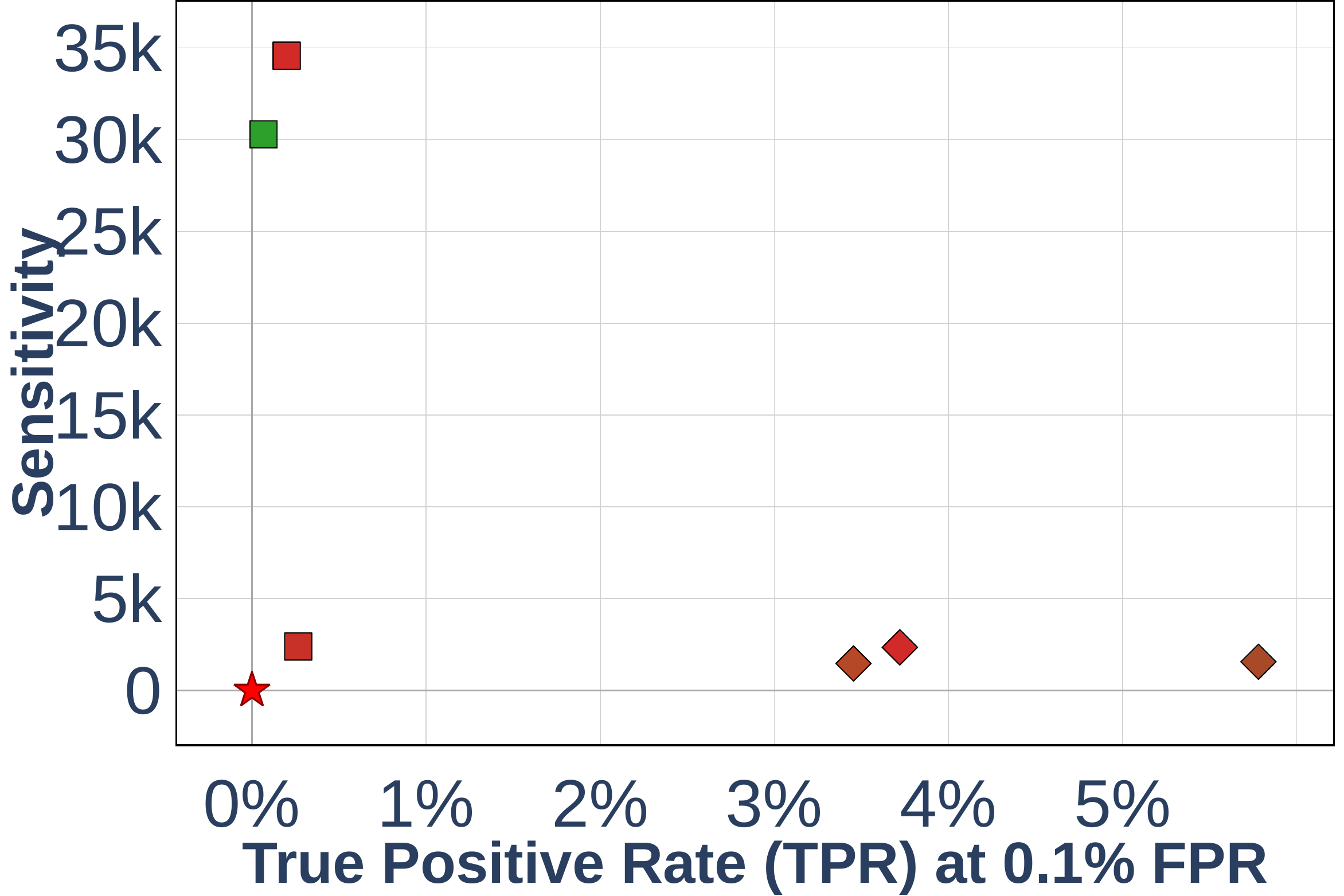}}
\subfloat[]{\includegraphics[width=0.17\textwidth]{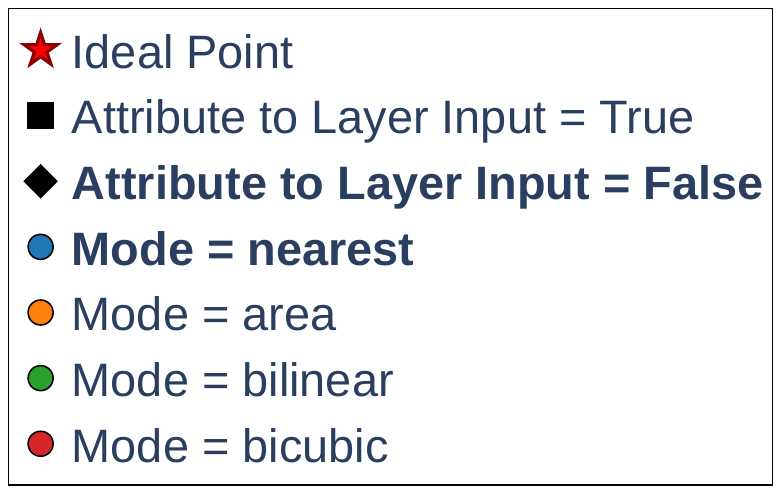}}
     \vspace{-0.8cm}
     \noindent
\rotatebox{90}{\makebox[3.5cm][c]{\textbf{\scriptsize LIME}}}
                    \subfloat[]{\includegraphics[width=0.27\textwidth]{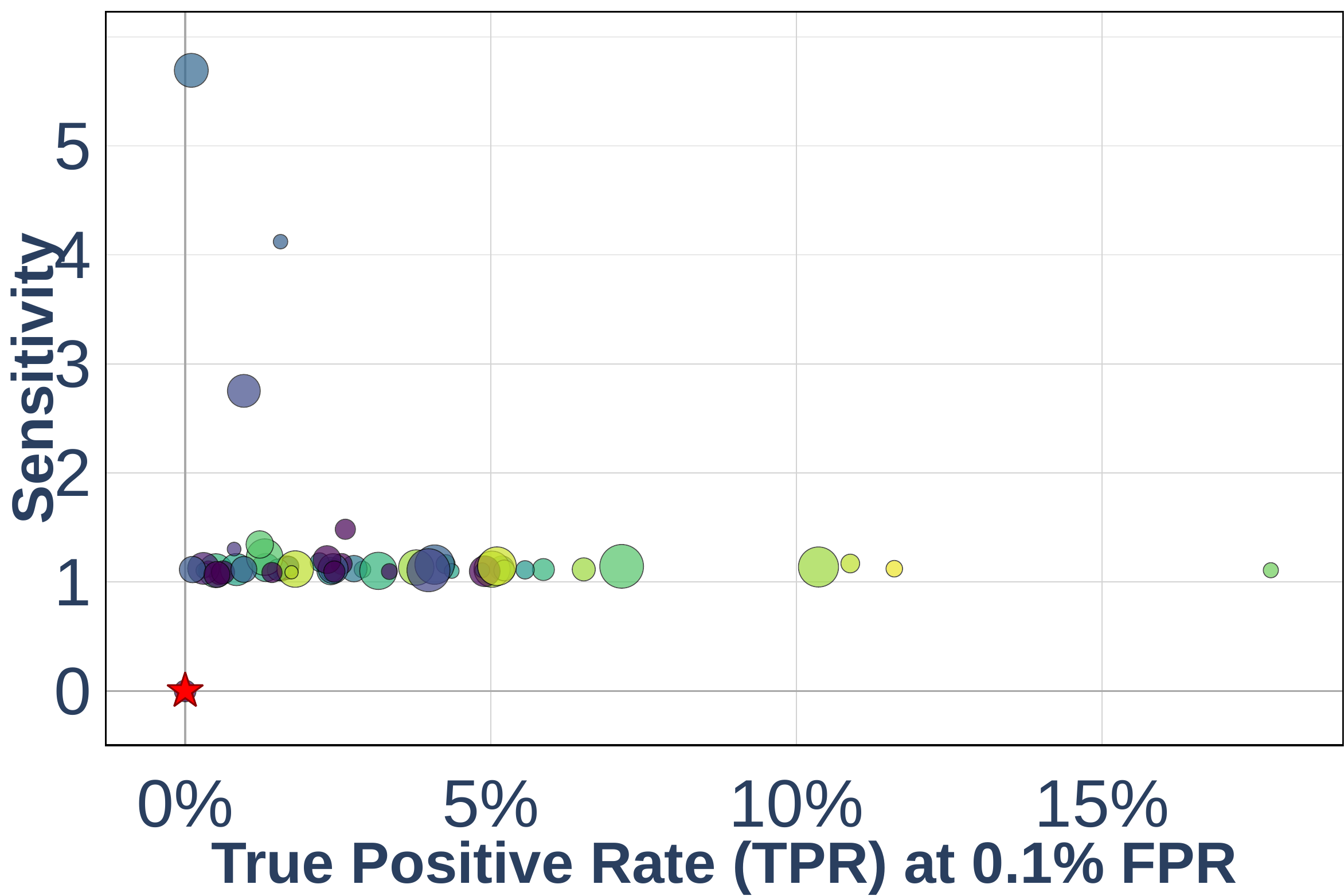}}
\subfloat[]{\includegraphics[width=0.27\textwidth]{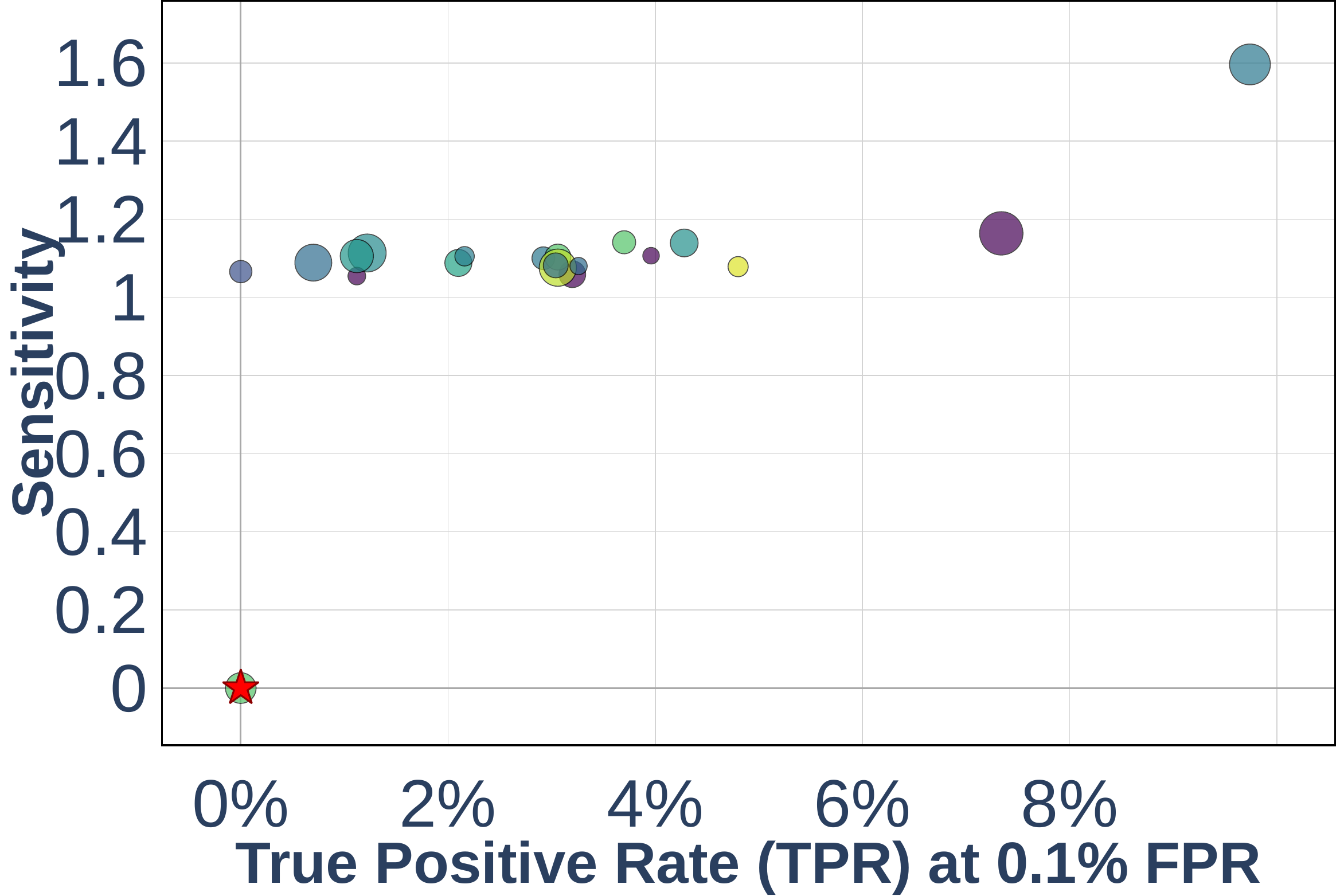}}
\subfloat[]{\includegraphics[width=0.27\textwidth]{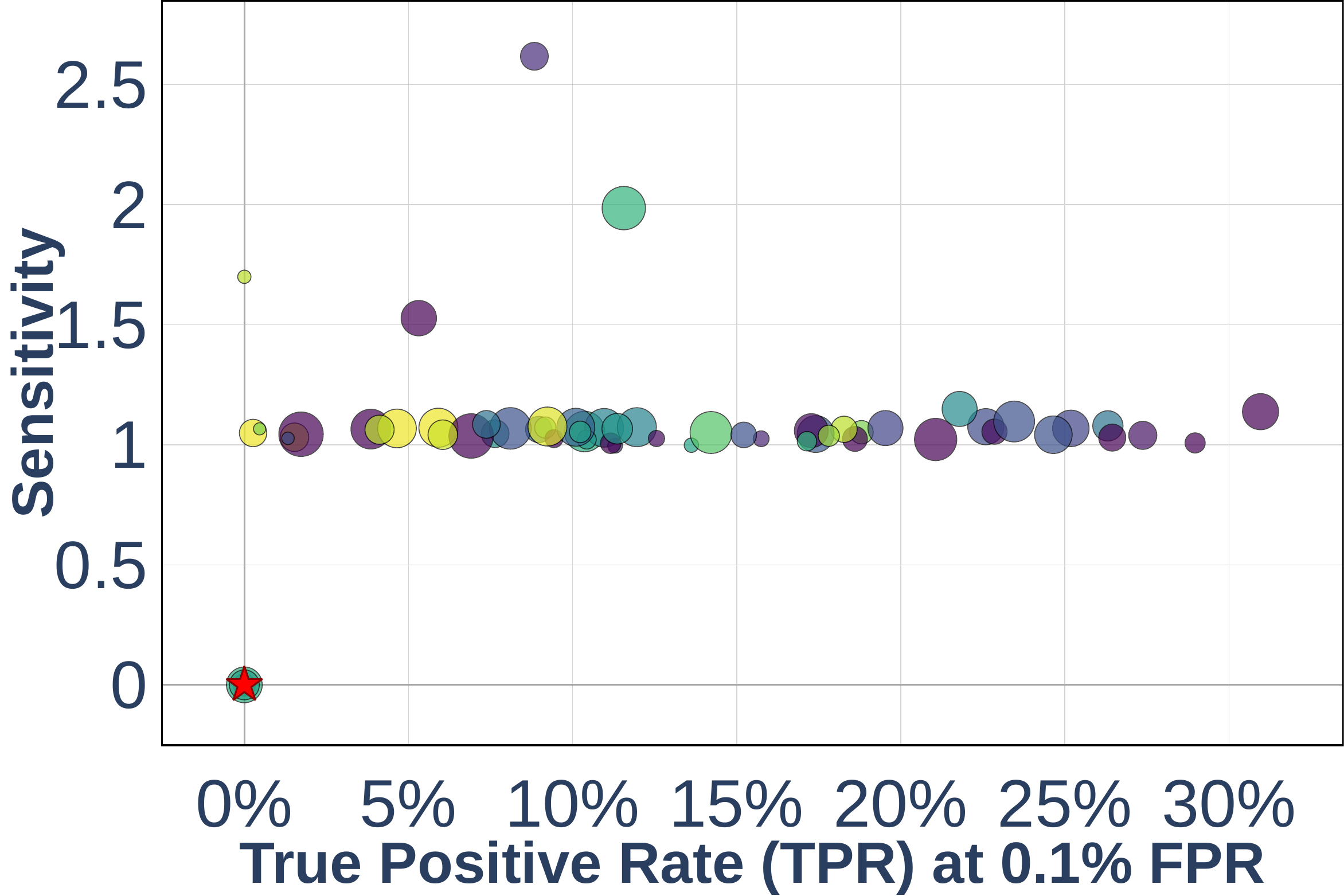}}
\subfloat[]{\includegraphics[width=0.17\textwidth]{figs/LIMElegend.pdf}} \\
\vspace{-0.7cm}
     \noindent
\rotatebox{90}{\makebox[3.5cm][c]{\textbf{\scriptsize ProtoDash}}}
      \subfloat[CIFAR-10]{\includegraphics[width=0.27\textwidth]{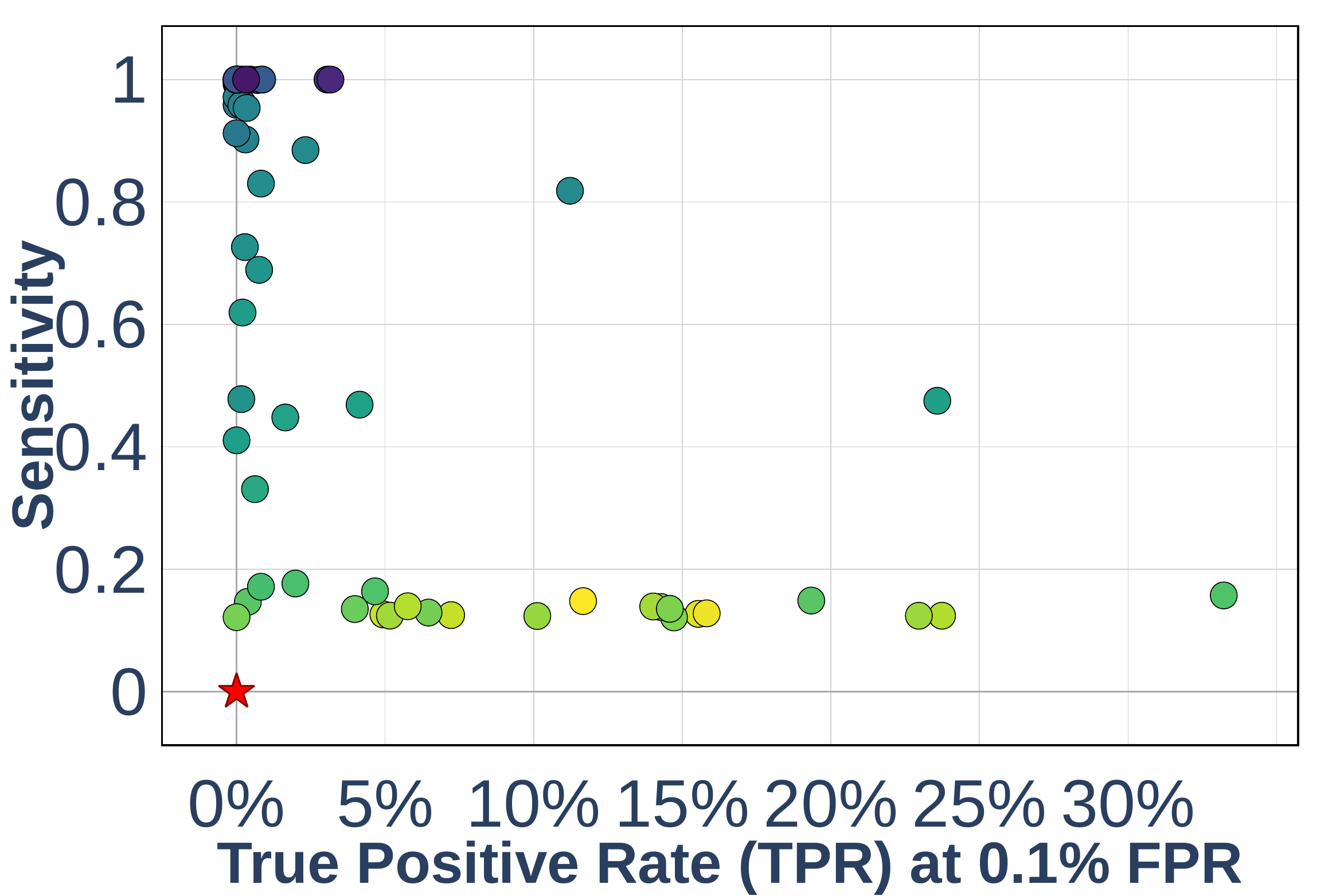}}
\subfloat[CIFAR-100]{\includegraphics[width=0.27\textwidth]{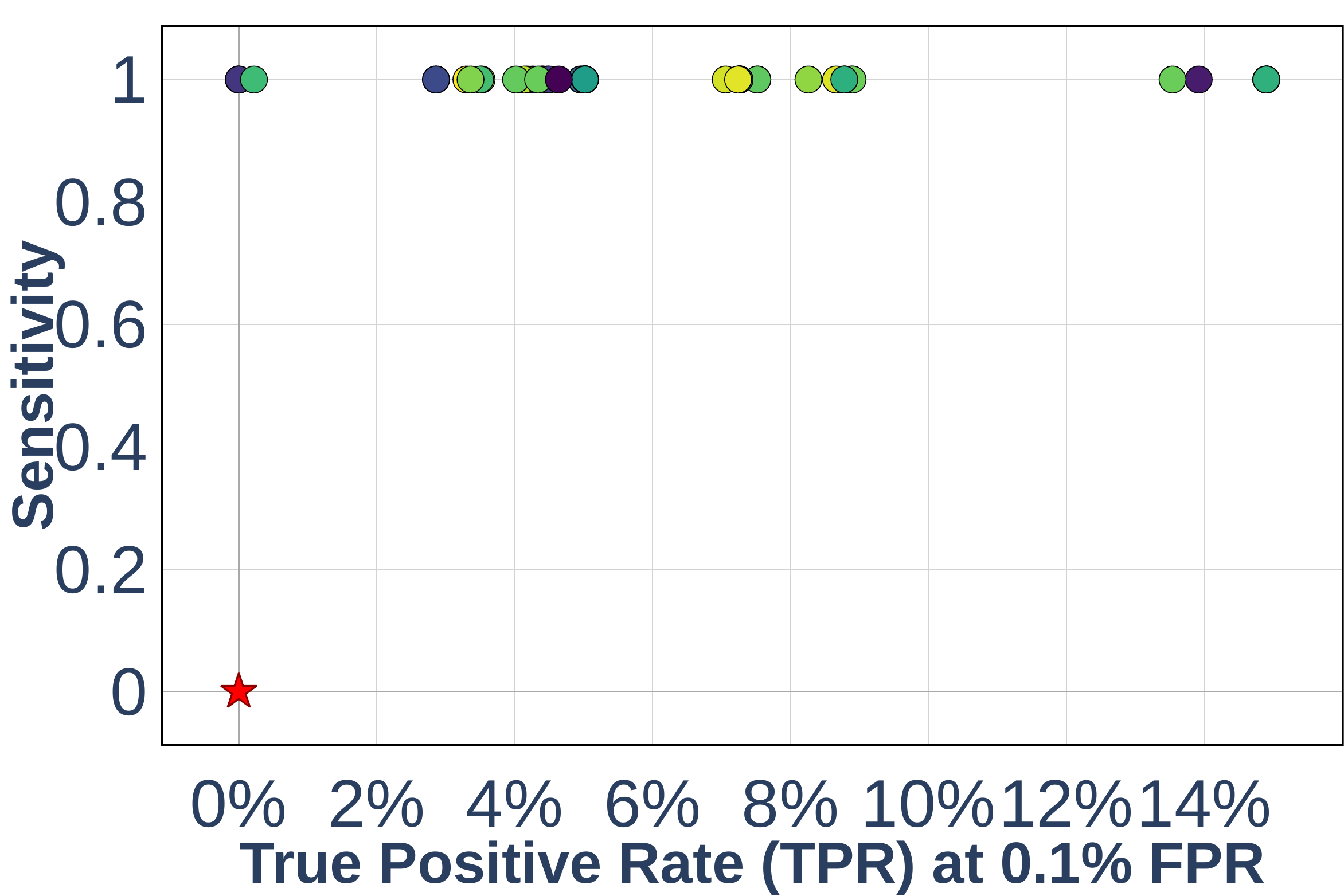}}
\subfloat[GTSRB]{\includegraphics[width=0.27\textwidth]{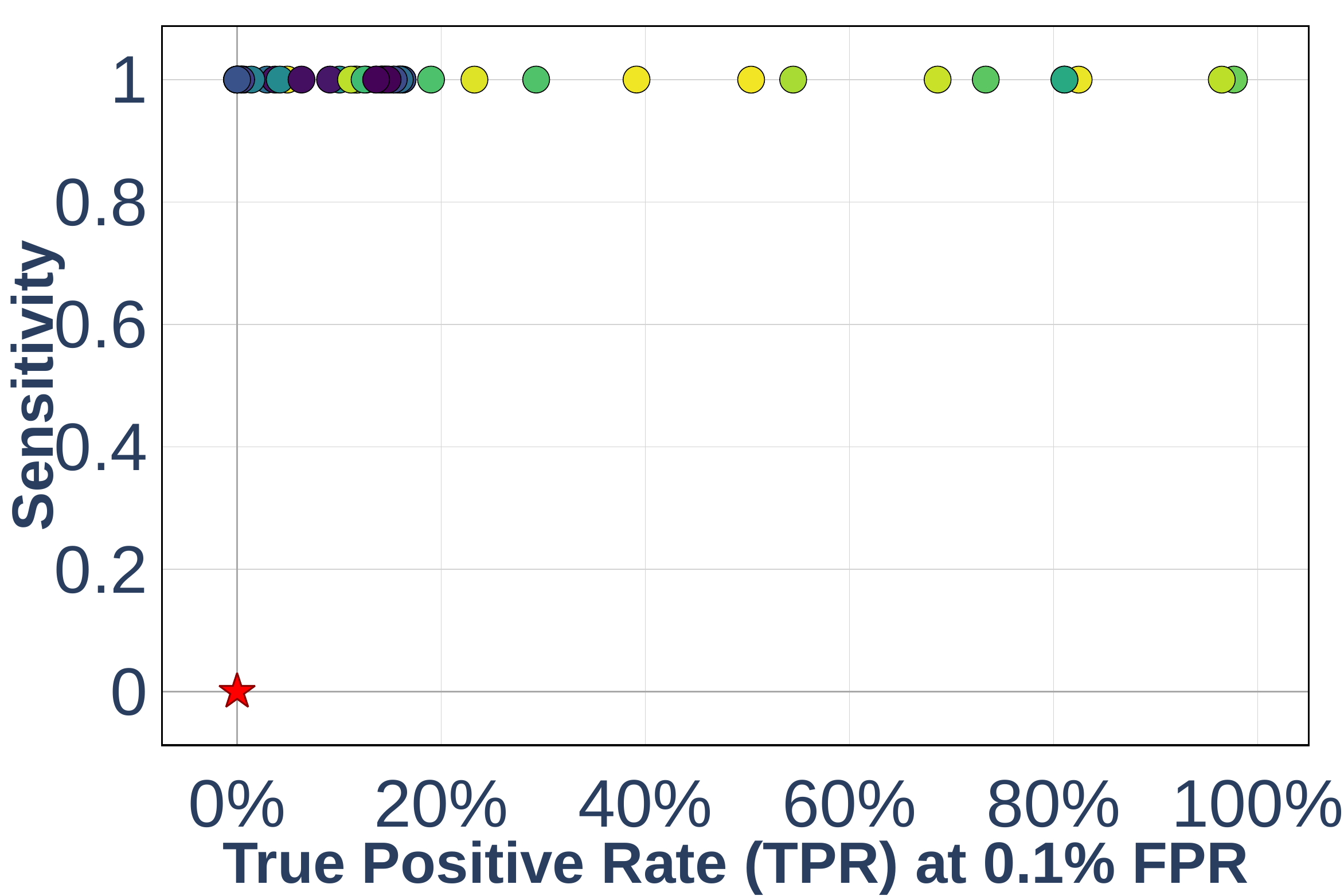}}
\subfloat[Legend]{\begin{minipage}[b]{0.17\textwidth}\centering\includegraphics[width=0.56\textwidth]{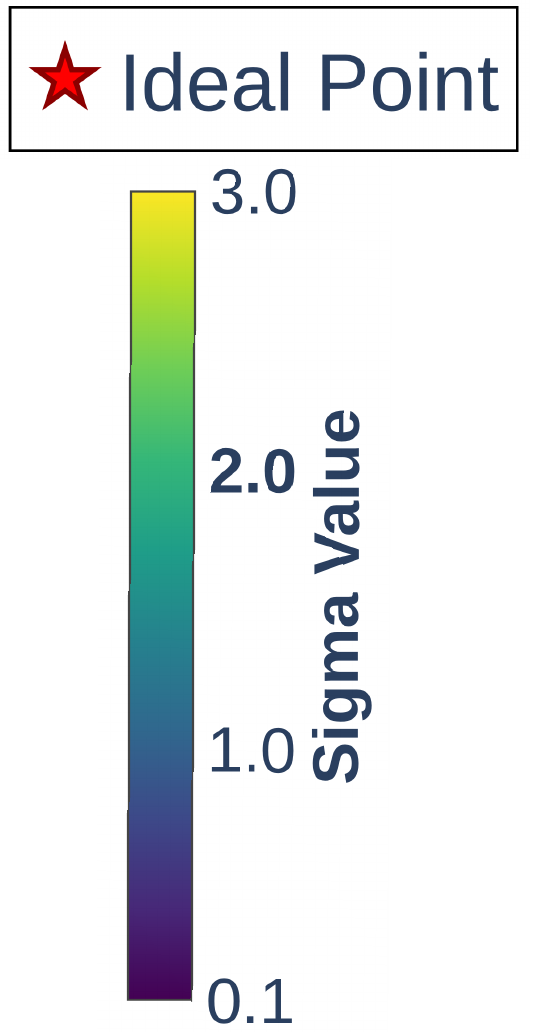}\end{minipage}}

      \caption{\textbf{\sysname{} hardening optimization Pareto fronts for Occlusion, GradCam++, LIME, and ProtoDash.}}
    \label{fig:XAI_pareto}
     \vspace{-1.5em}
\end{figure*}

To address RQ3, we analyze each explanation method family for membership leakage causes and mitigation.

\textbf{Gradient-based explanations.}
Methods like Guided Backpropagation, Integrated Gradients, SmoothGrad, Saliency Map, DeepLIFT, Deconvolution, and InputXGrad leak membership signals due to their direct access to model gradients. In particular, the attribution scores of these methods reflect output changes with respect to input, revealing sensitive patterns.

Hardening these methods (Table~\ref{tab:hardeningres}) focuses on improving the privacy of these methods while simultaneously improving the sensitivity, if possible, by tuning the methods' specific parameters. Starting with Integrated Gradients: increasing \texttt{n\_steps}, selecting a better integral approximation method, and using  \texttt{multiply\_by\_inputs=True} yields more global, less sensitive attributions (Figure~\ref{fig:pareto1}, row 2). More specifically, increasing the number of steps required for the integral approximation method makes the method resilient to leakage while also looking at different approximation methods to fit our goal of decreased sensitivity at little to no trade-off to privacy.  Furthermore, \texttt{multiply\_by\_inputs=True} sets the attributions to be global, as opposed to local, with global looking at the overall model's gradients rather than analyzing individual predictions, giving less risk of privacy and lower sensitivity due to averaging all the model's predictions (Figure \ref{fig:pareto1}, row 2). For this method in particular, we opted to portray the results (Table \ref{tab:hardeningres}) prioritizing sensitivity over membership leakage, setting this parameter to be True for all 3 datasets, despite the improved privacy seen with local attributions in CIFAR-100 (Figure \ref{fig:pareto1}: row 2, col 2).

For Saliency Map, Deconvolution, Guided Backpropagation, InputXGrad, and DeepLIFT, which do not have parameters to tune to reach our goals, we added the following parameters: clamp range (for clipping), mask threshold (for masking), and noise level (for calibrated noise). Interestingly, since we had control over the implementation of these parameters, the order in which they are applied (Algorithm~\ref{alg:transfor}) is representative of the results in Table \ref{tab:eval_alg2_xai_grouped}. Consequently, there is a significant enhancement in privacy but also a significant decrease in sensitivity for these methods (Table \ref{tab:hardeningres}). This decrease in sensitivity is explained by the added noise, which was initially added to enhance privacy.  However, since sensitivity measures robustness against small perturbations, this already added noise accounts for these small perturbations, nullifying them from the perspectives of the explanation methods. 

Lastly, for SmoothGrad: increasing \texttt{stdevs} (standard deviations of Gaussian noise added) and setting to \texttt{True} the parameter \texttt{draw\_baseline\_from\_distrib} (i.e., to randomly draw baseline samples from baseline distribution), shows improvements, but excessive noise through tuning \texttt{stdevs} leads to reduced utility (Figure~\ref{fig:XAI_pareto_appendix}, row 3).

\textbf{Perturbation-based explanations.}
These methods (e.g., SHAP, Anchors, Occlusion) modify inputs to observe output changes and determine feature importance. Despite the black-box setup, their design and inherent logic of finding patterns can still cause membership leakage in the explanations.

Occlusion, a method that perturbs a given rectangular region of an input, has the parameters \texttt{sliding\_window\_shapes} (rectangular region size) and \texttt{strides} (how much to shift the window per iteration). By increasing the window size, the method uses a larger region to occlude, making it harder to reveal patterns, enhancing privacy. Additionally, by having smaller strides, or as described in Figure \ref{fig:XAI_pareto} (row 1), a smaller stride ratio (ratio of stride size to window size), the large window being occluded shifts slightly. This allows for the method to better explain the model as smaller sections get revealed in correspondence with previously revealed sections, while still being private due to the large window size. By choosing the optimal selection of these parameters, Occlusion achieves a significant decrease in membership leakage with minimal fluctuation in sensitivity as shown in Table \ref{tab:hardeningres}.

Moving on to SHAP with two tuneable parameters, \texttt{n\_segments} (how many superpixels to split the input into) and \texttt{compactness} (the shape of the superpixels). The number of segments has a direct correlation to the goals in mind as a reduced number of segments provides considerably more privacy at the cost of utility. Inversely, an increased compactness parameter provides more utility at the cost of privacy due to having more uniform superpixels (Figure \ref{fig:pareto1} row 3). With these understandings, an ideal trade-off with the values of these parameters can represent an improvement in both directions as seen in our results shown in Table \ref{tab:hardeningres}.

Lastly, Anchors' parameters such as \texttt{threshold} (anchor precision threshold), \texttt{tau} (helps find best anchor), and \texttt{beam\_size} (number of anchors per iteration), all relate to the anchor component of this explanation method, influencing both the utility and privacy as seen in Figure~\ref{fig:XAI_pareto_appendix} (row 5). Similar to SHAP, the (\texttt{n\_segments}) parameter in Anchors provides more privacy at the cost of utility when given a smaller value.  Overall, tuning Anchors’ parameters reveals a clear utility–privacy trade-off, where smaller, coarser settings reduce utility but substantially improve privacy, while careful parameter hardening leads to a notable reduction in membership leakage and sensitivity (Table \ref{tab:hardeningres}).

\textbf{Representation-guided explanations.}
GradCAM++ uses internal features and gradients, making it inherently leaky despite high utility.
It offers a tunable interpolation mode (3-D `nearest' to 5-D `bicubic') and a boolean \texttt{attr\_to\_layer\_input} (compute attribution with respect to layer input/output). Higher interpolation complexity increases utility at a potential privacy cost, while setting \texttt{attr\_to\_layer\_input=True} improves privacy with some utility loss (Figure~\ref{fig:XAI_pareto}, row 2). Table~\ref{tab:hardeningres} shows that tuning these parameters substantially enhances privacy with only minor utility reduction.

\textbf{Approximation-based explanations.}
LIME, ProtoDash, and SHAP approximate the model with prototypes, making them relatively private yet vulnerable to leakage through pattern approximation.
For ProtoDash, the parameter \texttt{sigma} (kernel width) directly affects privacy–utility trade-offs, as shown in the CIFAR-10 experiment (Figure~\ref{fig:XAI_pareto}, row 4, col 1). A smaller \texttt{sigma} yields narrower, more selective prototypes that increase sensitivity while improving privacy. Hence, selecting an ideal \texttt{sigma} achieves a balanced trade-off (Table~\ref{tab:hardeningres}).

SHAP and LIME share tuning parameters \texttt{n\_segments} (number of superpixels) and \texttt{compactness} (superpixel shape), which directly impact sensitivity and leakage. More or fewer superpixels and their uniformity control this balance, as illustrated in Figure~\ref{fig:pareto1} (row 3) and Figure~\ref{fig:XAI_pareto} (row 3). With optimal parameter settings, both methods increase privacy with only marginal sensitivity loss.
{\small
\begin{summarybox}
\textbf{Summary:} Membership leakage stems primarily from attribution sparsity, gradient outliers, and high input sensitivity. 
Gradient-based methods are most susceptible, emphasizing the need for leakage-aware explanation method selection.
\end{summarybox}
}
\subsection{Ablation Studies and Discussion}
\label{sec:discussion}

\textbf{Sequence of Hardening Transformations.}
Using Algorithm~\ref{alg:transfor}, we exhaustively explored sequences of three privacy-enhancing transformations for non-parameterized explanation methods: Clipping attribution values (C), Threshold Masking of low signals (M), and Gaussian Noise addition (N) (Table~\ref{tab:eval_alg2_xai_grouped}). We evaluated each sequence by reduction in TPR and preservation of explanation utility. The optimal sequence C$\rightarrow$M$\rightarrow$N achieves 0\% TPR with utility gains of 88.98\% for Saliency Map and 89.49\% for Deconvolution. Other orders yield only marginally higher utility but less privacy. This ordering is intuitive: Clipping and Masking before Noise prevents the mechanisms from canceling each other out.

\begin{table}[t]
\centering
\caption{\textbf{Effectiveness of Algorithm~\ref{alg:transfor} in hardening non-parameterized explanation methods on GTSRB. In the ``Variant” column: C = Clipping, M = Masking, N = Gaussian Noise. TPR and Sensitivity (\%) are reported for SMAP (Saliency Map) and DeConv (Deconvolution)}.}
\label{tab:eval_alg2_xai_grouped}
\scalebox{0.73}{
\begin{tabular}{l|cc|cc|cc}
\toprule
\multirow{2}{*}{\textbf{Variant}} & \multicolumn{2}{c|}{\textbf{Pre-Hardening TPR}} & \multicolumn{2}{c|}{\textbf{Post-Hardening TPR}} & \multicolumn{2}{c}{\textbf{Sensitivity Change (\%)}} \\
& SMAP & DeConv & SMAP & DeConv & SMAP & DeConv \\
\midrule
C & 5.48 & 6.64 & 1.72 & 0.53 & \textcolor{blue}{$\downarrow$ 9.72\%} & \textcolor{blue}{$\downarrow$ 4.54\%} \\
M & 5.48 & 6.64 & 0.0 & 0.06 & \textcolor{red}{$\uparrow$ 113.93\%} & \textcolor{red}{$\uparrow$ 65.07\%} \\
N & 5.48 & 6.64 & 0.13 & 0.0 & \textcolor{blue}{$\downarrow$ 88.15\%} & \textcolor{blue}{$\downarrow$ 84.54\%} \\
C$\rightarrow$M & 5.48 & 6.64 & 0.13 & 0.06 & \textcolor{red}{$\uparrow$ 15.38\%} & \textcolor{red}{$\uparrow$ 61.20\%} \\
C$\rightarrow$N & 5.48 & 6.64 & 0.0 & 0.0 & \textcolor{blue}{$\downarrow$ 88.39\%} & \textcolor{blue}{$\downarrow$ 83.18\%} \\
M$\rightarrow$C & 5.48 & 6.64 & 0.33 & 0.13 & \textcolor{red}{$\uparrow$ 48.49\%} & \textcolor{red}{$\uparrow$ 2.12\%} \\
M$\rightarrow$N & 5.48 & 6.64 & 0.0 & 0.26 & \textcolor{blue}{$\downarrow$ 88.09\%} & \textcolor{blue}{$\downarrow$ 88.27\%} \\
N$\rightarrow$C & 5.48 & 6.64 & 0.06 & 0.0 & \textcolor{blue}{$\downarrow$ 87.09\%} & \textcolor{blue}{$\downarrow$ 88.21\%} \\
N$\rightarrow$M & 5.48 & 6.64 & 0.0 & 0.06 & \textcolor{blue}{$\downarrow$ 88.33\%} & \textcolor{blue}{$\downarrow$ 87.32\%} \\
\rowcolor{cyan!10}
C$\rightarrow$M$\rightarrow$N & 5.48 & 6.64 & 0.0 & 0.13 & \textcolor{blue}{$\downarrow$ 88.98\%} & \textcolor{blue}{$\downarrow$ 89.49\%} \\
C$\rightarrow$N$\rightarrow$M & 5.48 & 6.64 & 0.0 & 0.0 & \textcolor{blue}{$\downarrow$ 88.62\%} & \textcolor{blue}{$\downarrow$ 86.79\%} \\
M$\rightarrow$C$\rightarrow$N & 5.48 & 6.64 & 0.0 & 0.0 & \textcolor{blue}{$\downarrow$ 69.99\%} & \textcolor{blue}{$\downarrow$ 87.43\%} \\
M$\rightarrow$N$\rightarrow$C & 5.48 & 6.64 & 0.13 & 0.0 & \textcolor{blue}{$\downarrow$ 88.55\%} & \textcolor{blue}{$\downarrow$ 87.96\%} \\
N$\rightarrow$C$\rightarrow$M & 5.48 & 6.64 & 0.0 & 0.19 & \textcolor{blue}{$\downarrow$ 92.47\%} & \textcolor{blue}{$\downarrow$ 88.29\%} \\
N$\rightarrow$M$\rightarrow$C & 5.48 & 6.64 & 0.0 & 0.0 & \textcolor{blue}{$\downarrow$ 32.22\%} & \textcolor{blue}{$\downarrow$ 90.39\%} \\
\bottomrule
\end{tabular}}
 % \vspace{-1em}
\end{table}

\textbf{Disjoint Datasets.}
Under the common threat model, the attacker has an auxiliary dataset that may partly overlap with the target model’s training set. In practice, the adversary may collect data that is disjoint with the training data. We show that even in this case our attack outperforms state-of-the-art explanation-aware MIAs, with only minor performance impact. For this setup, we used disjoint CIFAR-10 subsets: 40K samples for $D^{\text{train}}_{\text{target}}$ and 10K for $D^{\text{train}}_{\text{shadow}}$, ensuring $D^{\text{train}}_{\text{target}} \cap D^{\text{train}}_{\text{shadow}} = \emptyset$ (Table~\ref{tab:data_splits}). Across 7 explanation methods, our attack consistently surpassed prior work~\cite{PleaseTellmeMore}, with a 0.5\% TPR gain on SHAP, IntegratedGradients, and VarGrad, and a 2\% TPR drop on the remaining methods.
\begin{table}[t!]
\centering
\caption{\textbf{Evaluation of our attack on disjoint datasets for shadow and target model training data on CIFAR-10.}}
\begin{tabular}{lcc}
\toprule
\textbf{Explanation Method
TPR@0.1\%FPR} &  \textbf{Subset} & \textbf{Disjoint Subset}  \\
\midrule
SmoothGrad   & 10.38 \% & 5.72\%  \\ 
VarGrad      & 6.78 \%  & 8.95\%  \\ 
IntegratedGradients          & 4.88 \% & 5.2\%  \\ 
GradCAM      & 0.95 \%  &0.93\%  \\ 
GradCAM++    & 6.7 \%  &5.46\%  \\ 
SHAP         & 5.04 \% &5.2\%  \\ 
LIME         &  10.26\%  &4.6\%  \\ 
\bottomrule\\
\end{tabular}
\label{tab:disdataset}
% \vspace{-1.5em}
\end{table}

\textbf{Different Model Architectures.}
We relax the assumption that the adversary knows the target architecture. On CIFAR-10, we vary the shadow model (ResNet18, MobileVNet2, DenseNet161) while fixing the target as MobileVNet2 (Table~\ref{tab:diffmodel}). IntegratedGradients, GradCAM++, Guided BackProp, and Deconvolution incur only a 1.5\% average TPR drop and still outperform same-architecture baselines. Using ResNet18 improves Saliency Map and IntegratedGradients by 2\%, reflecting architectural similarity. In contrast, the dissimilar DenseNet161 causes complete failure for Deconvolution, Guided BackProp, and GradCAM++, while IntegratedGradients, Saliency Map, and Input$\times$Grad improve by 1.7\% on average.

\begin{table}[h!]
\caption{\textbf{Attack evaluation with different shadow models on CIFAR-10 MobileNetV2.}}
\label{tab:diffmodel}
\centering
\scalebox{0.8}{
\begin{tabular}{lccc}
\toprule
\textbf{Explanation Method} & \multicolumn{3}{c}{\textbf{TPR@0.1\%FPR}} \\ 
\midrule
\textbf{Shadow Model Architecture} & MobileNetV2 (BaseLine) & DenseNet161 & ResNet18 \\ 
\midrule
Saliency Map              & 5.24\%   & 7.08\%   & 6.42\%  \\ 
Guided BackProp   & 10.94\%  & 0\%      & 10.8\%  \\ 
IntegratedGradients                & 4.88\%   & 6.76\%   & 3.98\%  \\ 
GradCAM++         & 6.7\%    & 0\%      & 5.40\%  \\ 
InputXGrad        & 2.4\%    & 3.86\%   & 4.3\%   \\ 
Deconvolution     & 16.08\%  & 0\%      & 13.46\% \\ 
\bottomrule
\end{tabular}}
\end{table}

%% file: conclusions.tex
\section{Conclusion}
This paper introduced \sysname{}, an end-to-end toolkit for auditing membership leakage in explanation methods and developing lightweight, model-agnostic defenses. By designing a stronger explanation-aware membership inference attack, we quantified leakage across 15 popular techniques, showing that default settings can expose far more information than previously reported. Our defenses, attribution clipping, sensitivity-aware noise injection, and masking, reduced leakage by up to 95\% with negligible utility loss. \sysname{} is the first to balance the privacy-utility trade-off in model explanations, offering practical guidance for privacy-enhancing deployment of explainable ML.

\section{Acknowledgments}
We thank the anonymous reviewers for their insightful feedback
that improved the paper. This work was supported by the National Science Foundation (NSF) award CNS-2238084.

%% file: appendix.tex
\appendix
\label{sec:appendix}

\subsection{Ablation Study on Generalization GAP}
Following~\cite{PleaseTellmeMore}, we used CIFAR-100, where the target model initially had the lowest test accuracy (45\%), and improved its training to 80\% to study the effect of the generalization gap. We then attacked this model across multiple explanation methods and observed only a 0.2\% average drop in attack performance (Table~\ref{tab:genGAP}), showing that leakage stems mainly from explanation methods rather than model generalization. This highlights the importance of preserving model utility while carefully hardening explanation methods to mitigate leakage.
\begin{table}[h!]
\centering
\caption{\textbf{Impact of generalization GAP on \sysname{} attack results on CIFAR-100 measured via sensitivity change $\Delta \mathbb{S}$. }}
\scalebox{0.9}{
\begin{tabular}{lccc}
\toprule
\textbf{Explanation Method
TPR@0.1\%FPR} & High GAP &  Low GAP & $\Delta \mathbb{S}$  \\

\midrule
SmoothGrad   & 6.81 \%  &6.59\%   &0.04\% \\ 
VarGrad      &  6.68 \%  &6.23\%   &0.1\%  \\ 
IntegratedGradients         & 16.42\%  &15.23\%  &0.07\%  \\ 
GradCAM      &  6.78 \%  &5.91\%   &0.02\%\\ 
GradCAM++    & 18.0 \% &6.12\%   &0.08\%\\ 
SHAP         & 11.8 \%  &12.02\%   &0.23\%\\ 
LIME         &  10.5 \% &9.85\%   &0.19\%\\ 
\bottomrule\\
\end{tabular}}
\label{tab:genGAP}
\vspace{-2em}
\end{table}

\subsection{Runtime Overhead of Hardening Strategies}
\begin{figure}[t!]
    \centering   
    \includegraphics[scale =0.27]{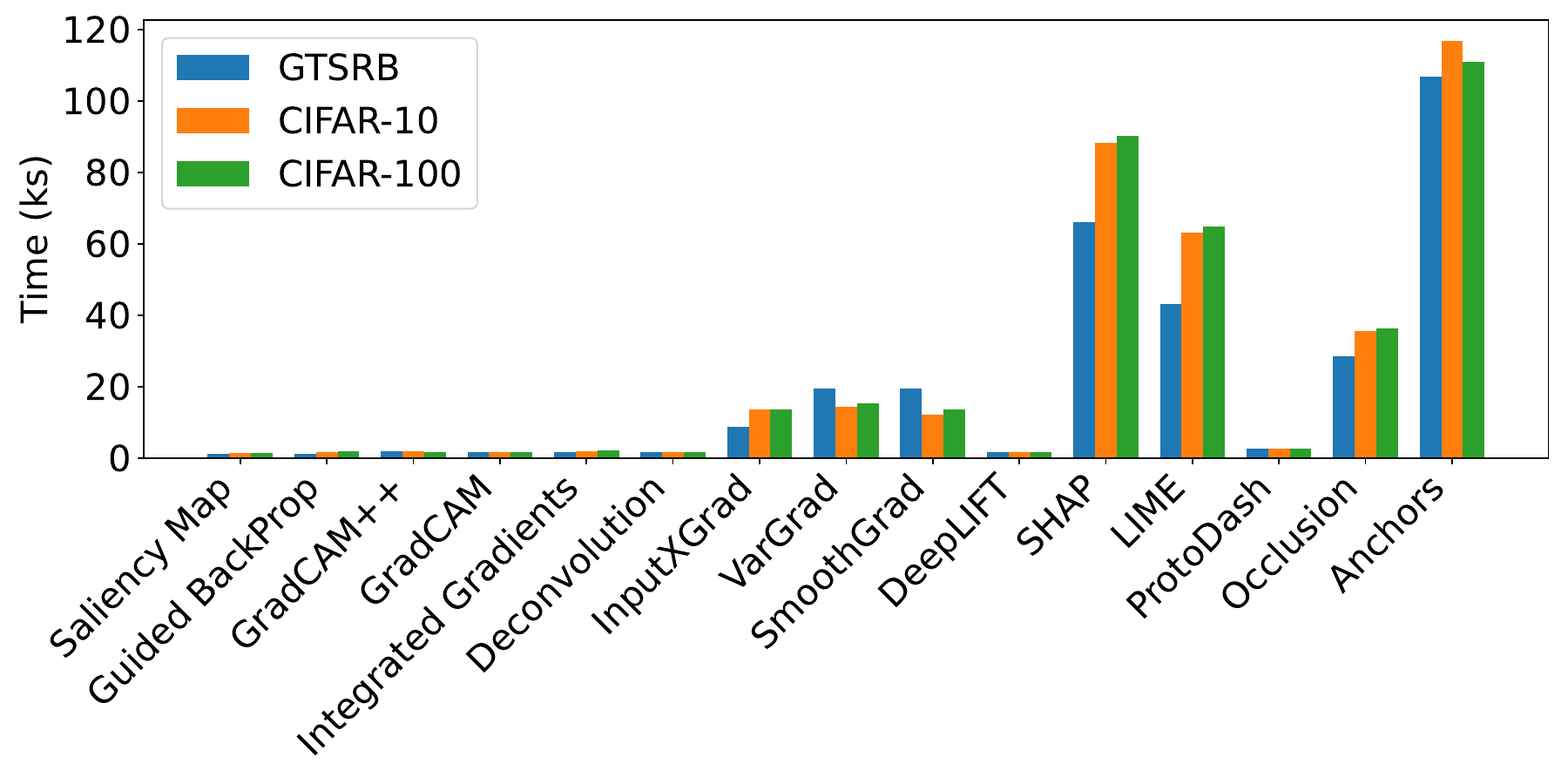}
     
    \caption{\textbf{Runtime overhead ($ks$) of hardening strategies across explanation methods and datasets.}}
    
    \label{fig:time1}
\end{figure}

\begin{figure}[h!]
    \centering   
    \includegraphics[scale =0.35]{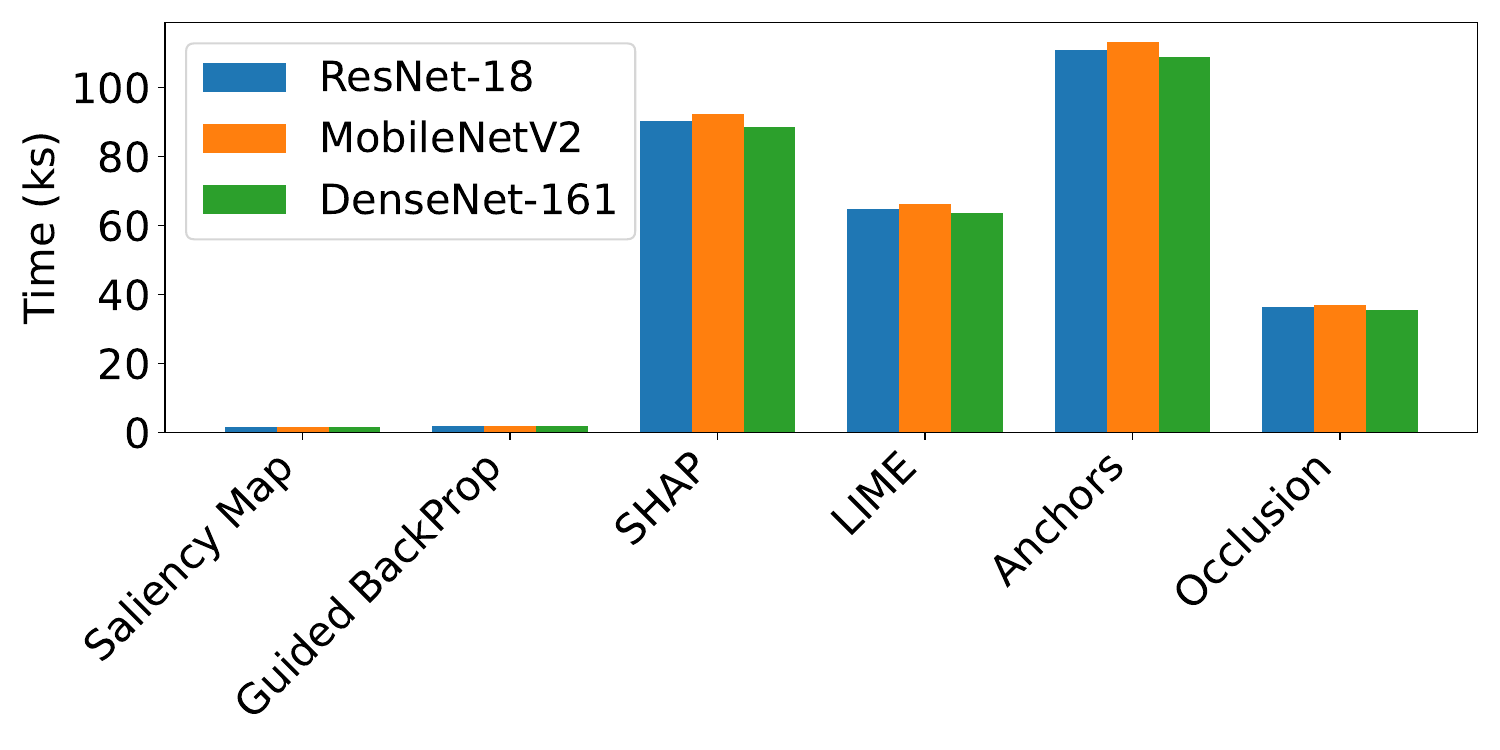}
    \caption{\textbf{Runtime overhead ($ks$) of hardening strategies across representative explanation methods and different model architectures.}}
    
    \label{fig:time2}
\end{figure}

We measure the runtime overhead of \sysname{} on a hardware setup of ubuntu 22.04 with AMD Ryzen ThreatReaper CPU (24 cores), 128GB RAM, and Double RTX 4090 GPUs with 24GB VRAM each. Figures \ref{fig:time1}
 and \ref{fig:time2} show runtime overhead across datasets and across model architectures, respectively. We measured it through 20 runs per explanation method, where overhead ranges from 1200 seconds to 100, 000 seconds. To get a sense of the average runtime per a single run, we divide by 20. "Time (ks)" values reported in Figure \ref{fig:time1} and Figure \ref{fig:time2}.
 Figure \ref{fig:time1} shows runtime overhead (in $ks$) of our hardening approaches (Algorithm \ref{alg:privacy_explanation} for parameterized methods and Algorithm \ref{alg:transfor} for non-parameterized methods) run 20 times for each explanation method across the three datasets. Gradient-based attribution techniques (e.g., Saliency Maps, Guided Backprop, GradCAM variants, Integrated Gradients) consistently exhibit near-negligible runtime, reflecting their efficiency and compatibility with modern deep networks. In contrast, perturbation-based and sampling-based explanation methods, including SmoothGrad, SHAP, LIME, and ProtoDash, incur substantially higher costs. Anchors is the most expensive method overall, reaching above 100 ks in the most complex setting. The figure underscores the significant scalability gap between gradient-based and perturbation-based explanation methods, highlighting computational overhead as well as the invariance of the explanation method hardening runtime for different datasets.

Figure \ref{fig:time2} shows runtime overhead (in $ks$) of our hardening approaches (Algorithm \ref{alg:privacy_explanation} for parameterized methods and Algorithm \ref{alg:transfor} for non-parameterized methods) run 20 times for each representative explanation method across the three model architectures for the CIFAR-100 dataset. Similar to the dataset-level comparison, gradient-based methods (Saliency Maps, Guided Backprop) maintain extremely low cost across all architectures, emphasizing their architectural robustness and efficiency. Across all explanation methods, we see invariance of computational time between different model architectures.

\subsection{Details of Parameterized Explanation Methods}
 Table \ref{tab:param} shows details of constructor and attribution parameters for explanation methods that have tunable parameters. 
 
\begin{table*}[t!]
\centering
\caption{\textbf{List of constructor parameters (used for initializing the method) and attribution parameters (used for generating attribution scores) for parameterized explanation methods in Captum v0.7.0, AIX360 v0.3.0, and Alibi v0.9.6}. (-) indicates no constructor parameter besides the model or no attribution parameter besides the target and inputs. A \textbf{bolded} parameter value indicates the default value, and any numerical parameter is either a float or integer indicated as so with a range and, if applicable, a default value.}

\scalebox{0.9}{

\begin{tabular}{lp{4cm}p{10cm}}
\toprule
\textbf{Explanation Method} & \textbf{Constructor Parameters} & \textbf{Attribution Parameters} \\
\midrule
Integrated Gradients & multiply by inputs[\textbf{T}/F] &  method[(\textbf{gausslegendre}, 
riemann\_right, 
riemann\_left, 
riemann\_middle, 
riemann\_trapezoid] \\
\midrule
SmoothGrad & - & stdevs[float $>$ 0, \textbf{1.0}], draw\_baseline\_from\_distrib[T/\textbf{F}] \\
\midrule

VarGrad & - & stdevs[float $>$ 0, \textbf{1.0}], draw\_baseline\_from\_distrib[T/\textbf{F}] \\
\midrule
GradCAM++ & - & interpolation\_mode[\textbf{nearest},area,linear,bi-linear,bicubic,trilinear], 
attr\_to\_layer[T/\textbf{F}] \\
\midrule
GradCAM & - & interpolation\_mode[\textbf{nearest},area,linear,bi-linear,bicubic,trilinear], 
attr\_to\_layer[T/\textbf{F}] \\
\midrule
Occlusion & - & sliding\_window\_shapes[(3,1,1)-(3,32,32)], strides[(3,1,1)- (3,32,32)] \\
\midrule
SHAP & - & 
n\_segments[int $>$ 0, \textbf{50}], compactness [int $>$ 0] \\
\midrule
Anchors & - & threshold[float 0-1, \textbf{0.95}], 
tau[float 0-1, \textbf{0.15}], 
delta[float 0-1, \textbf{0.1}], 
batch\_size[int $>$ 0, \textbf{100}], 
coverage\_samples[int $>$ 0, \textbf{10000}], 
beam\_size[int $>$ 0, \textbf{1}], 
segmentation\_fn[\textbf{slic}, felzenszwalb,quickshift], 
segmentation\_kwargs[n\_segments[ int $>$ 0, \textbf{15}], 
compactness[int $>$ 0, \textbf{20}], 
sigma(\textbf{0.5})], 
p\_sample[float 0-1, \textbf{0.5}] \\
\midrule
Lime & - & 
n\_segments[int $>$ 0, \textbf{50}], compactness [int $>$ 0] \\
\midrule
ProtoDash & - &  
sigma[float $>$ 0, \textbf{2.0}], 
kernel[\textbf{'other'},'Gaussian'] \\
\bottomrule
\end{tabular}
}
\label{tab:param}
\end{table*}

\subsection{Evaluation Setup}
Table \ref{tab:data_splits} shows how we split each dataset into training and testing data for the target and shadow model. Table \ref{tab:model_acc} shows train and test accuracies of the three models across the three datasets.
\begin{table}[H]
\centering
\caption{\textbf{Dataset splits on different datasets.}}
\label{tab:data_splits}
\begin{tabular}{lcccc}
\toprule
\textbf{Dataset} & $D_{\text{train}}^{\text{target}}$ & $D_{\text{test}}^{\text{target}}$ & $D_{\text{train}}^{\text{shadow}}$ & $D_{\text{test}}^{\text{shadow}}$ \\
\midrule
CIFAR-10  & 50000 & 10000 & 10000 & 10000 \\
CIFAR-100 & 50000 & 10000 & 10000 & 10000\\ 
GTSRB     & 7500  & 1500  & 1500  & 1500\\  
\bottomrule
\end{tabular}
\end{table}
Table~\ref{tab:model_acc} reports the target model’s performance on training and test sets
\begin{table}[H]
\centering
\caption{\textbf{Training and testing accuracy for the target model.}}

\label{tab:model_acc}
\begin{tabular}{lccccc}
\toprule
\textbf{Dataset} & \textbf{Model} &\textbf{ Train Acc} & \textbf{ Test Acc}  \\
\midrule
CIFAR-10  & MobileVNET2 & 0.998 & 0.771\\
CIFAR-100 & MobileVNET2 & 1.000 & 0.454\\
GTSRB     & ResNet18 & 0.997 & 0.745 \\
\bottomrule
\end{tabular}

\end{table}

% \begin{figure*}

% \rotatebox{90}{\makebox[3.5cm][c]{\textbf{\scriptsize GuidedBackProp}}}
%       \subfloat[CIFAR-10]{\includegraphics[width=0.27\textwidth]{figs/C10_GuidedBackProp.pdf}}
% \subfloat[CIFAR-100]{\includegraphics[width=0.27\textwidth]{figs/C100_GuidedBackProp.pdf}}
% \subfloat[GTSRB]{\includegraphics[width=0.27\textwidth]{figs/GTSRB_GuidedBackProp.pdf}}
% \subfloat[Legend]{\includegraphics[width=0.16\textwidth]{figs/Smaplegned.pdf}}
%       \caption{\textbf{\sysname{} hardening optimization Pareto front  illustrations for GuidedBackProp}}
%     \label{fig:XAI_pareto_appendix}
% \end{figure*}

\begin{figure*}[h]
    % \vspace{-0.5cm}
    \centering
     % Set caption format for subfloats in this figure
    \captionsetup[subfloat]{labelformat=empty}
    
    % Set subrefformat for subcaptions in this figure
\noindent

\rotatebox{90}{\makebox[3.5cm][c]{\textbf{\scriptsize Input $\times$ Grad }}}
     \subfloat[]{\includegraphics[width=0.27\textwidth]{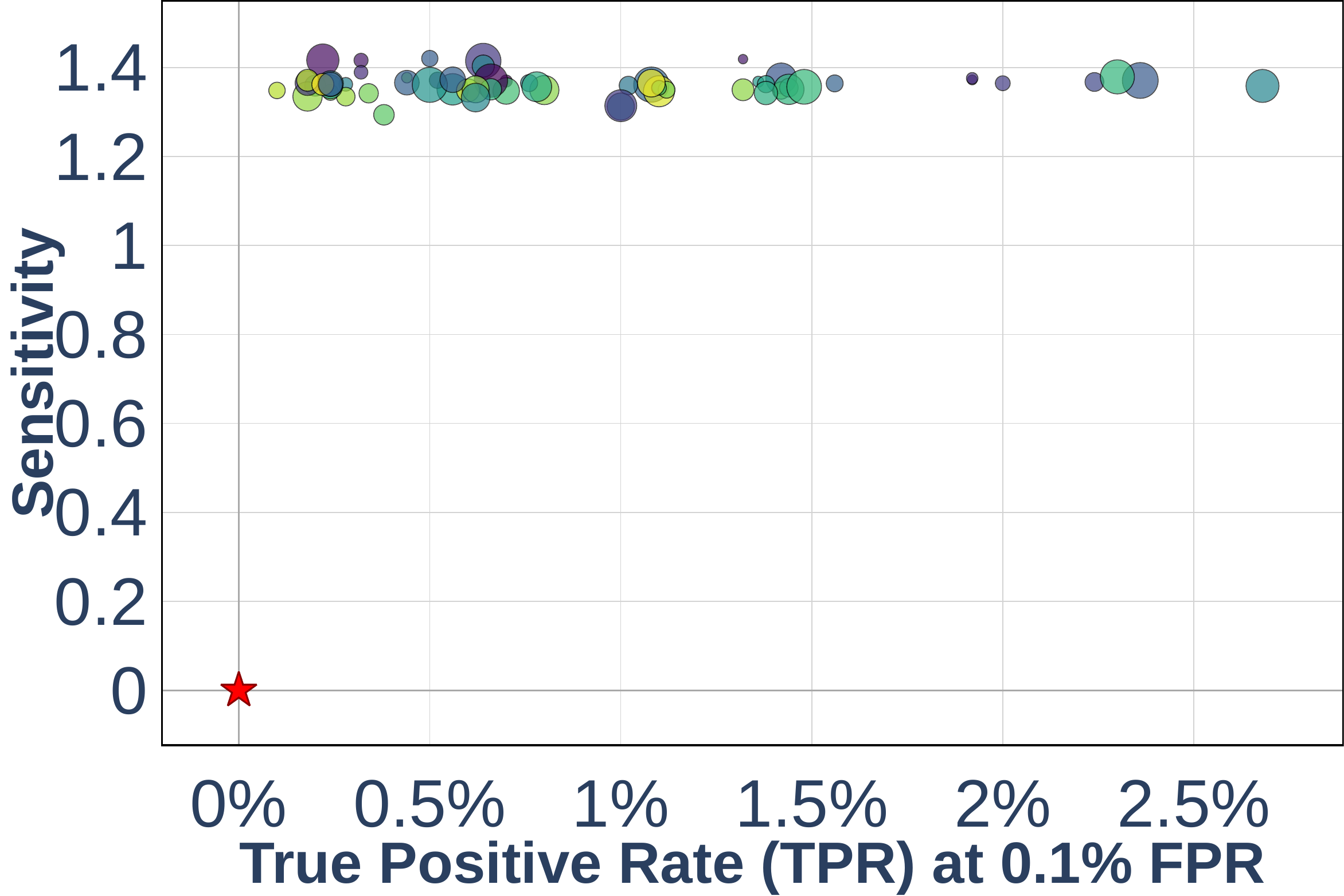}}
\subfloat[]{\includegraphics[width=0.27\textwidth]{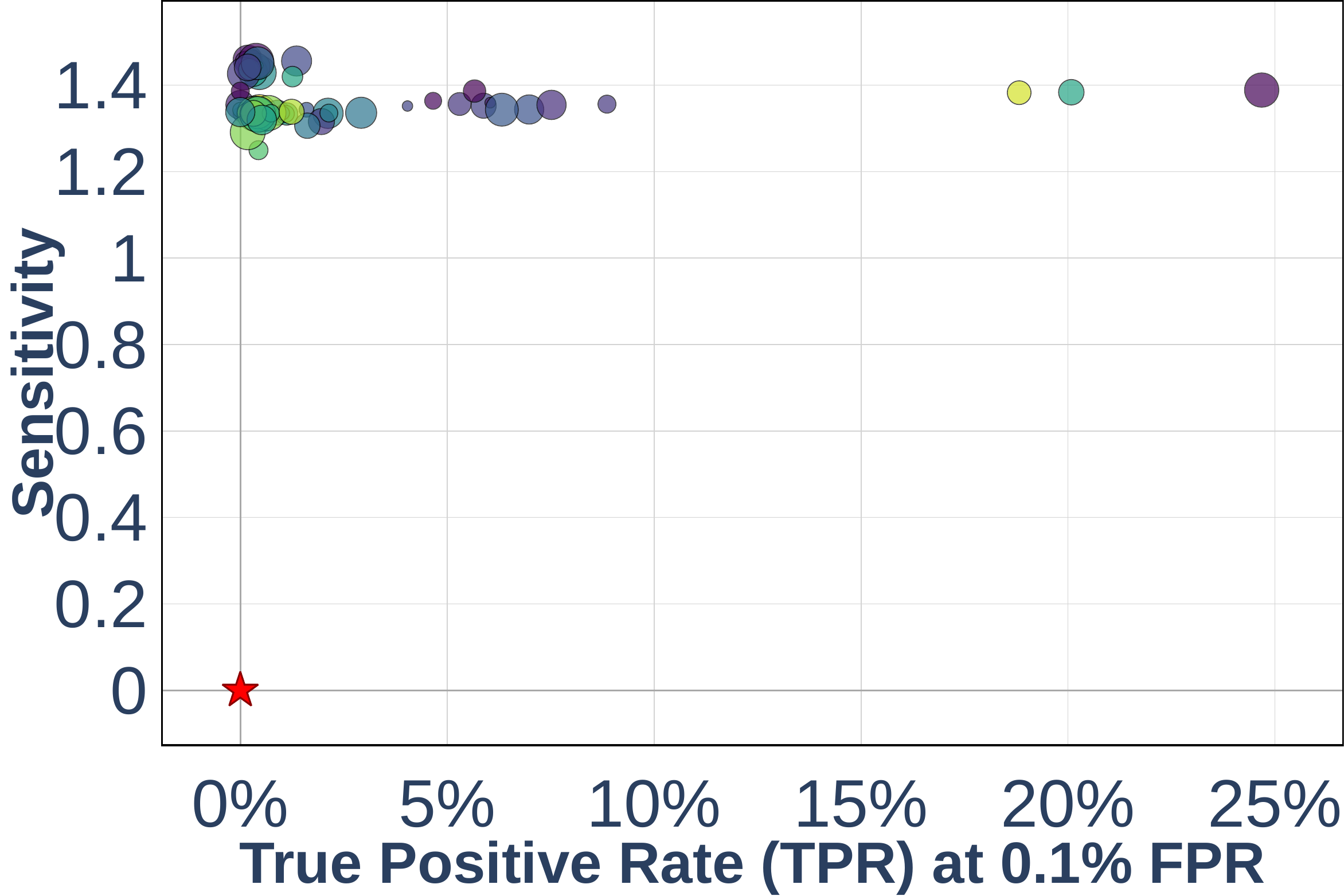}}
\subfloat[]{\includegraphics[width=0.27\textwidth]{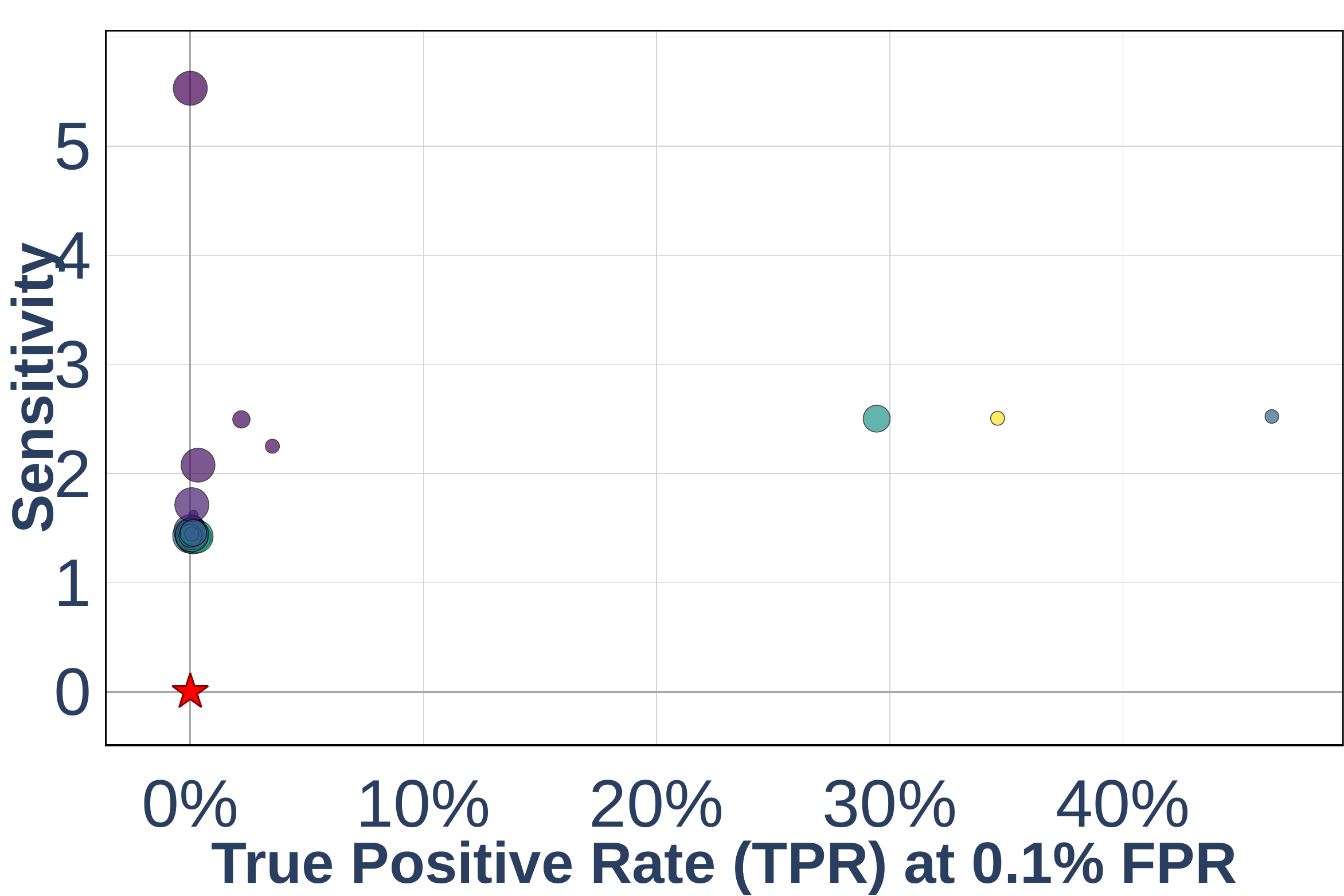}}
\subfloat[]{\includegraphics[width=0.16\textwidth]{figs/Smaplegned.pdf}}
     \\
     \vspace{-0.8cm}

     \noindent
\rotatebox{90}{\makebox[3.5cm][c]{\textbf{\scriptsize Deconvolution}}}
          \subfloat[]{\includegraphics[width=0.27\textwidth]{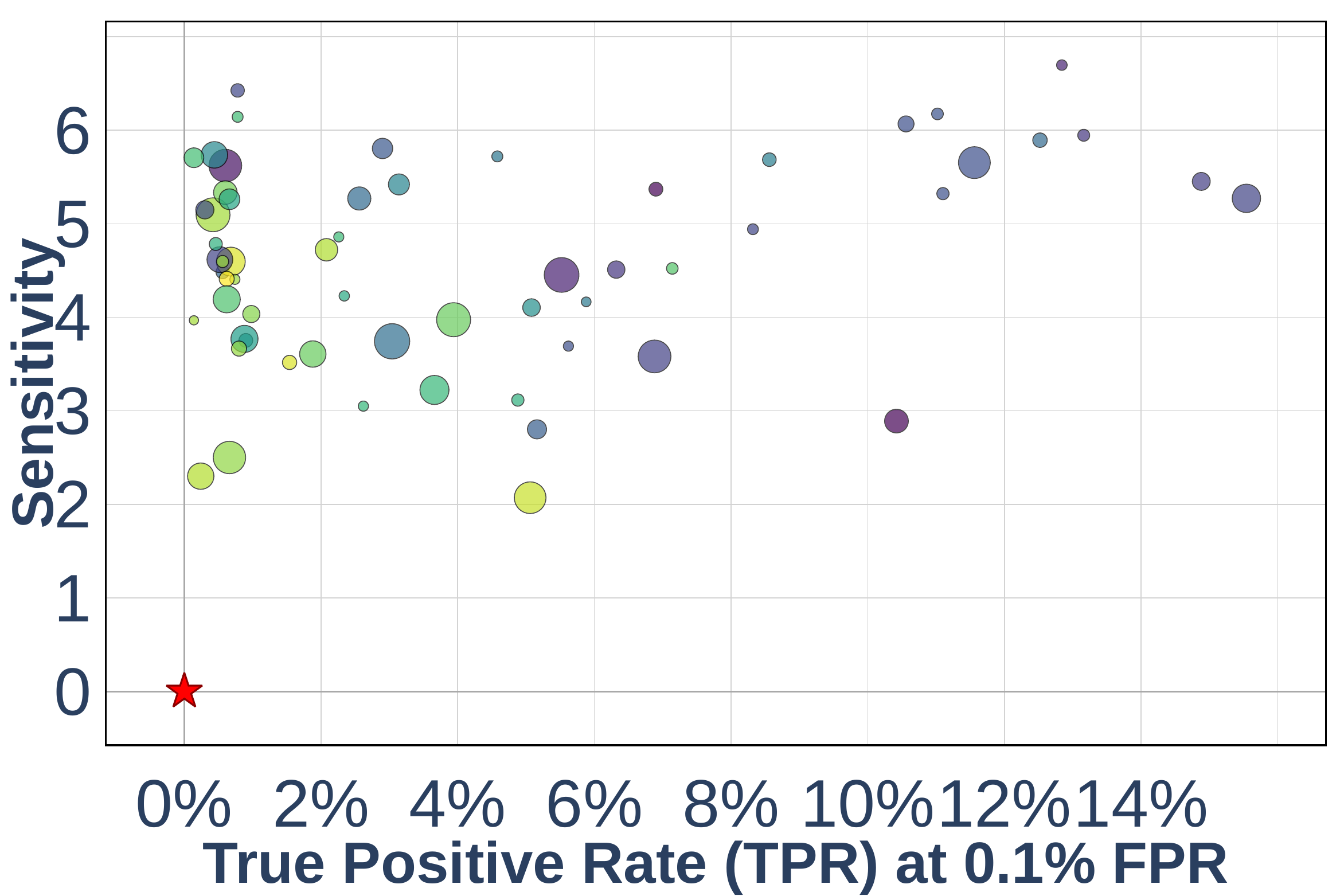}}
\subfloat[]{\includegraphics[width=0.27\textwidth]{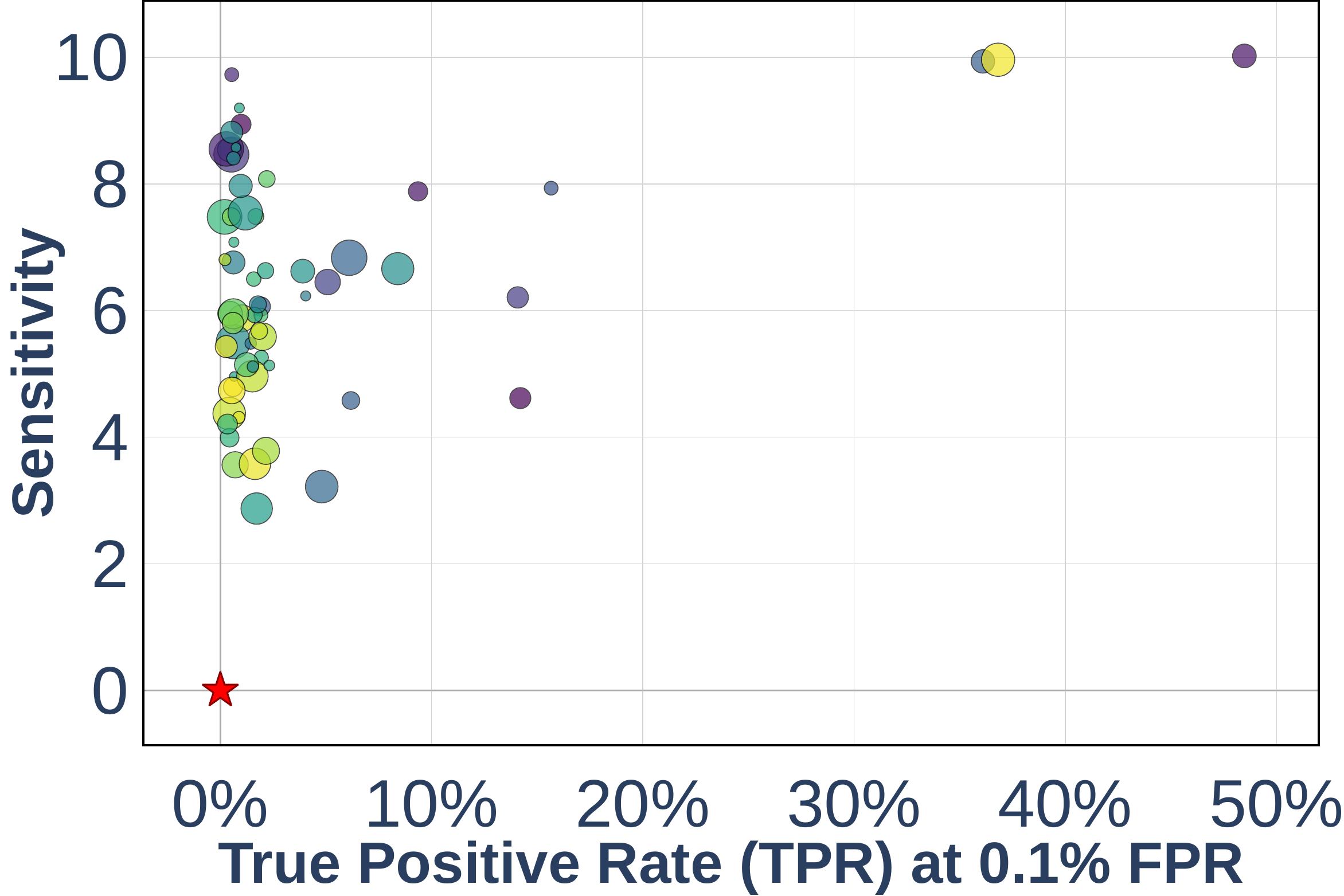}}
\subfloat[]{\includegraphics[width=0.27\textwidth]{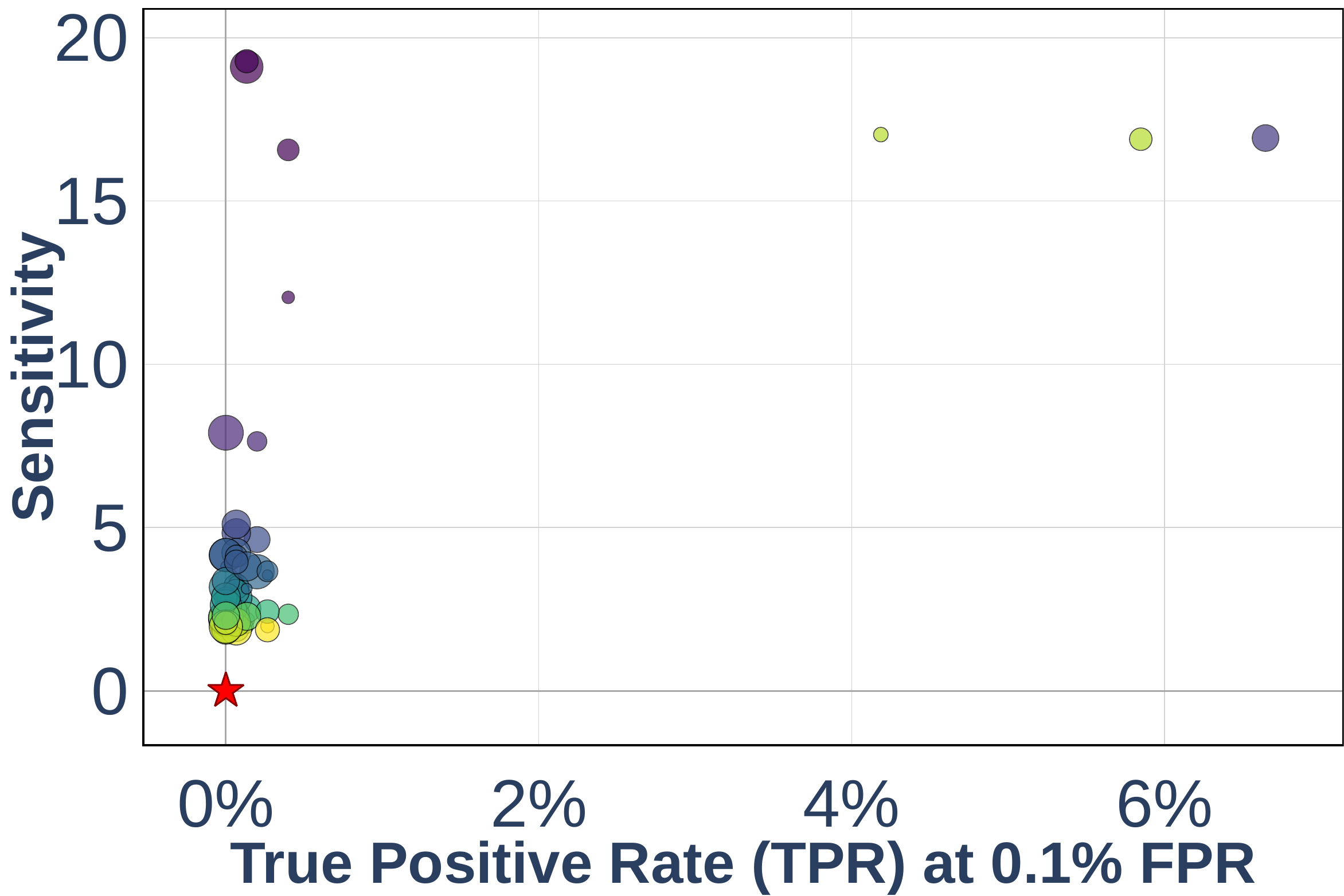}}
\subfloat[]{\includegraphics[width=0.16\textwidth]{figs/Smaplegned.pdf}}
         \\
         \vspace{-0.8cm}
     \noindent
\rotatebox{90}{\makebox[3.5cm][c]{\textbf{\scriptsize SmoothGrad}}}
          \subfloat[]{\includegraphics[width=0.27\textwidth]{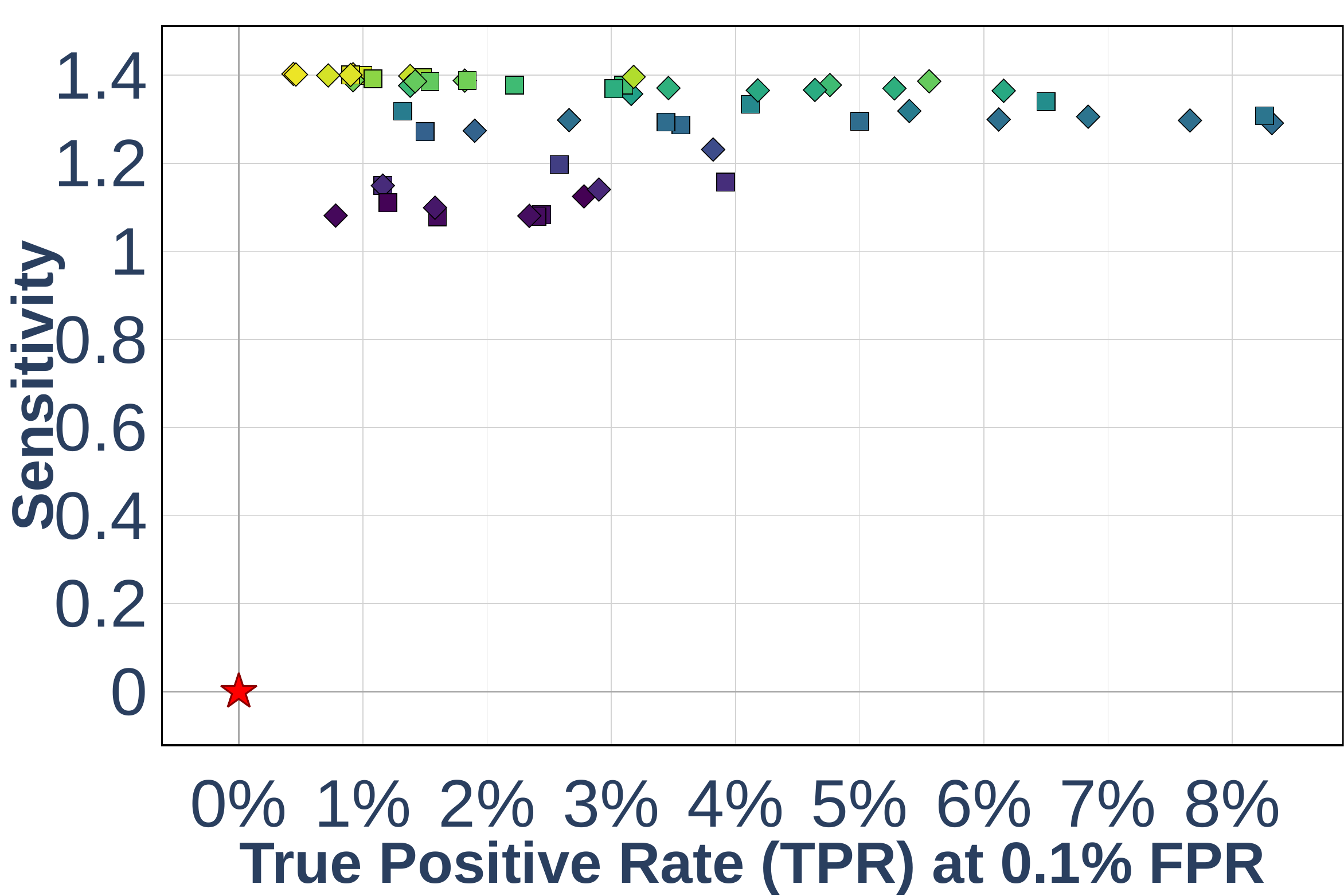}}
\subfloat[]{\includegraphics[width=0.27\textwidth]{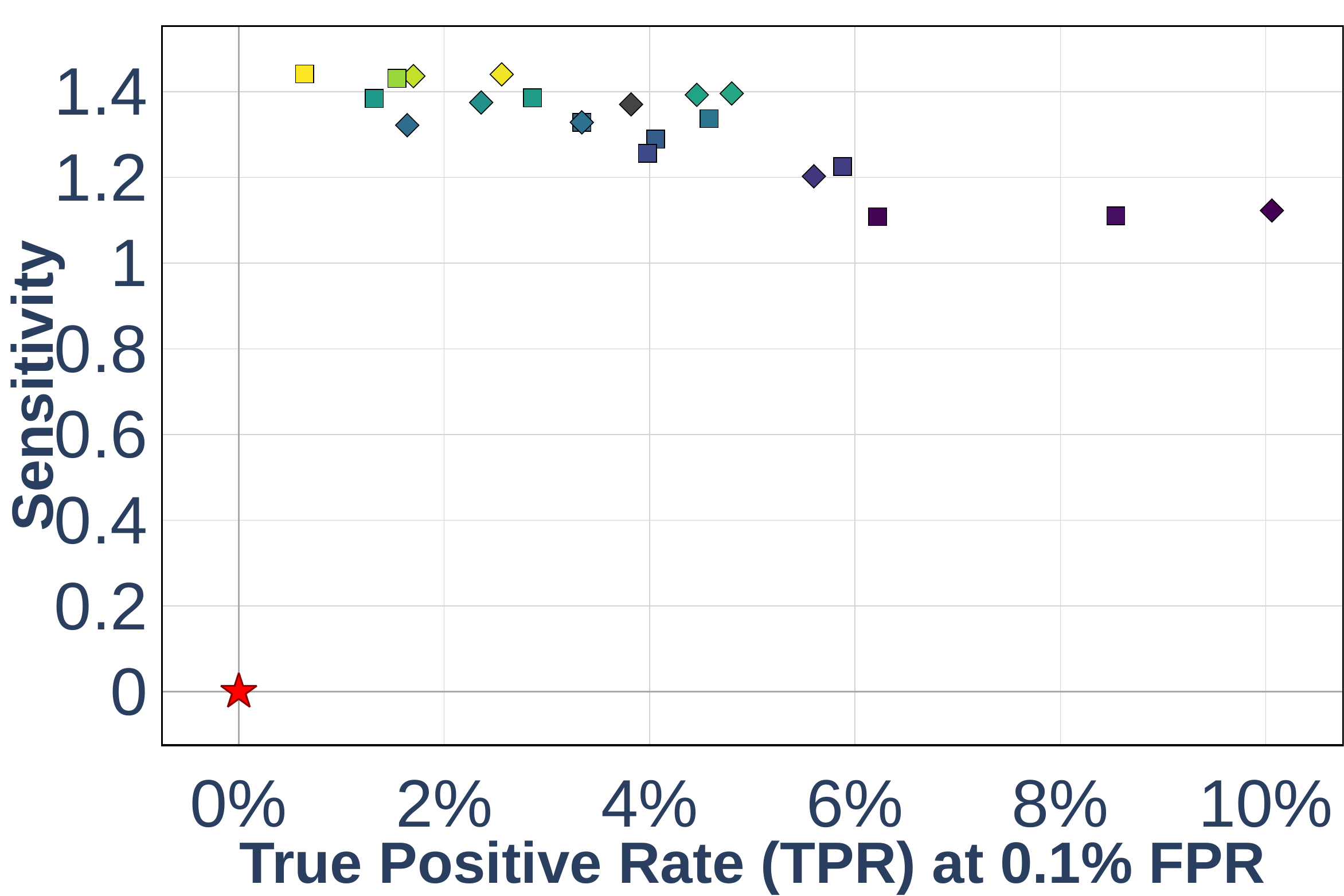}}
\subfloat[]{\includegraphics[width=0.27\textwidth]{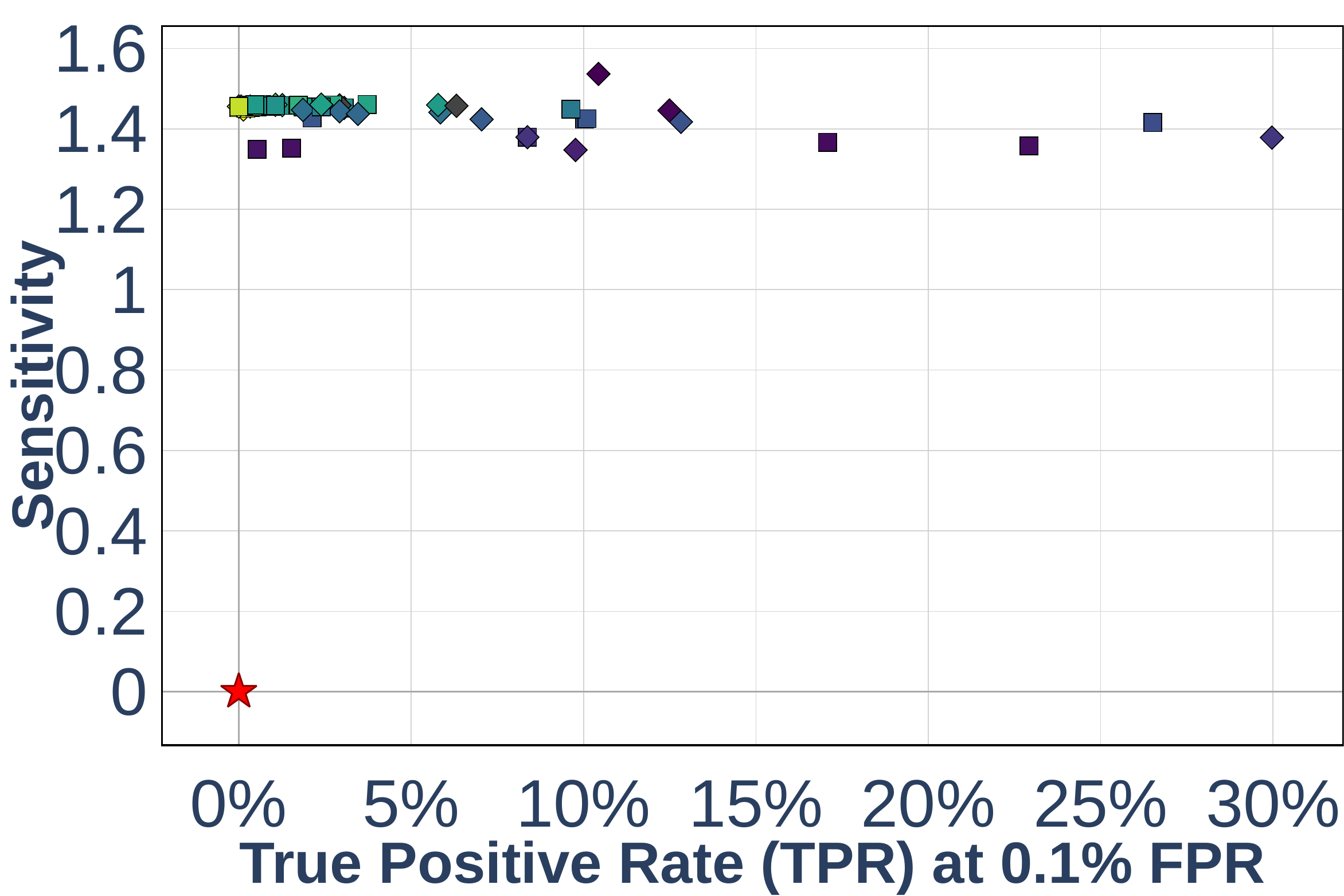}}
\subfloat[]{\includegraphics[width=0.17\textwidth]{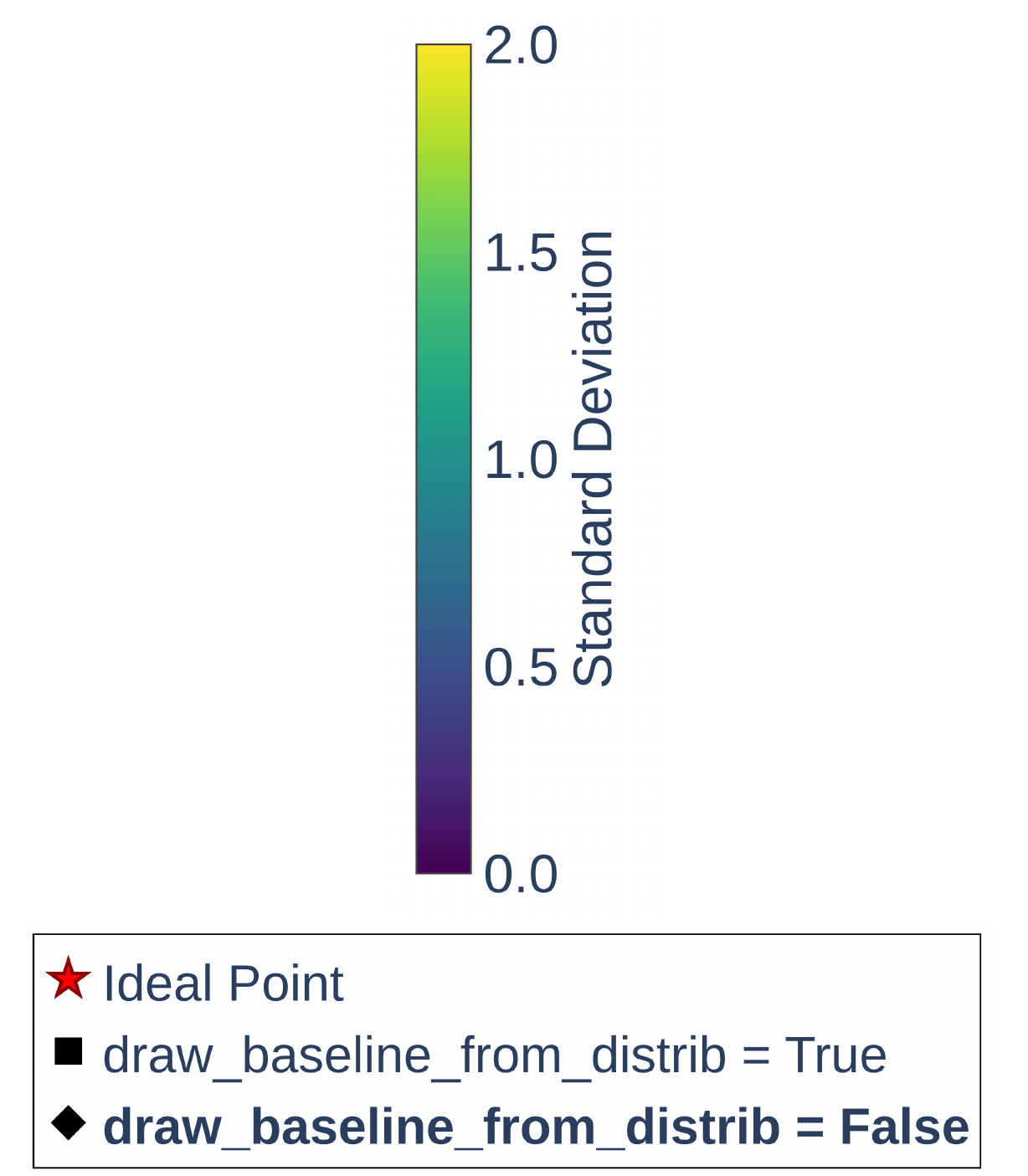}}
     \\
     \vspace{-0.8cm}
     \noindent
\rotatebox{90}{\makebox[3.5cm][c]{\textbf{\scriptsize DeepLift}}}
          \subfloat[]{\includegraphics[width=0.27\textwidth]{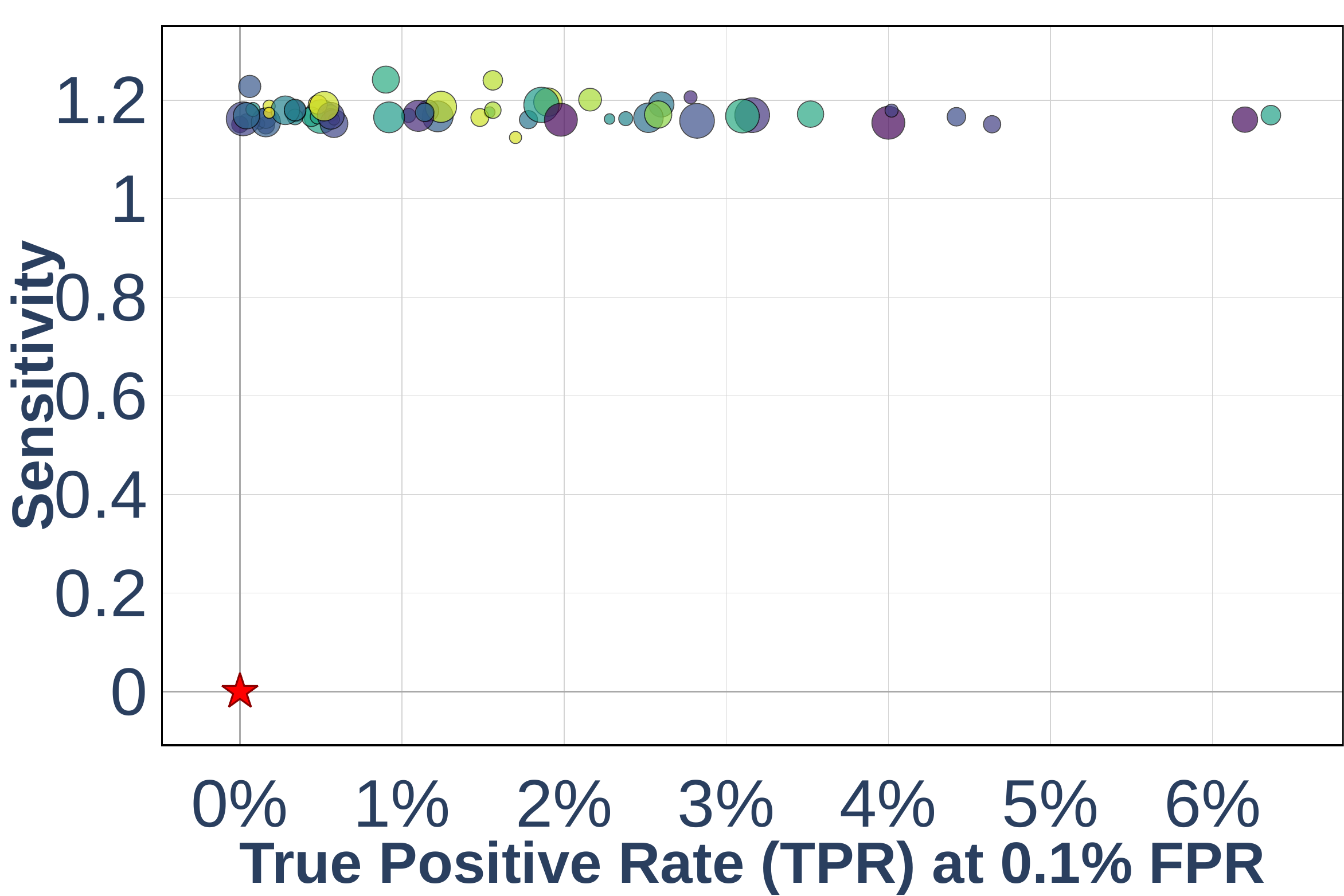}}
\subfloat[]{\includegraphics[width=0.27\textwidth]{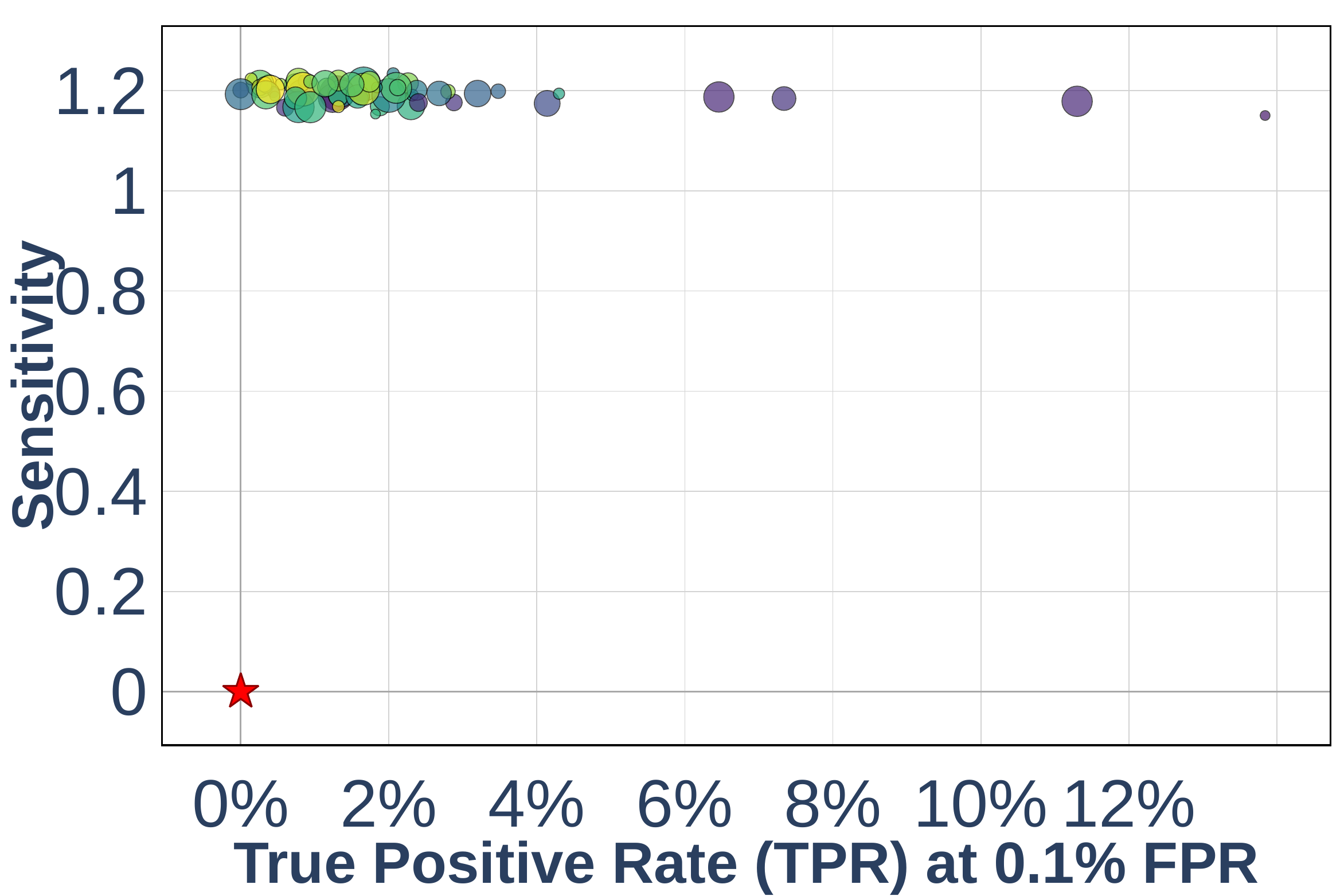}}
\subfloat[]{\includegraphics[width=0.27\textwidth]{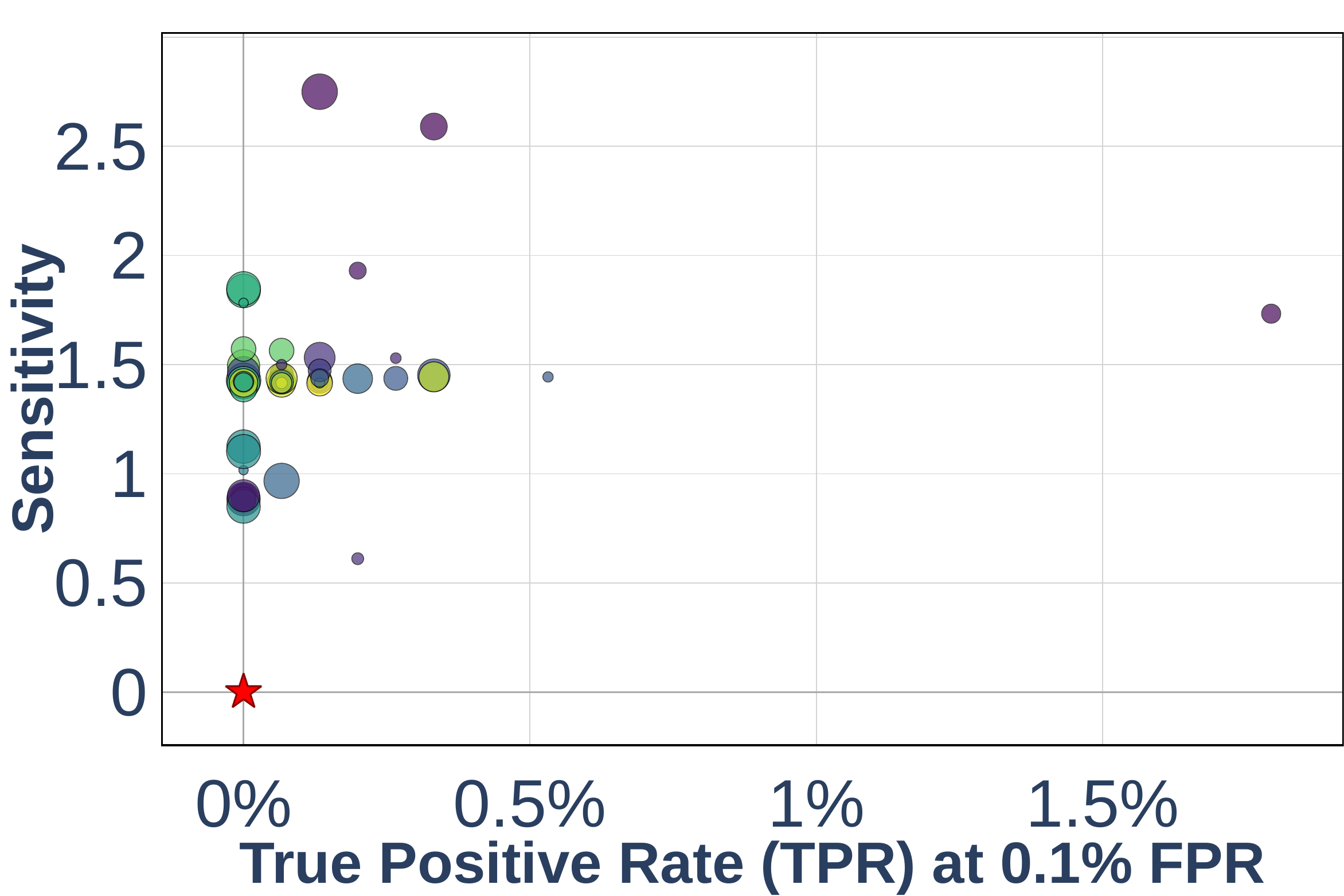}}
\subfloat[]{\includegraphics[width=0.16\textwidth]{figs/Smaplegned.pdf}}
     \\
     \vspace{-0.8cm}
     \noindent
\rotatebox{90}{\makebox[3.5cm][c]{\textbf{\scriptsize Anchors}}}
               \subfloat[]{\includegraphics[width=0.27\textwidth]{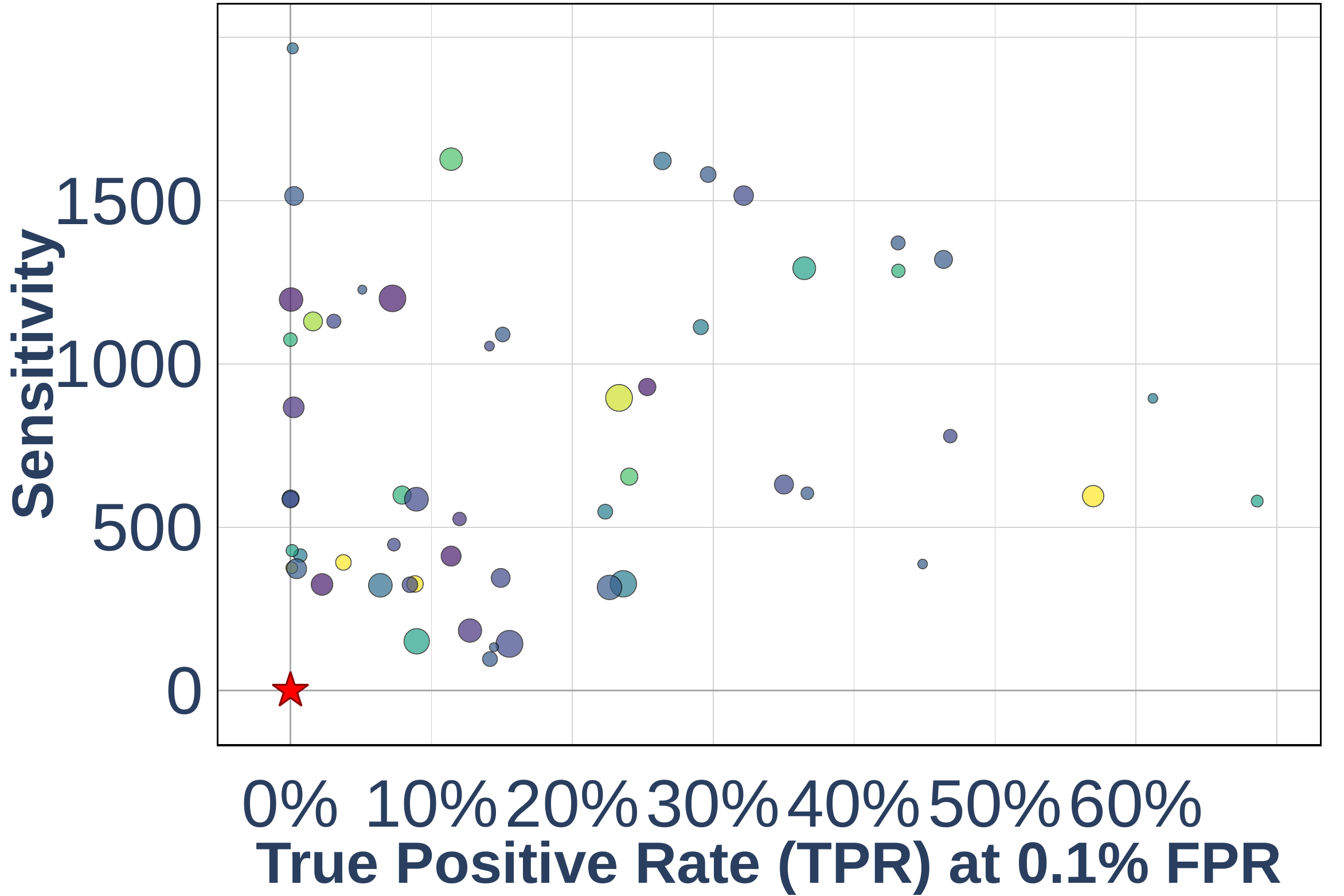}}
\subfloat[]{\includegraphics[width=0.27\textwidth]{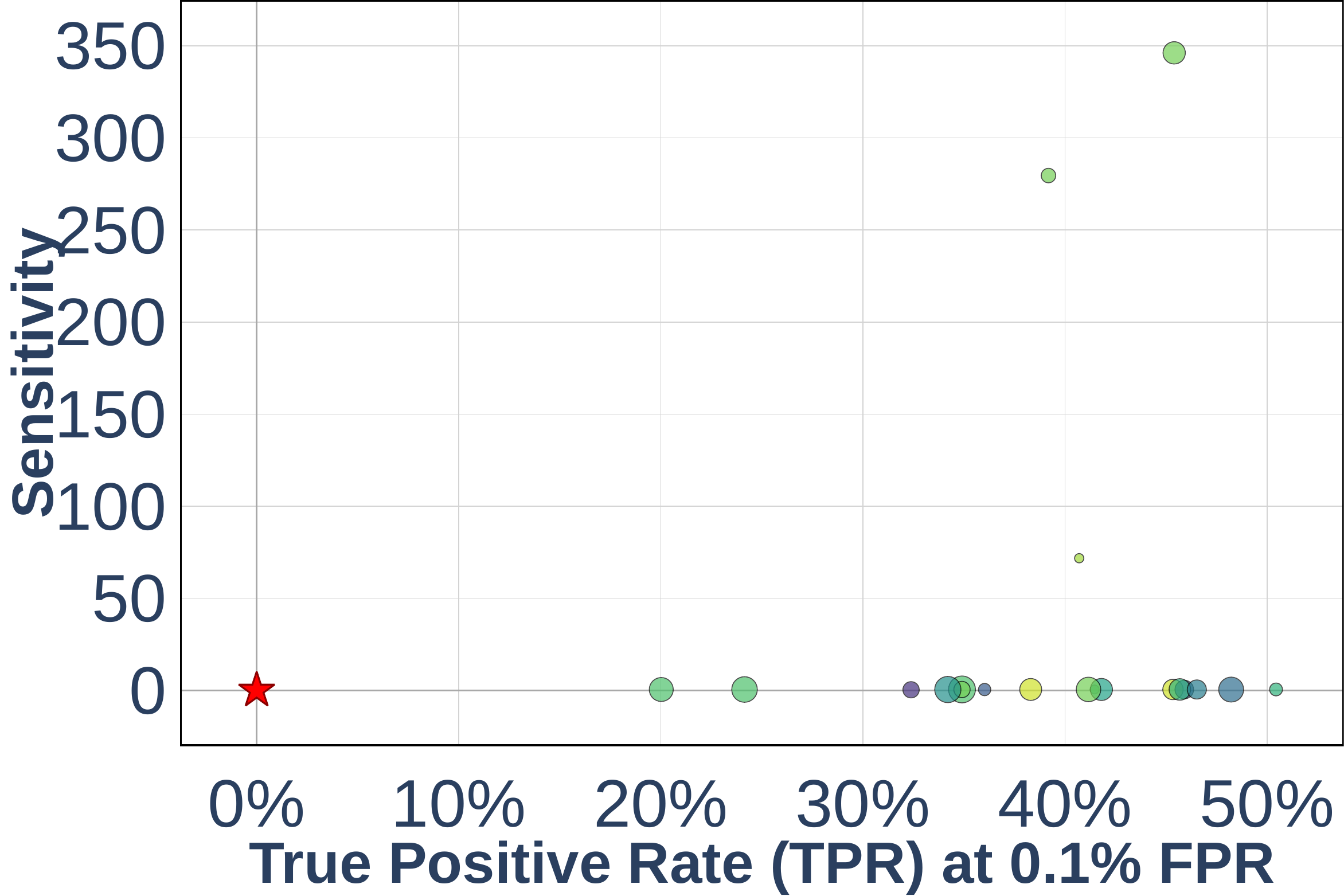}}
\subfloat[]{\includegraphics[width=0.27\textwidth]{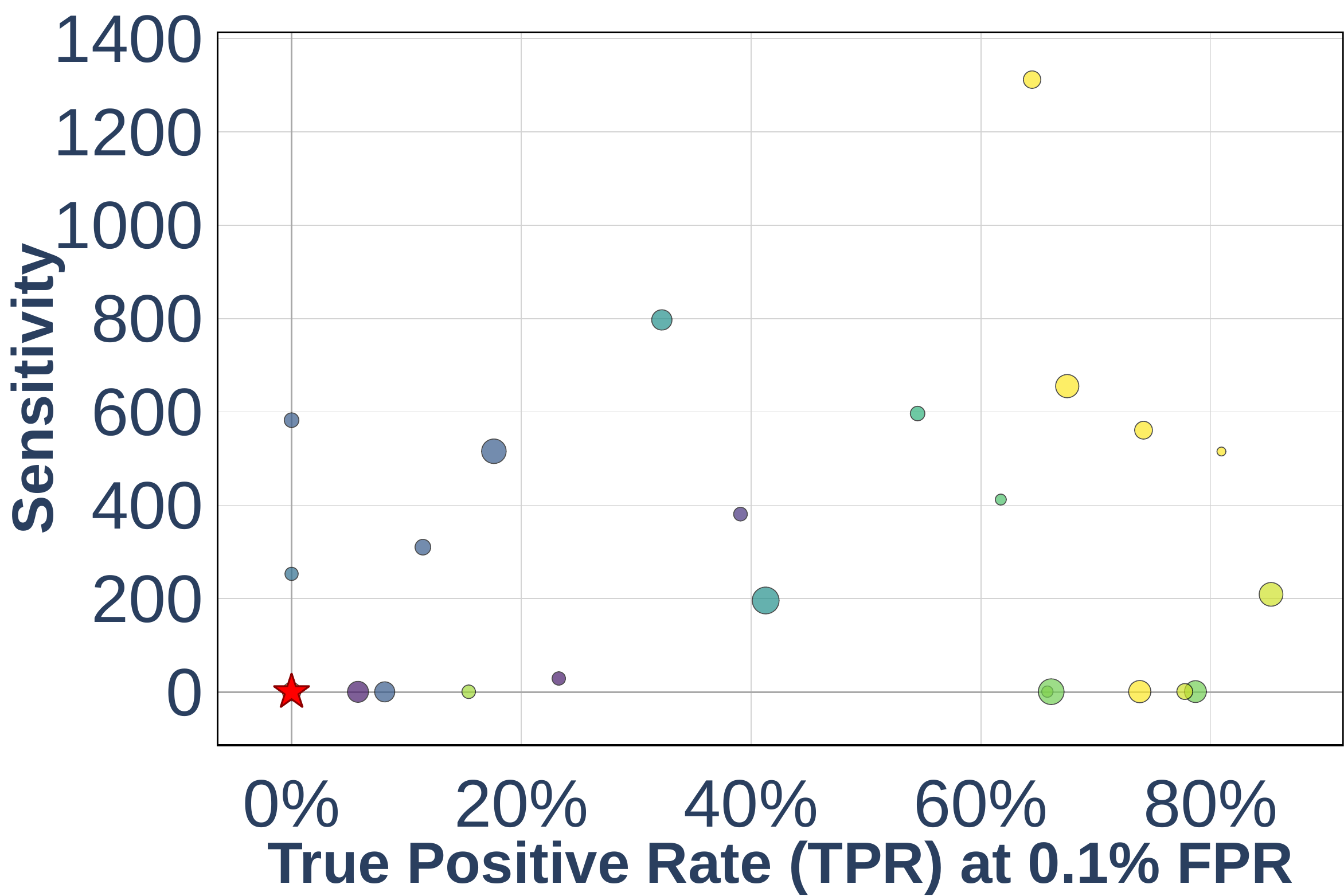}}
\subfloat[]{\begin{minipage}[b]{0.16\textwidth}\centering\includegraphics[width=0.56\textwidth]{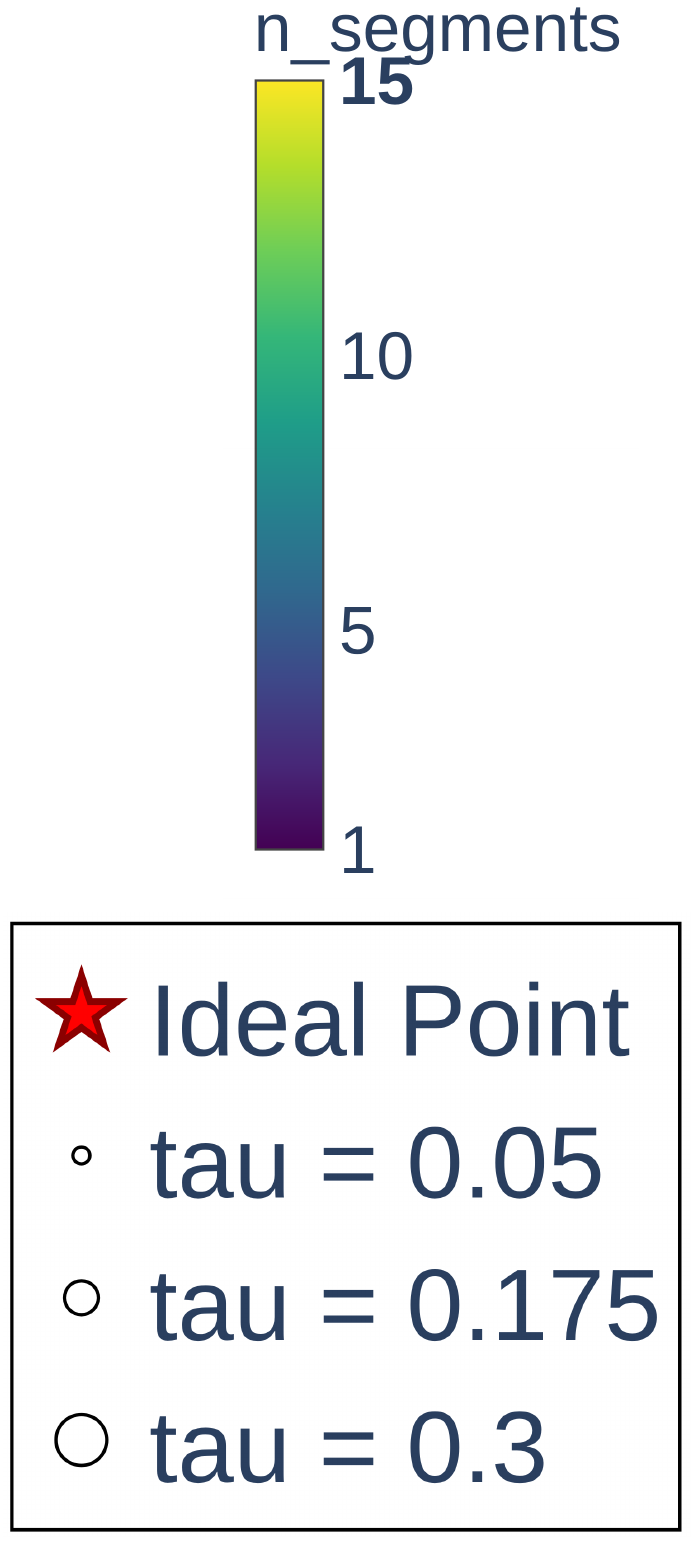}\end{minipage}}

     % \subfloat[]{\includegraphics[width=0.32\textwidth]{evaluation_figs/GTSRB/.pdf}}
     
     \vspace{-0.8cm}
     \noindent
\rotatebox{90}{\makebox[3.5cm][c]{\textbf{\scriptsize VarGrad}}}
    \subfloat[CIFAR-10]{\includegraphics[width=0.27\textwidth]{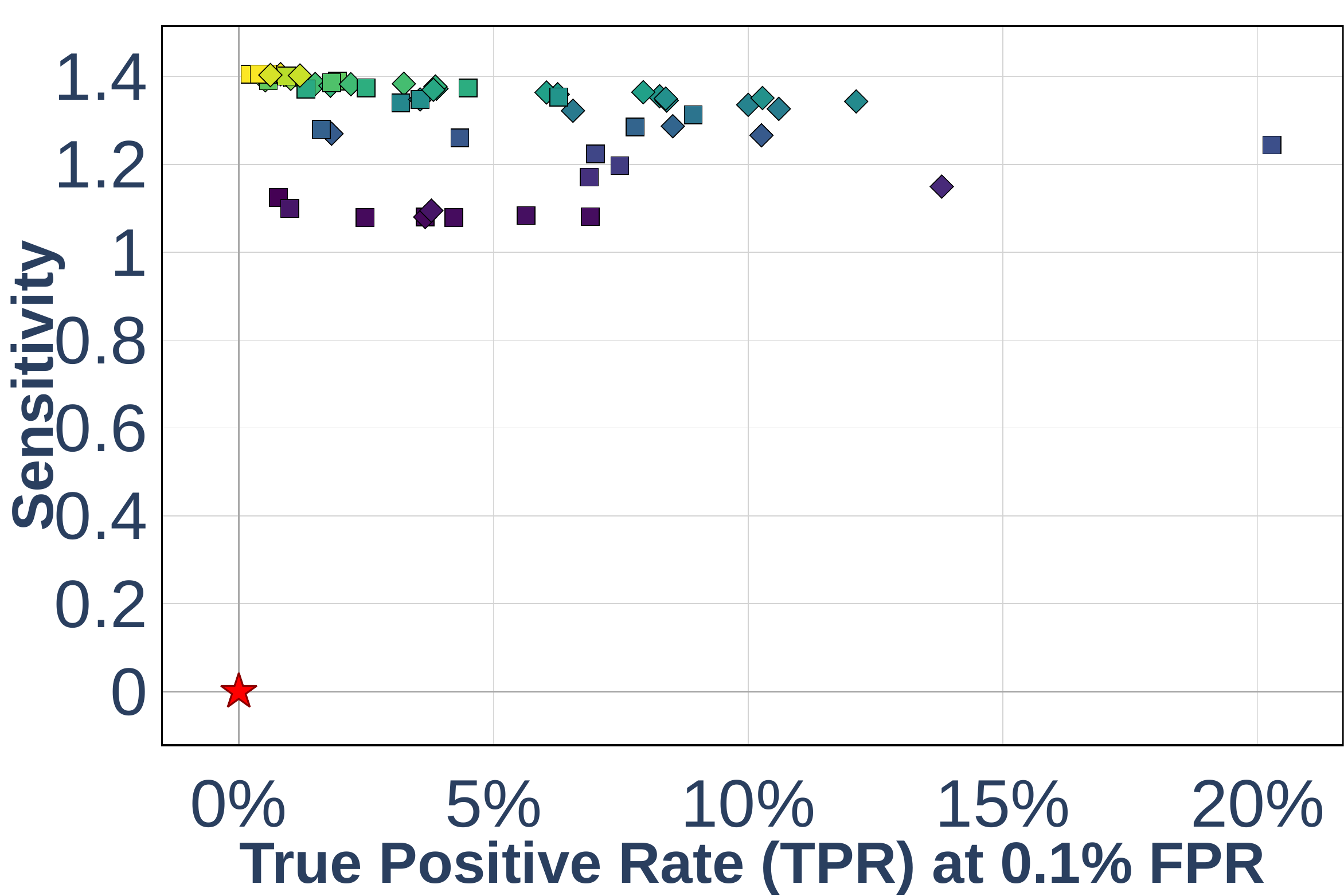}}
\subfloat[CIFAR-100]{\includegraphics[width=0.27\textwidth]{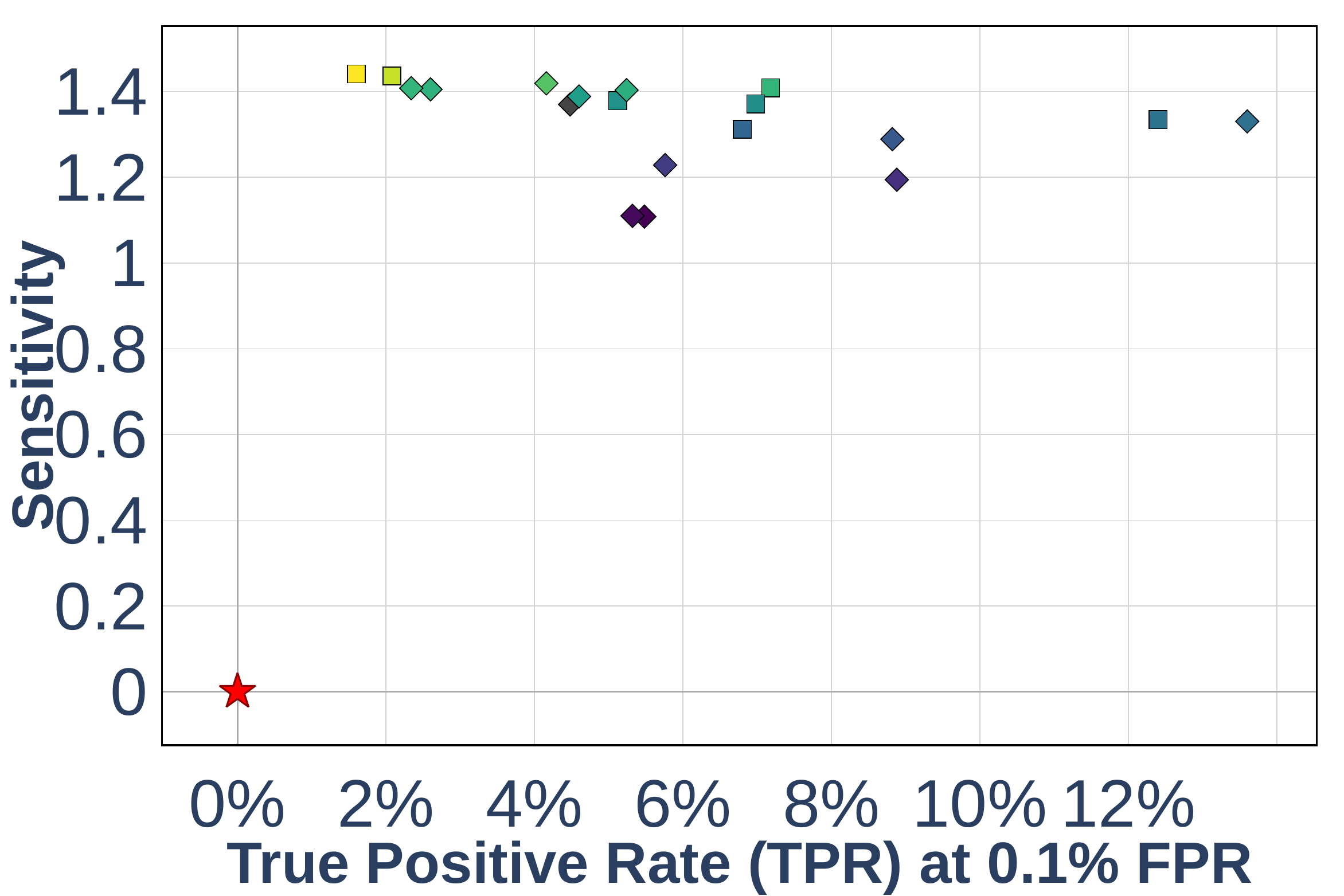}}
\subfloat[GTSRB]{\includegraphics[width=0.27\textwidth]{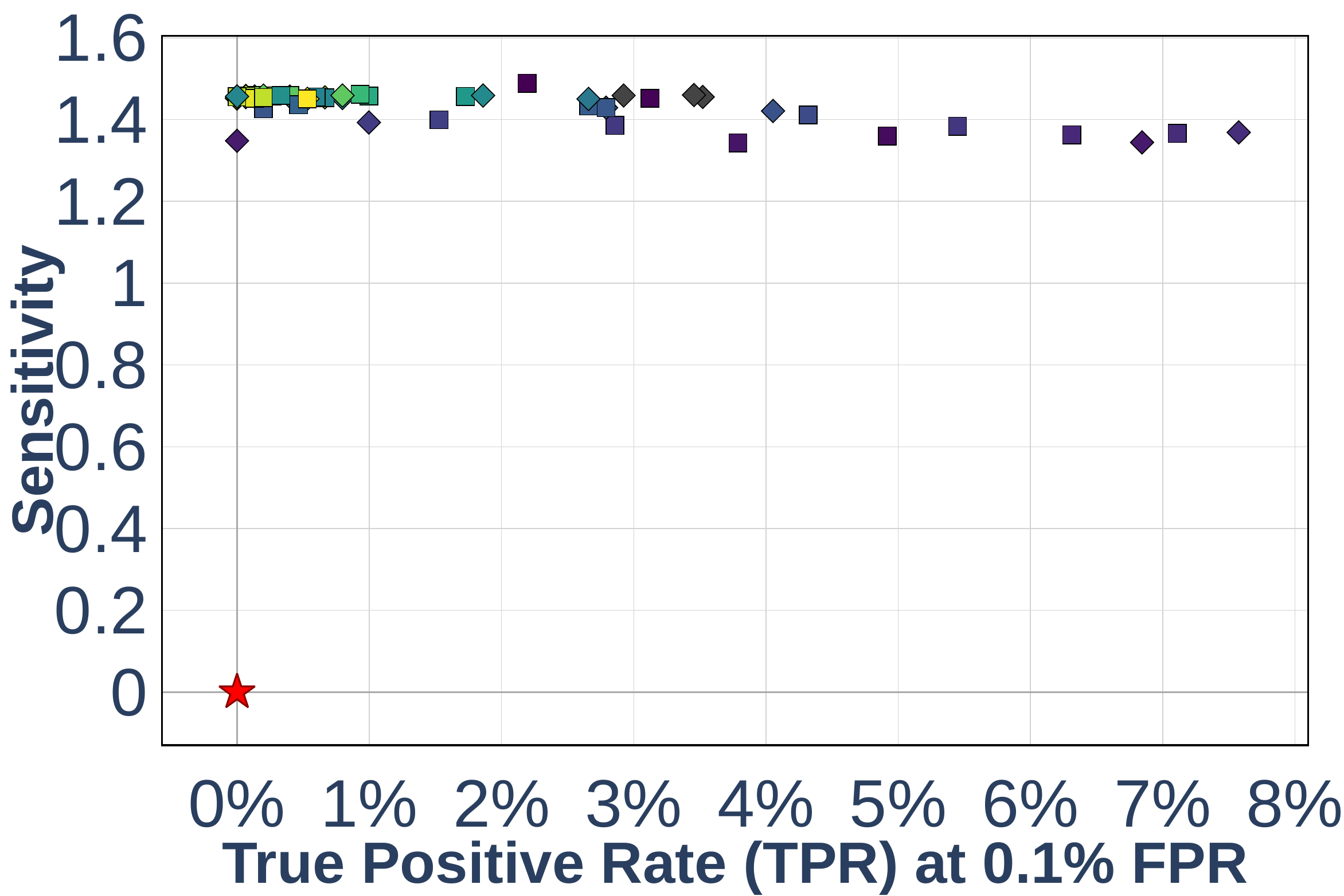}}
\subfloat[Legend]{\includegraphics[width=0.17\textwidth]{figs/vargradlegend.pdf}}\\
%      \noindent
% \rotatebox{90}{\makebox[3.5cm][c]{\textbf{\scriptsize GuidedBackProp}}}
%       \subfloat[CIFAR-10]{\includegraphics[width=0.27\textwidth]{figs/C10_GuidedBackProp.pdf}}
% \subfloat[CIFAR-100]{\includegraphics[width=0.27\textwidth]{figs/C100_GuidedBackProp.pdf}}
% \subfloat[GTSRB]{\includegraphics[width=0.27\textwidth]{figs/GTSRB_GuidedBackProp.pdf}}
% \subfloat[Legend]{\includegraphics[width=0.16\textwidth]{figs/Smaplegned.pdf}}

     % \vspace{-0.1cm}
      \caption{\textbf{\sysname{} hardening optimization Pareto front  illustrations for representative explanation methods.}}
    \label{fig:XAI_pareto_appendix}
     % \vspace{-1.5em}
     
\end{figure*}